\documentclass[preprint,12pt]{elsarticle}

\usepackage{color} 
\usepackage{soul} 
\usepackage{amssymb}
\usepackage{graphicx}
\usepackage{dcolumn}
\usepackage{bm}
\usepackage{pifont}
\usepackage{amsmath}
\usepackage{times}

\journal{Journal of Theoretical Biology}

\begin{document}

\begin{frontmatter}





\title{Stochastic multi-scale models of competition within heterogeneous cellular populations: simulation methods and 
mean-field analysis}

\address[1]{Centre de Recerca Matem\`atica, Edifici C, Campus de Bellaterra, 08193 Bellaterra (Barcelona), Spain.}
\address[2]{Departament de Matem\`atiques, Universitat Aut\`onoma de Barcelona, 08193 Bellaterra (Barcelona), Spain.}
\address[3]{Department of Mathematics, University College London, Gower Street, London WC1E 6BT, UK}
\address[4]{Department of Mechanical Engineering, Massachusetts Institute of Technology, 77 Massachusetts Avenue, Cambridge, MA 02139, USA.}
\address[5]{Department of Biomedical Engineering, Boston University, 44 Cummington Street, Boston MA 02215, USA.}
\address[6]{ICREA (Instituci\'o Catalana de Recerca i Estudis Avan\c{c}ats), Barcelona, Spain.}
\address[7]{Barcelona Graduate School of Mathematics (BGSMath), Barcelona, Spain.}
\author[1,2]{Roberto de la Cruz}
\author[3]{Pilar Guerrero}
\author[4,5]{Fabian Spill}
\author[6,1,2,7]{Tom\'as Alarc\'on}

\date{\today}

\begin{abstract}
We propose a modelling framework to analyse the stochastic behaviour of heterogeneous, multi-scale cellular populations. We illustrate our methodology with a particular example in which we study a population with an oxygen-regulated proliferation rate. Our formulation is based on an age-dependent stochastic process. Cells within the population are characterised by their age (\emph{i.e.} time elapsed since they were borne). The age-dependent (oxygen-regulated) birth rate is given by a stochastic model of oxygen-dependent cell cycle progression. Once the birth rate is determined, we formulate an age-dependent birth-and-death process, which dictates the time evolution of the cell population. The population is under a feedback loop which controls its steady state size (carrying capacity): cells consume oxygen which in turns fuels cell proliferation. We show that our stochastic model of cell cycle progression allows for heterogeneity within the cell population induced by stochastic effects. Such heterogeneous behaviour is reflected in variations in the proliferation rate. Within this set-up, we have established three main results. First, we have shown that the age to the G1/S transition, which essentially determines the birth rate, exhibits a remarkably simple scaling behaviour. Besides the fact that this simple behaviour emerges from a rather complex model, this allows for a huge simplification of our numerical methodology. A further result is the observation that heterogeneous populations undergo an internal process of quasi-neutral competition. Finally, we investigated the effects of cell-cycle-phase dependent therapies (such as radiation therapy) on heterogeneous populations. In particular, we have studied the case in which the population contains a quiescent sub-population. Our mean-field analysis and numerical simulations confirm that, if the survival fraction of the therapy is too high, rescue of the quiescent population occurs. This gives rise to emergence of resistance to therapy since the rescued population is less sensitive to therapy.        
\end{abstract}


\end{frontmatter}

\section{Introduction}\label{sec:intro}

Global cell traits and behaviour in response to stimuli, the so-called phenotype, results from a complex network of interactions between genes and gene products 
which ultimately regulates gene expression. These networks of gene regulation constitute non-linear, high-dimensional dynamical systems whose 
structure has been shaped up by evolution by natural selection, so that they exhibit properties such as robustness (\emph{i.e.} resilience of the phenotype against genetic alterations) and canalisation (\emph{i.e.} the ability for phenotypes to increase their robustness as time progresses). These properties are 
exploited by tumours to increase their proliferative potential and resist to therapies \cite{kitano2004}. In addition to complex, non-linear 
interactions within individual cells, there exist intricate interactions between different components of the biological systems at all levels: From 
complex signalling pathways and gene regulatory networks to complex non-local effects where perturbations at whole-tissue level induce changes at the 
level of the intra-cellular pathways of individual cells \cite{alarcon2005,ribba2006,macklin2009,osborne2010,deisboeck2011,powathil2013,jagiella2016}. 
These and other factors contribute towards a highly complex dynamics in biological tissues which is an emergent property of all the layers of complexity involved.

To tackle such complexity, multi-scale models of biological systems as diverse as cardiac systems \cite{smith2004,mcculloch2009,hand2010,land2013}, systems of developmental biology 
\cite{schnell2008,oates2009,hester2011,setty2012,walpole2013}, and tumour growth systems
\cite{alarcon2005,jiang2005,ribba2006,macklin2009,owen2009,preziosi2009,tracqui2009,byrne2010,lowengrub2010,osborne2010,rejniak2010,deisboeck2011,
perfahl2011,travasso2011,durrett2013,powathil2013,szabo2013,chisholm2015,curtius2015,scott2016,jagiella2016}
have been developed. In parallel to the model development, algorithms and analytic methods are being developed to allow for more efficient analysis and simulation of such models \cite{alarcon2014,spill2015,delacruz2015,spill2016optimisation}. 

In the case of cancer biology, the multi-scale interactions of intracellular changes at the genetic or molecular pathway level and tissue-level heterogeneity can conspire to generate unfortunate effects such as resistance to therapy 
\cite{maley2006a,gillies2012,greaves2012,chisholm2015,asatryan2016}. Heterogeneity plays a major role in the emergence of drug resistance within tumours and
can be of diverse types. There is heterogeneity in cell types due to increased gene mutation rate as a consequence of genomic instability and other 
factors \cite{maley2006a,greaves2012,chisholm2015,asatryan2016}. 
Heterogeneity can also be caused by the complexity of the tumour microenvironment \cite{alarcon2003,gillies2012,chisholm2015}, in which diverse factors such as tumoural or immune cells \cite{kalluri2006fibroblasts,grivennikov2010immunity}, or the extracellular matrix and its physical properties \cite{spill2016impact}, strongly influence cancer cell behaviour. Note that hypoxia is also known to change the tumour microenvironment \cite{spill2016impact}. In either case, heterogeneity within the tumour creates the necessary conditions 
for resistant varieties to emerge and be selected upon the administration of a given therapy. 

The main aim of this paper is to analyse the properties 
of heterogeneous populations under the effects of fluctuations both within the intracellular pathways which regulate (individual) cell behaviour and 
those associated to intrinsic randomness due to finite size of the population. To this purpose, we expand upon the stochastic multi-scale methodology developed in \cite{guerrero2013}, where it was shown that such a system can be 
described by an age-structured birth-and-death process. The coupling between intracellular and the birth-and-death dynamics is carried out through a 
novel method to obtain the birth rate from the stochastic cell-cycle model, based on a mean-first passage time approach. Cell proliferation is assumed 
to be activated when one or more of the proteins involved in the cell-cycle regulatory pathway hit a threshold. This view allows us to calculate the 
birth rate as a function of the age of the cell and the extracellular oxygen in terms of the associated mean-first passage time (MFPT) problem 
\cite{redner2001}. The present approach differs from that in \cite{guerrero2013} in that our treatment of the intracellular MFPT is done in terms of a 
large deviations approach, the so-called optimal path theory \cite{bressloff2014b}.

This methodology allows us to explore the effects of intrinsic fluctuations within the intracellular dynamics, in particular a model of the 
oxygen-regulated G1/S which dictates when cells are prepared to divide, as a source of heterogeneous behaviour: fluctuations induce variability in 
the birth rate within the population (even to the point of rendering some cells quiescent, \emph{i.e.} stuck in G0) upon which a cell-cycle dependent 
therapy acts as a selective pressure. 

This paper is organised as follows. Section \ref{sec:summary} provides a summary of the structure of the multi-scale. In Section \ref{sec:intra}, we give a detailed discussion of the intracellular dynamics, \emph{i.e.} the stochastic model of the 
oxygen-regulated G1/S transition, and its analysis. In Section \ref{sec:MSME}, we summarise the formulation of the age-structured birth-and-death 
process, the numerical simulation technique, and the mean-field analysis of a homogeneous population. In Section \ref{sec:iec}, we discuss how 
noise within the intracellular dynamics induces heterogeneity in the population and analyse the stochastic dynamics of competition for a limited 
resource within such heterogeneous populations. In Section \ref{sec:therapy} we further study the effects of noise-induced heterogeneity on the 
emergence of drug resistance upon administration of a cell cycle-specific therapy. Finally, in Section \ref{sec:discussion} we summarise our results 
and discuss our conclusions as well as avenues for future research.

\section{Summary of the multi-scale model}\label{sec:summary}

\begin{figure}
\begin{center}
\includegraphics[scale=0.3]{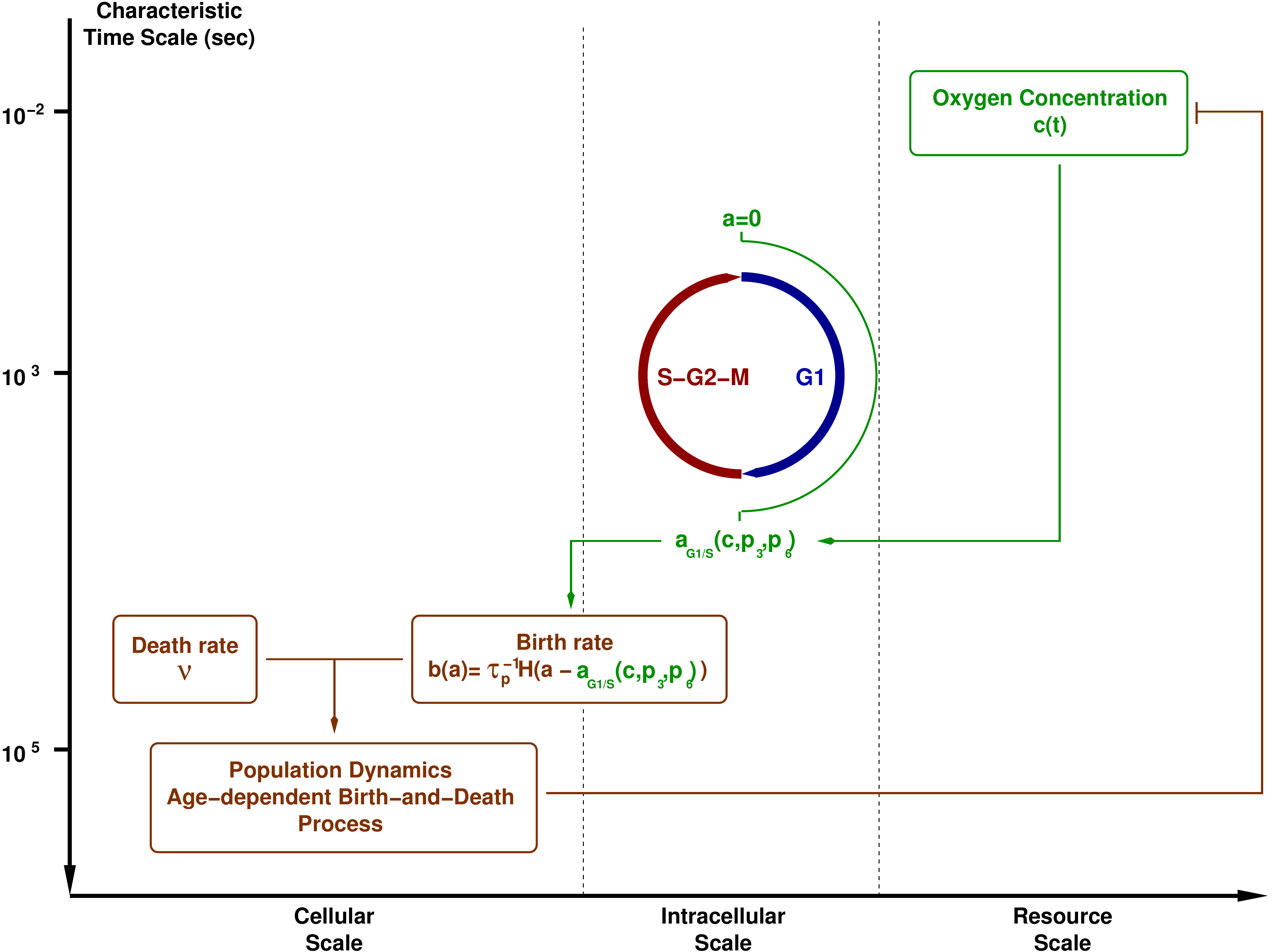}
\caption{Schematic representation of the different elements that compose our multi-scale model. We show the different levels of biological 
organisation as well 
as associated characteristic time scales \cite{guerrero2013} associated to each of these layers: resource scale, \emph{i.e.} oxygen which is supplied at a 
constant 
rate and consumed by the cell population, cellular scale, \emph{i.e.} oxygen-regulated cell cycle progression which determines the age-dependent birth rate 
into the cellular layer, and, finally, the cellular scale, which is associated to the stochastic population dynamics.}\label{fig:scheme}
\end{center} 
\end{figure}

Before proceeding with a detailed discussion of the different elements involved in the formulation of the stochastic multi-scale model, it is useful 
to provide a general overview of the overall structure of the model, which is closely related to that of the model proposed in \cite{alarcon2005}.

The model we present in this article integrates phenomena characterised by different time scales, as schematically shown 
in Fig. \ref{fig:scheme}. This model intends to tackle the growth and competition of cellular populations under the restriction of finite 
amount of available resources (in this case, oxygen) supplied at a finite rate, $\bar{S}$.  
    
The general approach used in this model is a natural generalisation of the standard continuous-time birth-and-death Markov process and its 
description via a Master Equation \cite{gardiner2009}. As we will see, the consideration of the multi-scale character of the system, \emph{i.e.} the 
inclusion of the physiological structure associated with the cell-cycle variables, introduce an age-structure within the population: the birth rate 
depends on the age of cell (\emph{i.e.} time elapsed since last division) which determines, through the corresponding cell-cycle model, the cell-cycle 
status of the corresponding cells.  

The evolution of the concentration of oxygen, $c(t)$, (\emph{resource scale}, see Fig. 
\ref{fig:scheme}) is modelled by:

\begin{equation}\label{eq:oxygen}
\frac{d c}{dt}=\bar{S}-\bar{k} c\sum_{i=1}^{N_T}N_i(t) 
\end{equation}
 
\noindent where $N_T$ is the number of cellular types consuming the resource $c$, and $N_i(t)$, $i=1,\dots,N_T$, is the number of cells of type $i$ 
at time $t$. Note that, in general, $N_i(t)$ is a stochastic process, and, therefore, in principle Eq. (\ref{eq:oxygen}) should be treated as a 
stochastic differential equation \cite{Oksendal2003}.

The second sub-model considered in our multi-scale model, associated with the intracellular scale (see Fig. \ref{fig:scheme}), is an stochastic 
model of oxygen-regulated cell-cycle progression. This sub-model is formulated using the standard techniques of chemical 
kinetics modelling \cite{gillespie1976} so that the mean-field limit of the stochastic model corresponds to the deterministic cell-cycle model 
formulated in \cite{bedessem2014}. This model provides the physiological state of the cell in terms of the number of molecules 
of each protein involved in the cell-cycle of a cell of a given age, $a$. From such physiological state, we derive whether the G1/S transition 
has occurred. The cell-cycle status of a cell of age $a$ is determined in terms of whether the abundance of certain proteins which activate the 
cell-cycle (cyclins) have reached a certain threshold. In our particular case, if at age $a$, the cyclin levels are below the corresponding threshold, 
the cell is still in G1. If, on the contrary, the prescribed threshold level has been reached, the cell has passed 
onto S, and therefore is ready to divide. This implies that the probability of a cell having crossed the threshold of cyclin levels at age $a$ can be 
formulated in terms of a mean first-passage time problem (MFTP) in which one analyses the probability of a Markov process to hit a certain boundary 
\cite{redner2001,gardiner2009}. Unlike our approach in \cite{guerrero2013}, based on approximating the full probability distribution of the 
stochastic cell 
cycle model, in the present approach,  passage time is (approximately) solved in terms of an optimal trajectory path approach \cite{bressloff2014b}. 

At the interface between the intracellular and cellular scales sits our model of the (age-dependent) birth rate, which defines the probability of 
birth per unit time (cellular scale) in terms of the cell cycle variables (intracellular scale). The rate at which our cell-cycle model hits the 
cyclin activation threshold, \emph{i.e.} the rate at which cells undergo G1/S transition, is taken as proportional to the birth 
rate. In particular, the birth rate is taken to be function of the age of the cell as well as the concentration of oxygen, as the oxygen abundance 
regulates the G1/S transition age, $a_{G1/S}(c)$, \emph{i.e.} the time (age) elapsed between the birth of a cell and its G1/S transition:

\begin{equation}\label{eq:birthratesummary}
b(a)=\tau_p^{-1}H(a-a_{G1/S}(c)). 
\end{equation}

\noindent \emph{i.e.} cell division occurs at a constant rate, $\tau_p^{-1}$, provided cells have undergo the G1/S transition and H is the Heaviside function. In other words, we 
consider that the duration of the G1 phase is regulated by the cell cycle model, whereas the duration of the S-G2-M is a random variable, 
exponentially distributed with average duration equal to $\tau_p$ (see Fig. \ref{fig:scheme}).  

The third and last sub-model is that associated with the cellular scale. It corresponds to the dynamics of the cell population and it is governed the 
Master Equation for the probability density function of the number of cells \cite{gardiner2009}. The stochastic process that describes the dynamics of 
the population of cells is an age-dependent birth-and-death process where the birth rate is given by Eq. (\ref{eq:birthratesummary}) where 
$a_{G1/S}(c)$ is provided by the intracellular model. The death rate is, for simplicity, considered constant. As a consequence of the fact that the 
birth rate is age-dependent, our Multi-Scale Master Equation (MSME) does not present the standard form for unstructured populations, rather, it is an 
age-dependent Master Equation.

A detailed description of each sub-model and its analysis is given in Sections \ref{sec:intra} and \ref{sec:MSME}.

\section{Intracellular scale: Stochastic model of the oxygen-regulated G$_1$/S transition}\label{sec:intra}

\subsection{Biological background}

Cell proliferation is orchestrated by a complex network of protein and gene expression regulation, the so-called cell cycle, which accounts for the 
timely coordination of proliferation with growth and, by means of signalling cues such as growth factors, tissue function \cite{yao2014}.  Dysregulation of such orderly organisation of cell proliferation is one of the main contributors to the aberrant behaviour observed in tumours 
\cite{weinberg2007}.

The cell cycle has the purpose of regulating the successive activation of the so-called cyclin-dependent 
kinases (CDKs) which control the progression along the four phases of cycle: G1 (first gap phase), S (DNA 
replication), G2 (second gap phase), and M (mitosis) \cite{gerard2009,gerard2011,gerard2015}. This four phases must be supplemented with a fifth, G0, 
which account for cells that are quiescent due to lack of stimulation (\emph{i.e.} absence of growth factors, lack of basic nutrients, etc.) to 
proliferate. Recent models of the cell cycle organise the complex regulatory network into CDK modules, each centred around a cyclin-CDK complex 
which is key for the transition between the cell cycle phases (see, for example, \cite{gerard2009}): cyclin D/CDK4-6 and cyclin E/CDK2 regulate 
progression during the G1 phase and elicit the G1/S transition, cyclin A/CDK2 promote progression during S phase and orchestrates the S/G2 
transition, and, finally, cyclin B/CDK2 brings about the G2/M transition. The activity of each of these cyclin-CDK complexes is regulated in a timely 
manner, so that each phase of the cell cycle ensues at the proper time, by means of transcriptional regulation, postranscriptional modifications 
(e.g. phosphorylation), and degradation (via ubiquitination) in which a large number of other components participate, including transcription 
factors, enzymes, ubiquitins, etc. 

In the present paper, we propose a coarse-grained description of the cell cycle phases by lumping S, G2, and M into one phase, so that we consider 
a two-phase model G1 and S-G2-M, as shown in Fig. \ref{fig:scheme}, \cite{alarcon2004}. In particular, we consider that cells can only divide once 
they have entered the S-G2-M phase at a constant rate. Entry in S-G2-M is regulated by a (stochastic) model of the G1/S transition which takes into 
account the regulation of the duration of the G1 phase by hypoxia (lack of oxygen).      

The abundance of oxygen is known to be one of the factors that regulate the duration of the G$_1$ phase of the cell cycle. The issue of the 
regulation of the G$_1$/S transition by the oxygen concentration has been the subject of several modelling studies \cite{alarcon2004,bedessem2014}. 
These models focus on the hypoxia-induced delay of the activation of the cyclins either through activation of cyclin inhibitors \cite{alarcon2004} or 
via up-regulation of the HIF-1$\alpha$ transcription factor \cite{bedessem2014}. From the modelling point of view, both of them are mean-field models, 
thus neglecting fluctuations. In this Section, we formulate a stochastic version of the model of Bedessem \& Stephanou \cite{bedessem2014}, of which 
an schematic representation is shown in Fig. \ref{fig:cellcyclescheme}.

HIF-1 mediates adaptive responses to lack of oxygen \cite{semenza2013}. HIF-1 is a heterodimer consisting of two sub-units: HIF-1$\alpha$ and 
HIF-1$\beta$. Whilst the latter is constitutively expressed, HIF-1$\alpha$ is O$_2$-regulated. In the presence of adequate oxygen availability is 
negatively regulated by the von Hippel-Lindau (VHL) tumour suppressor protein, which allows HIF-1$\alpha$ for degradation. VHL loss-of-function 
mutations are usual in many types of tumours, which allows for de-regulated HIF-1$\alpha$ degradation \cite{semenza2013}. HIF-1 is involved in a 
number of cellular responses including switch from oxidative phosphorylation to glycolysis, activation of angiogenic pathways, and inhibition of cell 
cycle progression \cite{bristow2008,semenza2013}.

\subsection{Mean-field model of the regulation of the G1/S transition by hypoxia}\label{sec:meanfieldg1s}

Bedessem \& Stephanou formulate a model of the effect of hypoxia (mediated by HIF-1) on the timing of the G1/S transition \cite{bedessem2014}. The 
involvement of HIF-1 in cell cycle regulation is complex and not completely  understood. There is evidence that HIF-1 delays entry into S(-G2-M) 
phase by activating p21, a CDK inhibitor, \cite{koshiji2004,ortmann2014}. HIF-1 up-regulation of p21 mediates indirect down-regulation of cyclin E 
\cite{goda2003}. Further to HIF-1 regulation of the cell cycle, there exists a feedback regulation of cell cycle proteins of HIF-1. Hubbi et al. 
\cite{hubbi2014} report that CDK1 and CDK2 physically and functionally interact with HIF-1$\alpha$: CDK1 down-regulates lysosomal degradation of 
HIF-1$\alpha$, thus consolidating its stability and promoting its transcriptional activity. On the contrary, CDK2 activates lysosomal degradation
of HIF-1$\alpha$ and promotes G1/S transition. Bedessem \& Stephanou \cite{bedessem2014} do not take into account all these issues and, for 
simplicity, chose to focus on the well-documented effect of HIF-1 on cyclin D \cite{wen2010,ortmann2014}. 

Bedessem \& Stephanou \cite{bedessem2014} model formulation is based on the following assumptions:

\begin{enumerate}
\item The G1/S is modelled by a biological switch which emerges from the mutual inhibition between a cyclin (in this case, cyclin E) and an 
ubiquitin complex (SCF complex): The latter marks the former for degradation whereas cyclin E mediates inactivation of the SCF complex. This mutual 
inhibition gives rise to a bistable situation in which two stable fixed points coexist, each associated with the G1 phase (high SCF activity, low 
cyclin E concentration) and the S-G2-M phase (low SCF activity, high cyclin E concentration). Activation and inactivation of the SCF complex are 
assumed to follow Michaelis-Menten kinetics. 
\item As in previous models \cite{tyson2001,novak2004,alarcon2004}, the G1/S transition is brought about by triggering a saddle-node 
bifurcation, whereby the G1 phase fixed point collides with the unstable saddle, driving the system into S-G2-M fixed point. Two factors drive 
the system through this bifurcation: cell growth (logistic increase of the cell mass \cite{tyson2001}) and activation of the E2F transcription factor. In 
\cite{bedessem2014}, both factors are taken to up-regulate the transcription of cyclin E. 
\item Activation of E2F is modelled in terms of the E2F-Retinoblastoma protein (RBP) switch \cite{lee2010}. Briefly, E2F is captured (bound) by 
unphosphorylated RBP. Upon phosphorylation, RBP releases E2F which activates transcription of G1/S-transition promoting cyclins, such as cyclin E 
\cite{alberts2002}. Further, E2F can be in unphosphorylated (active) form and phosphorylated (inactive) form. Following \cite{novak2004}, Bedessem \& 
Stephanou assume that fraction of active E2F and RBP-bound E2F are in adiabatic equilibrium with unphosphorylated RBP and total free E2F 
\cite{bedessem2014}. Furthermore, \cite{bedessem2014} takes cyclin D  to phosphorylate RBP.
\item Last, Bedessem \& Stephanou \cite{bedessem2014} assume that HIF-1$\alpha$ inhibits synthesis of cyclin D. Following experimental evidence 
reported in \cite{wen2010}, they assume that the level of HIF-1$\alpha$ depends exponentially of the oxygen concentration.    
\end{enumerate}

\begin{figure}
\begin{center}
\includegraphics[scale=0.5]{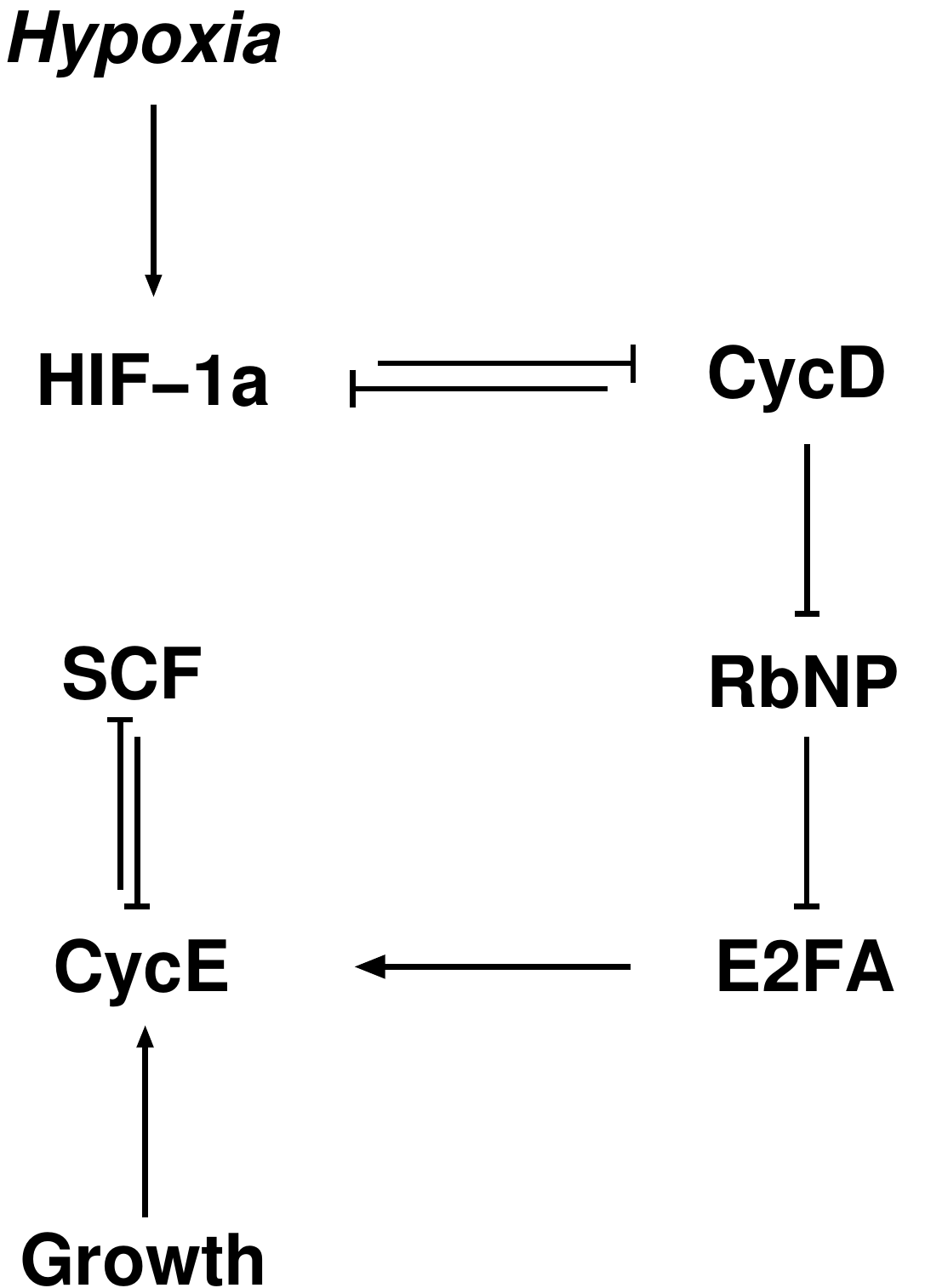}
\caption{Schematic representation of the elements involved in the model of hypoxia-regulated G$_1$/S transition proposed by Bedessem \& Stephanou 
\cite{bedessem2014}. Within the framework of this model, the negative-feedback between CycE and SCF is the key modelling ingredient for the system to emulate the behaviour of a cell during the G1/S transition. The relative balance between CycE (which promotes the G1/S transition) and SCF (G1/S transition inhibitor) is regulated by growth and hypoxia, so that the timing of the transition depends upon these two factors.}\label{fig:cellcyclescheme} 
\end{center}
\end{figure}

\noindent These basic hypotheses are primarily based on previous models \cite{tyson2001,novak2004,alarcon2004}. The resulting pathway is 
schematically represented in Fig. \ref{fig:cellcyclescheme}.

\subsection{Stochastic G1/S transition model}

\begin{table}
\begin{center}
\begin{tabular}{ll}
Variable & Description \\\hline
$X_1$, $X_8$ & Number of CyclinD and CyclinE molecules (respectively)\\
$X_2$, $X_5$ & Number of inactive and active SCF molecules (respectively)\\
$X_3$, $X_6$ & Number of SCF-activating and SCF-inactivating enzyme molecules (respectively)\\
$X_4$, $X_7$ & Number of enzyme-inactive SCF and enzyme-active SCF complexes (respectively)\\
$X_9$, $X_{10}$ & Number of free RbP and E2F molecules (respectively)\\\hline
\end{tabular} 
\begin{tabular}{ll}
Reaction probability p.u.t & $r_{i}$         \\\hline
$W_{1}=k_{1}-k_{2}[H]$    			& $(1,0,0,0,0,0,0,0,0,0)$ 	\\
$W_{2}=k_{3} X_1$	& $(-1,0,0,0,0,0,0,0,0,0)$	\\
$W_{3}=k_{4} X_2 X_{3}$ 		& $(0,-1,-1,1,0,0,0,0,0,0)$ \\
$W_{4}=k_{5}  X_{4}$ 						& $(0,1,1,-1,0,0,0,0,0,0)$  \\
$W_{5}=k_{9} X_8 X_{7}$ 						& $(0,1,0,0,0,1,-1,0,0,0)$  \\
$W_{6}=k_{6} X_4$ 						& $(0,0,1,-1,1,0,0,0,0,0)$  \\
$W_{7}=k_{7} X_5 X_8 X_6$ 		& $(0,0,0,0,-1,-1,1,0,0,0)$ \\
$W_{8}=k_{8} X_8 X_7$ 					& $(0,0,0,0,1,1,-1,0,0,0)$  \\
$W_{9}=
k_{10}m \left(1-\frac{X_9}{[E2F]_{tot}}\right)X_{10}$    			& $(0,0,0,0,0,0,0,1,0,0)$ 	\\
$W_{10}=(k_{11}+k_{12}X_5)X_8$    			& $(0,0,0,0,0,0,0,-1,0,0)$ 	\\
$W_{11}=k_{13}$    			& $(0,0,0,0,0,0,0,0,1,0)$ 	\\
$W_{12}=(k_{14}+k_{15}X_1)X_9$	& $(0,0,0,0,0,0,0,0,-1,0)$\\
$W_{13}=k_{16}$    			& $(0,0,0,0,0,0,0,0,0,1)$ 	\\
$W_{14}=k_{17} X_{10}$	& $(0,0,0,0,0,0,0,0,0,-1)$	
\end{tabular}
\caption{Reaction probability per unit time, $W_{i}\equiv W(X,r_i,t),\mbox{ }i=1,..,14$. The mass is assumed to grow following a logistic 
law: $m(a)=\frac{m_{*}Ke^{\eta a}}{Ke^{\eta a}-1}=\frac{m_{*}}{1-Ke^{-\eta a}}$, where $K=1-\frac{m_{*}}{m(0)}$ and $a$ is the age of the cell 
(\emph{i.e.} the time elapsed since birth). According to \cite{bedessem2014}, the level of active HIF-1$\alpha$, $[H]$, depends exponentially on the 
extracellular oxygen concentration, $c$: $[H]=H_0e^{\beta_1(1-c)}$. Furthermore, Following Novak and Tyson \cite{novak2004}, we assume that at each 
time, the active E2F, $[E2F_A]$, 
is the fraction of unphosphorylated free E2F factor, $[E2F]$: $[E2F_{A}]=\frac{([E2F]_{tot}-[E2F_{Rb}])[E2F]}{[E2F]_{tot}}$. The equilibrium between 
E2F-Rb complexes; free E2F and free Rb is given by \cite{novak2004}: 
$[E2F_{Rb}]=[Rb]=\frac{2[E2F]_{tot}[Rb]}{[E2F]_{tot}+[Rb]+\sqrt{([E2F]_{tot}+[Rb])^{2}-4[E2F]_{tot}[Rb]}}$}
\label{table:Wr2}
\end{center}
\end{table}

\begin{figure}
\begin{center}
\includegraphics[scale=0.3]{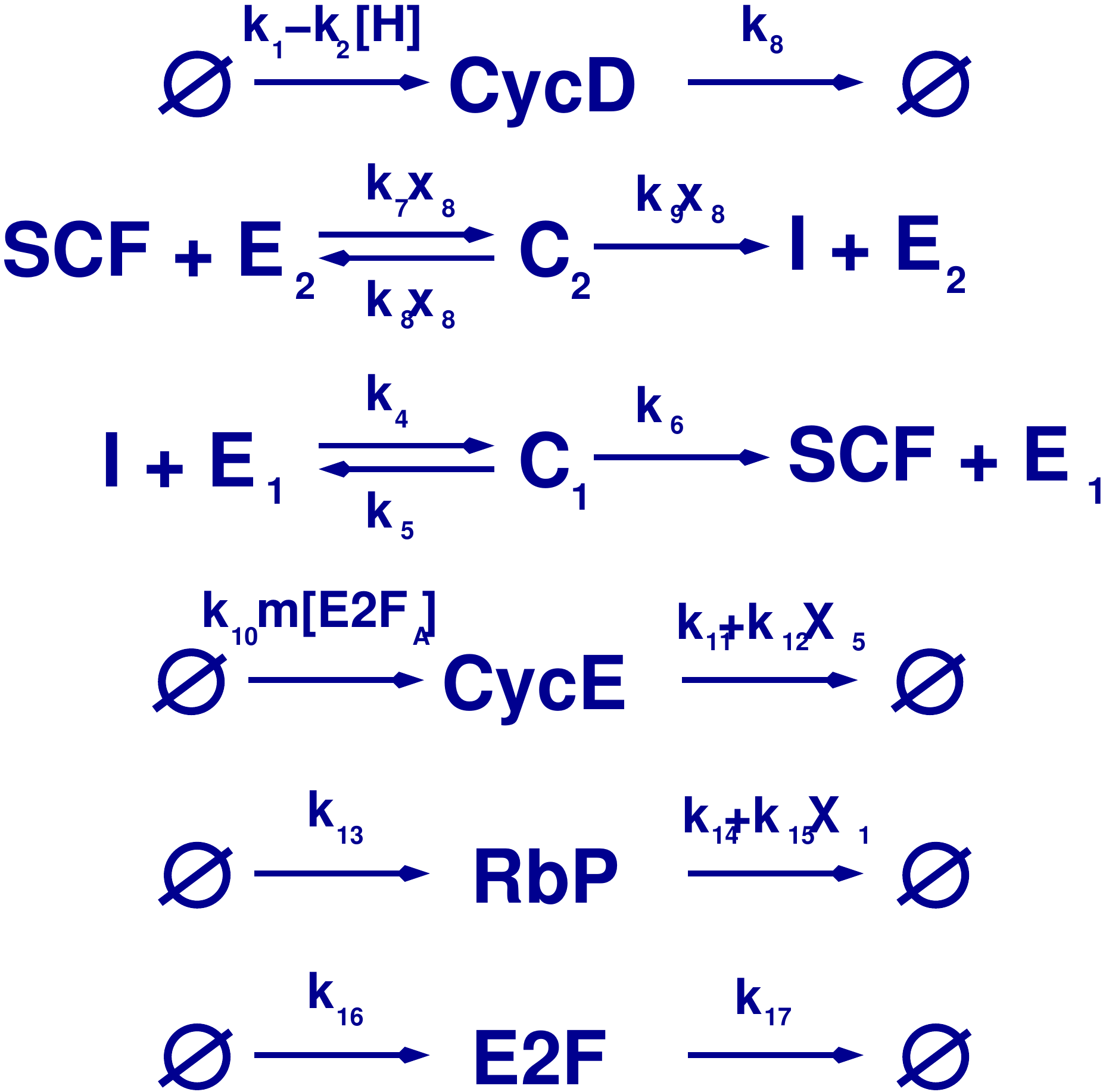} 
\caption{Reactions for the stochastic version of the model proposed by Bedessem \& Stephanou \cite{bedessem2014}. The 
reactions correspond (from top to bottom): hypoxia-inhibited synthesis and degradation of CycD, enzyme-catalysed, CycE-mediated inactivation 
of SCF, enzyme-catalysed activation of SCF, synthesis (regulated by growth and active E2F) and degradation (up-regulated by 
active SCF) of CycE, synthesis and degradation of Rb, and, last, synthesis and degradation of E2F. The negative feedback 
(mutual inhibition) between SCF and CycE mediates bistable behaviour in this 
model of the G1/S transition. Some of the transition rates associated to the reactions shown in are not constant but depend on the number of molecules of chemical species $i$ present at a particular time, $X_i$. For the definition of the quantities $X_i$, we refer the reader to Table \ref{table:Wr2}. }\label{fig:bedessemreactions}
\end{center}
\end{figure}

Based on the hypotheses summarised in Section \ref{sec:meanfieldg1s}, Bedessem \& Stephanou \cite{bedessem2014} formulated a mean-field model which is 
able to reproduce such behaviours as delay of the G1/S due lo lack of oxygen as well as hypoxia-induced quiescence. Here, we present a stochastic 
model (see Fig. \ref{fig:bedessemreactions} and Table \ref{table:Wr2}), whose mean field limit is the model formulated in \cite{bedessem2014}. We 
analyse this model using the stochastic quasi-steady state approximation we have developed in \cite{alarcon2014,delacruz2015}.

The deterministic model formulated in \cite{bedessem2014}, which we have briefly described in Section \ref{sec:meanfieldg1s}, is based on a series of 
reactions shown in Fig. \ref{fig:bedessemreactions}, which include Michaelis-Menten kinetics for activation and inactivation of SCF complexes. Our 
stochastic model of the G1/S transition builds upon the stochastic (Markovian) description of the same set of reactions.  

Our model is predicated on the stochastic dynamics of the state vector being described by a point process \cite{grimmett2001,vankampen2007}, whereby 
the state of the system, $X$, changes by an amount $r_i$ when the elementary reaction $i$ occurs. If we further assume that the waiting time between 
events is exponentially distributed, the process is characterised by the associated transition rates, \emph{i.e.} $P(X(a+\Delta a)=X+r_i\vert 
X(a)=X)=W_i(X)\Delta a+O(\Delta a^2)$. Using law of mass action \cite{gillespie1976} as our basic modelling framework, the transition rates of each 
elementary process are given in Table \ref{table:Wr2}. Once we have determined the transition rates associated with each elementary reaction (or 
channel), the dynamics of the system is given by the Chemical Master Equation of a (non-structured) Markov Process, $X(a)$:

\begin{equation}\label{eq:cmebedessem}
\frac{\partial P(X,a)}{\partial a}=\sum_{i=1}^R\left(W_i(X-r_i)P(X-r_i,a)-W_i(X)P(X,a)\right)
\end{equation}  

\noindent where $P(X,a)$ is the probability of the state vector of the system to be $X$ at age, \emph{i.e.} the time reckoned from the last division, $a$. The transition rates, $W_i(X)$, the 
vectors $r_i$ (whose components are the variation of the number of each chemical species upon occurrence of reaction $i$) are given in Table 
\ref{table:Wr2} and determine the dynamics of the system.

Even for moderately complex models, Eq. (\ref{eq:cmebedessem}) has no solution in closed form. Therefore, in order to study the properties of the 
system one must resort to numerical simulation (Monte Carlo) or asymptotic approximations. In the next section, we present an asymptotic analysis 
based on a recently developed form of stochastic quasi-steady state approximation. 

\subsection{Semi-classical quasi-steady state analysis of the stochastic G1/S transition model}\label{sec:semiclassg1s}

In \cite{alarcon2014,delacruz2015,sanchez2015}, we have developed a stochastic version of the classical QSS approximation, the so-called semi-classical QSS 
approximation (SCQSSA) which, within the framework of the optimal path theory, allows us to tackle systems which exhibit separation of time scales, 
such as enzyme-catalysed reactions. Since these type of reaction features prominently in our stochastic model of the hypoxia-regulated G1/S 
transition (see Fig. \ref{fig:bedessemreactions}), we will use the SCQSSA to analyse the effects of intrinsic noise on the stochastic model of the 
hypoxia-regulated G1/S transition (as determined by the transition rates shown in Table \ref{table:Wr2}). This approximation allows us to study
noise-induced phenomena which are relevant for the timing of the G1/S transition and, therefore, bear upon the population dynamics. 

Following the SCQSSA methodology summarised in \ref{sec:scqssa}, we derive the following set of equations which describe the optimal path associated with the stochastic G1/S transition model, Table \ref{table:Wr2}:
\begin{eqnarray} \label{eq:finalsystem1xmaintext}
 \frac{dq_{1}}{da}&=&\kappa_{1}-\kappa_{2}-\kappa_{3}q_{1}
\\
 \frac{dq_{5}}{da}&=&\frac{\kappa_{6}p_{3}p_{e_1}(p_c-q_{5})}{(p_c-q_{5})+\frac{\kappa_{5}+\kappa_{6}}{\kappa_{4}}}-\frac{\kappa_{9}p_{6}p_{e_2}q_{5}q_{8}}{q_{5}+(\kappa_{8}+\kappa_{9})}  \label{eq:scf2}
 \\
\frac{dq_{8}}{da}&=&\kappa_{10}mq_{10}\left(1-\frac{q_{9}}{\widehat{[e2f]}_{tot}}\right)-\kappa_{11}q_{8}-\kappa_{12}q_{5}q_{8}   \label{eq:cycE2}
\\
\frac{dq_{9}}{da}&=&\kappa_{13}-\kappa_{14}q_{9}-\kappa_{15}q_{1}q_{9}
\\
 \frac{dq_{10}}{da}&=& \kappa_{16}-\kappa_{17}q_{10}
\\
 q_{2}&=&p_c-q_{5}
\\
 q_{4}&=&\frac{p_{e_1}q_{2}}{q_{2}+\frac{\kappa_{5}+\kappa_{6}}{\kappa_{4}}}, \quad q_{3}+q_{4}=p_{e_1}
\\
q_{7}&=&\frac{p_{e_2}q_{5}}{q_{5}+(\kappa_{8}+\kappa_{9})}, \quad q_{6}+q_{7}=p_{e_2}
\\
\label{eq:finalsystemxmaintext}
p_{3}&=&\mbox{cnt.}\,p_{6}=\mbox{cnt.}
\end{eqnarray}

\noindent where $a$ is the rescaled age variable $a=k_7ESt$ (see \ref{sec:appscqssa}, Table \ref{tab:rescale}). The oxygen dependences enters the model though the  $[H]$  dependent parameter $\kappa_2$ (see Table \ref{tab:rescale}-\ref{table:parametersBedessem1}). The reader is referred to \ref{sec:scqssa} for a summary of the method and \ref{sec:appscqssa} for a detailed derivation of Eqs. (\ref{eq:finalsystem1xmaintext})-(\ref{eq:finalsystemxmaintext}) which hereafter is referred to as the semi-classical quasi-steady state approximation (SCQSSA) of the stochastic cell-cycle model and to \cite{alarcon2014,delacruz2015} for a full account of the SCQSSA methodology. Furthermore, since we will be interested in the random effects associated to the enzyme-regulated dynamics of SCF, encapsulated in the parameters $p_3$ and $p_6$, we have taken $p_i(a)=1$ for all $i\neq 3,6$. This corresponds to analysing the marginal distribution integrating out all the stochastic effects associated to all $X_i$ with $i\neq 3,6$. 

\subsection{Stochastic behaviour of the G1/S model}

We now proceed to study the behaviour of the stochastic model of the oxygen-regulated G1/S transition. We pay special attention to those aspects in which we observe a departure of the stochastic system from the mean-field behaviour. In particular, we highlight the effects of modifying the relative abundance of SCF-activating and inactivating enzymes, including the ability of inducing oxygen-independent quiescence.  

\begin{figure}
\begin{center}
$\begin{array}{c}
\includegraphics[width=.5\textwidth]{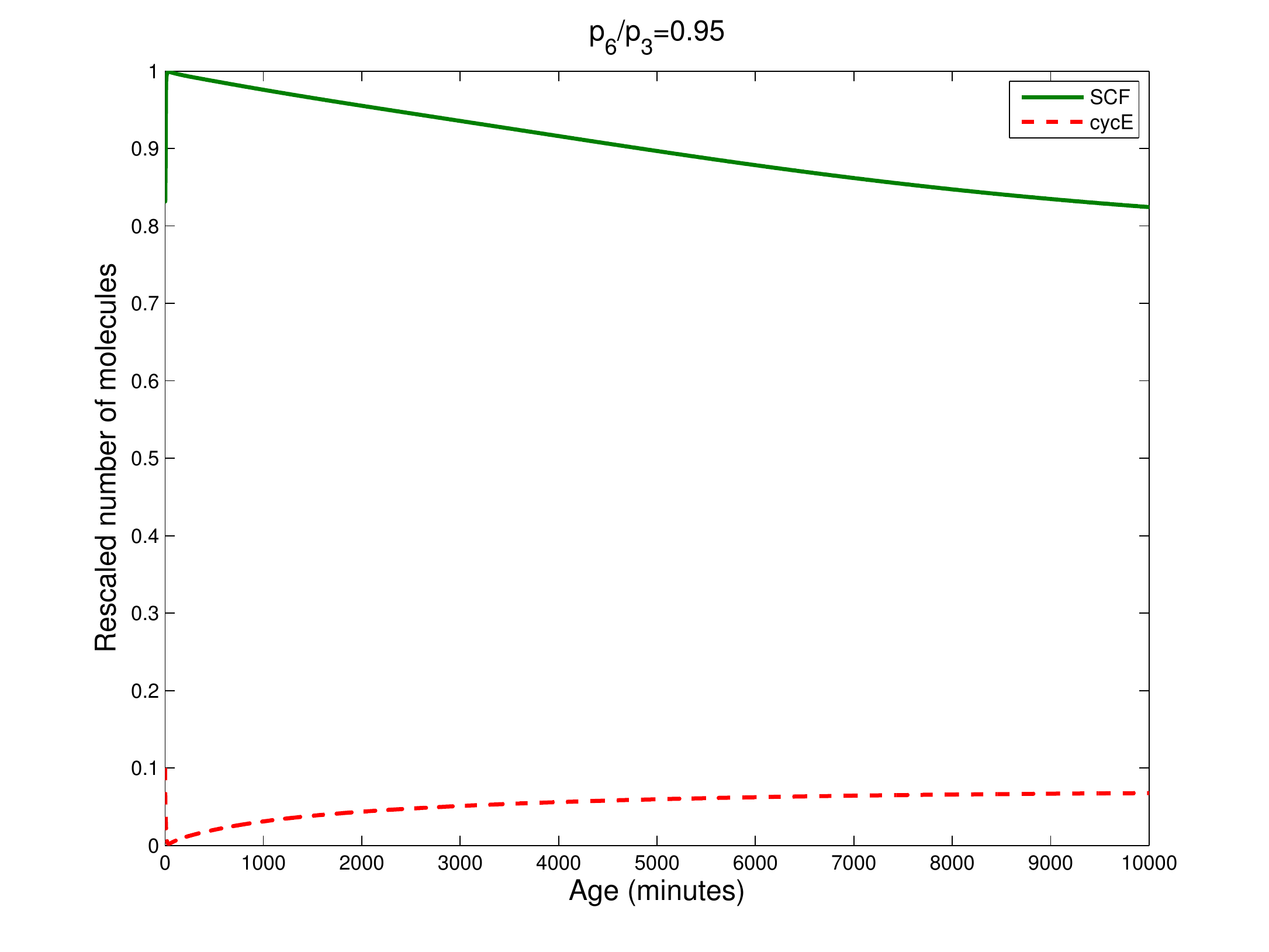} \\
\includegraphics[width=.5\textwidth]{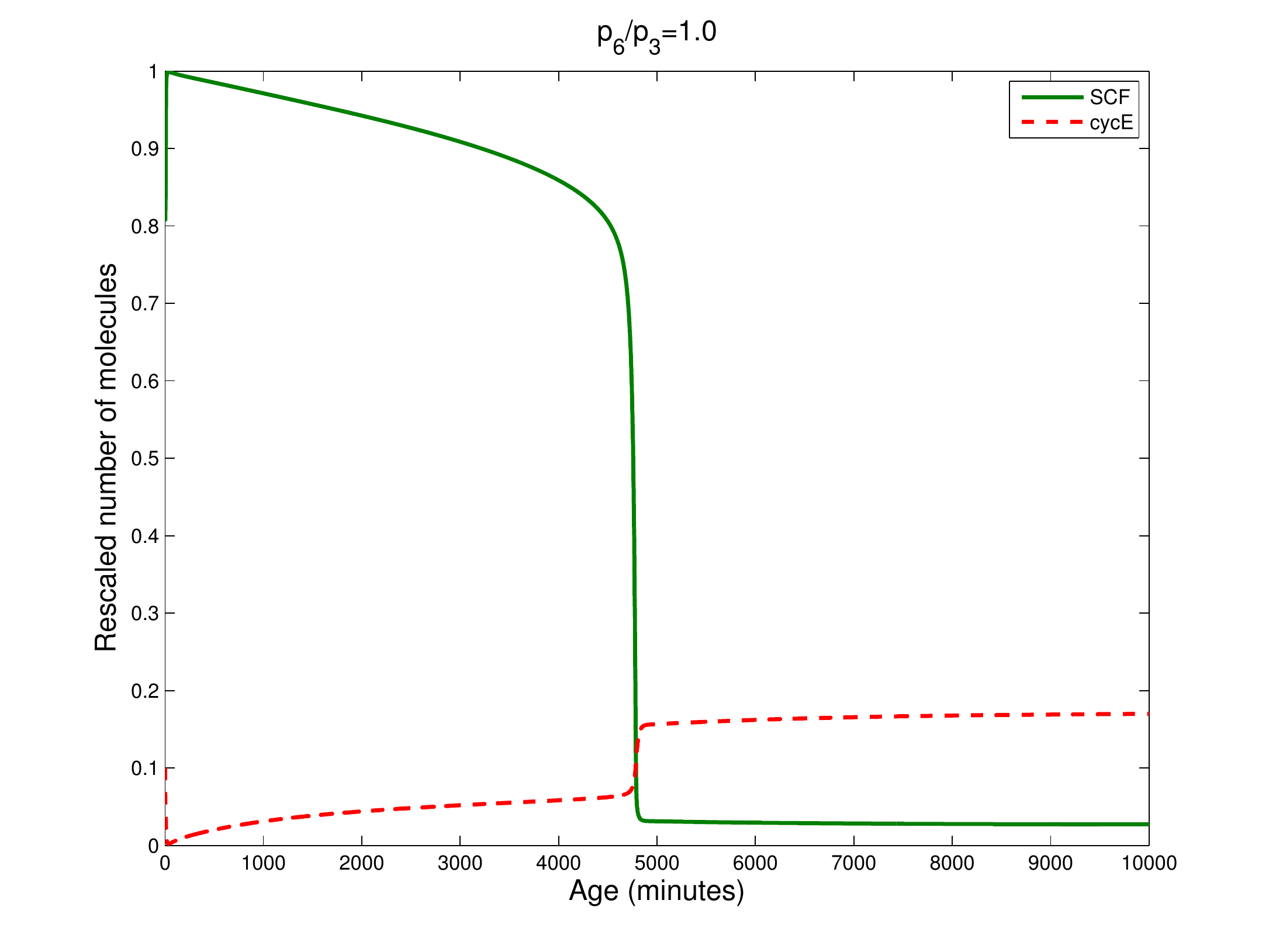} \\
\includegraphics[width=.5\textwidth]{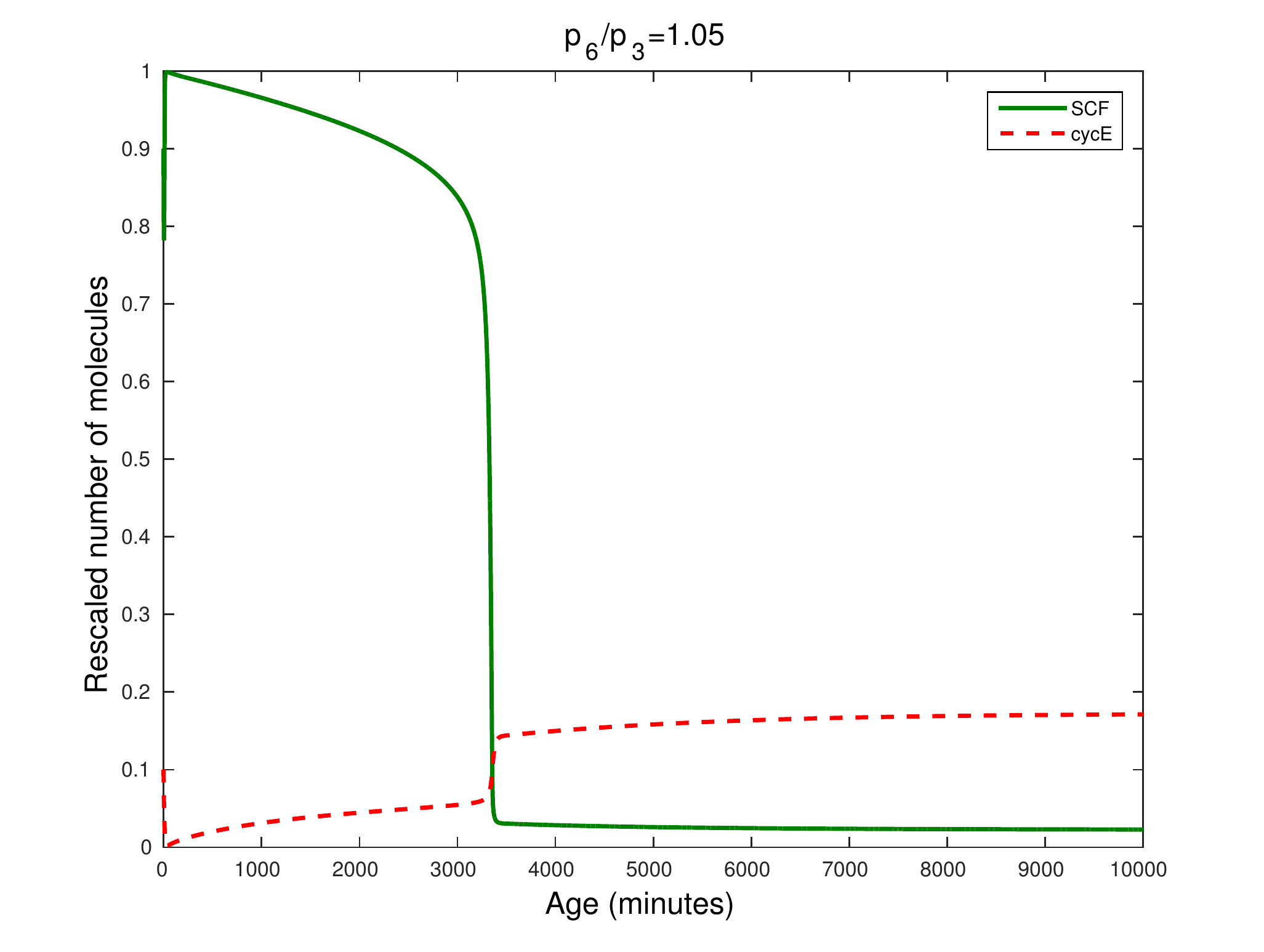}
\end{array}$
\caption{Series of plots illustrating how the ratio $p_6/p_3$, which is associated with the ratio of the number of SCF-inactivating and SCF-activating enzymes, modulates the timing of the G1/S transition. Parameter values as given in Table \ref{table:parametersBedessem1}. Initial conditions are provided in Table \ref{table:initialBedessem}} \label{fig:g1stransition}
\end{center}
\end{figure}

\subsubsection{The relative abundance of the SCF activating and inactivating enzymes controls the timing of the G1/S transition}\label{sec:timingg1s}

In references \cite{alarcon2014,delacruz2015} we have shown that, under SCQSSA conditions, the momenta $p_3$ and $p_6$, \emph{i.e.} the momenta coordinates associated with the SCF-activating and inactivating enzymes, respectively, are determined by the probability distribution of their initial (conserved) number. In particular, if we assume that the initial number of SCF-activating and inactivating enzyme molecules, $E_1$ and $E_2$, is distributed over a population of cells following a Poisson distribution with parameter $E$, we have shown that \cite{delacruz2015}:

\[p_3=\frac{e_1}{E}=p_{e_1},\mbox{ }p_6=\frac{e_2}{E}=p_{e_2}.\]

\begin{figure}
\begin{center}
\includegraphics[scale=0.5]{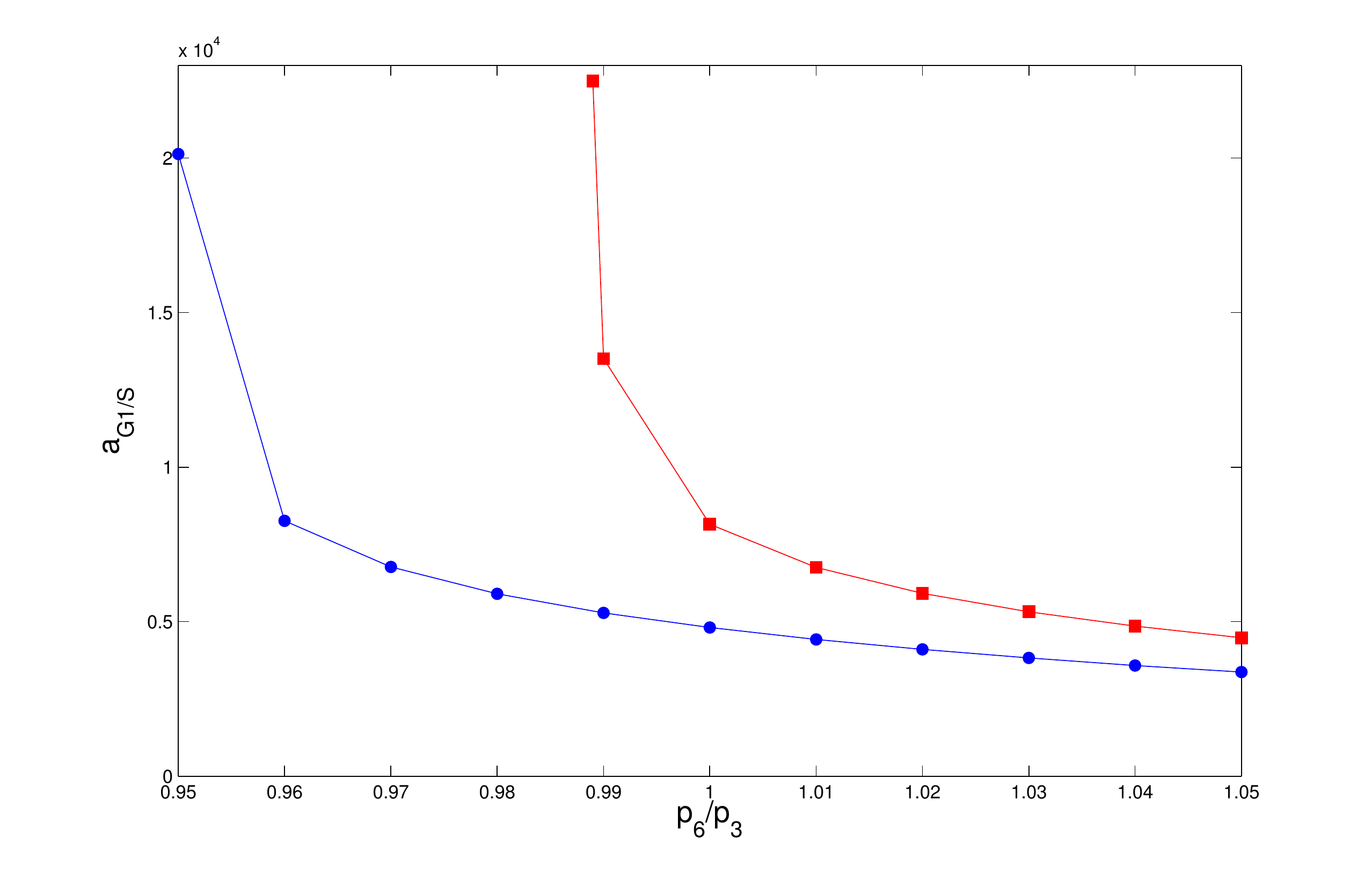}
\caption{Plot showing how the G1/S transition age, $a_{G1/S}(c,p_3,p_6)$, changes as the ratio $p_3/p_6$, which is determined by the ratio of the 
(conserved) amounts of SCF activating and inactivating enzymes \cite{alarcon2014,delacruz2015}, varies. We show $a_{G1/S}(c,p_3,p_6)$ for $c=1$ (blue 
circles) and $c=0.1$ (red squares). Parameter values as given in Table \ref{table:parametersBedessem1}.} \label{fig:g1stransitiontime}
\end{center}
\end{figure}

With this in mind, we can analyse the effect of changing the relative concentration of SCF-activating and inactivating enzymes on the timing of the 
G1/S transition. Our results are shown in Figs. \ref{fig:g1stransition} and \ref{fig:g1stransitiontime}. Fig. \ref{fig:g1stransition} illustrates that, 
for a fixed oxygen concentration, the G1/S transition is delayed by depriving the system of SCF-activating enzyme: as the ratio $p_3/p_6=e_1/e_2$ of SCF activating and deactivating enzyme increases, the G1/S transition takes longer to occur. Then, Fig. 
\ref{fig:g1stransitiontime} shows that the G1/S transition age $a_{G1/S}(c,p_3,p_6)$ decreases when $p_3/p_6$ increases. Furthermore, increasing the oxygen concentration $c$ from $c=0.1$ to $c=1$ shifts the curve towards lower transition ages $a_{G1/S}(c,p_3,p_6)$. Note that this prediction is beyond the reach of the mean-field limit \cite{bedessem2014}.  

\subsubsection{Induction of quiescence}\label{sec:quiescence}

\begin{figure}
\centering
\includegraphics[scale=0.7]{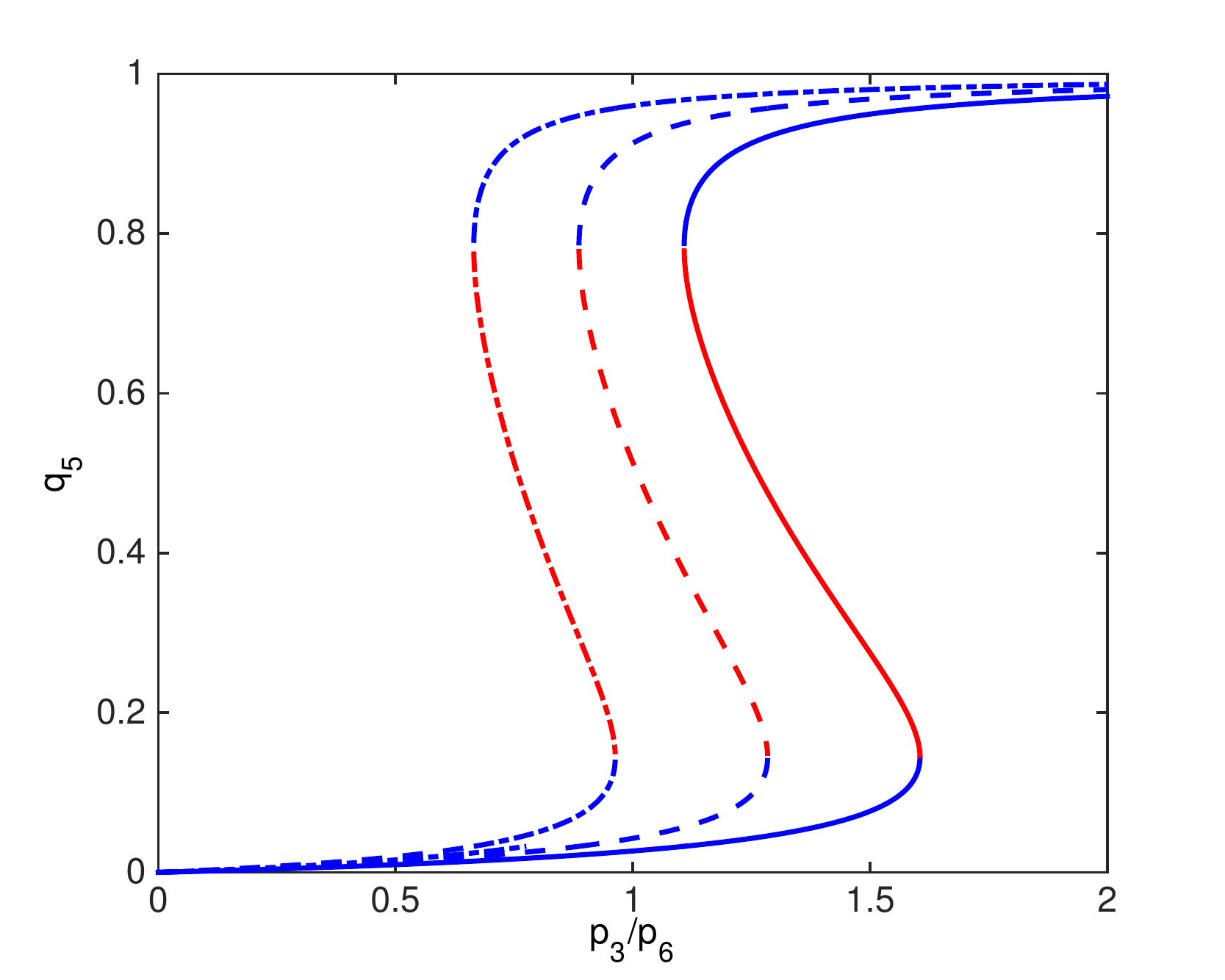}
\caption{This figure shows the bifurcation diagram of the SCQSSA of the stochastic cell-cycle model for different values of the parameter $m$. The ratio $p_3/p_6$ is the control parameter. The order parameter is the steady state value of the generalised coordinate associated with active SCF, $q_5$. Solid line corresponds to $m=10$, dash lines to $m=8$ and dotted lines to $m=6$. Parameter values as give in Table \ref{table:parametersBedessem}, $p_i=1,\,i\neq 3,6$, $p_c=1$ and $c=1$. Blue lines indicate stable steady state and red lines indicate unstable steady state.}
\label{fig:bif_pe1_pe2}
\end{figure}

In view of the results of Section \ref{sec:timingg1s}, we have proceeded to a more thorough analysis of the effect of varying the ratio $p_3/p_6$, which we recall that, within the SCQSSA, is equal to the ratio between the abundance of SCF-activating and inactivating enzymes, on the behaviour of the SCQSSA system Eqs. (\ref{eq:finalsystem1xmaintext})-(\ref{eq:finalsystemxmaintext}). In particular, we have investigated the bifurcation diagram of Eqs. (\ref{eq:finalsystem1xmaintext})-(\ref{eq:finalsystemxmaintext}) with $p_3/p_6$ as the control parameter. Our results are shown in Fig. \ref{fig:bif_pe1_pe2}. We observe that, regardless of the value of $m$, there exists a range of values of the control parameter for which the saddle-node bifurcation, which gives rise to the G1/S transition, does not occur (\emph{i.e.} only the G1-fixed point is stable). This result implies that depletion of SCF-activating enzyme, or, equivalently, over-expression of SCF-inactivating enzyme can stop cell-cycle progression by locking 
cells into the so-called G0 state, \emph{i.e.} quiescence. 

\begin{figure}
\begin{center}
\includegraphics[scale=0.35]{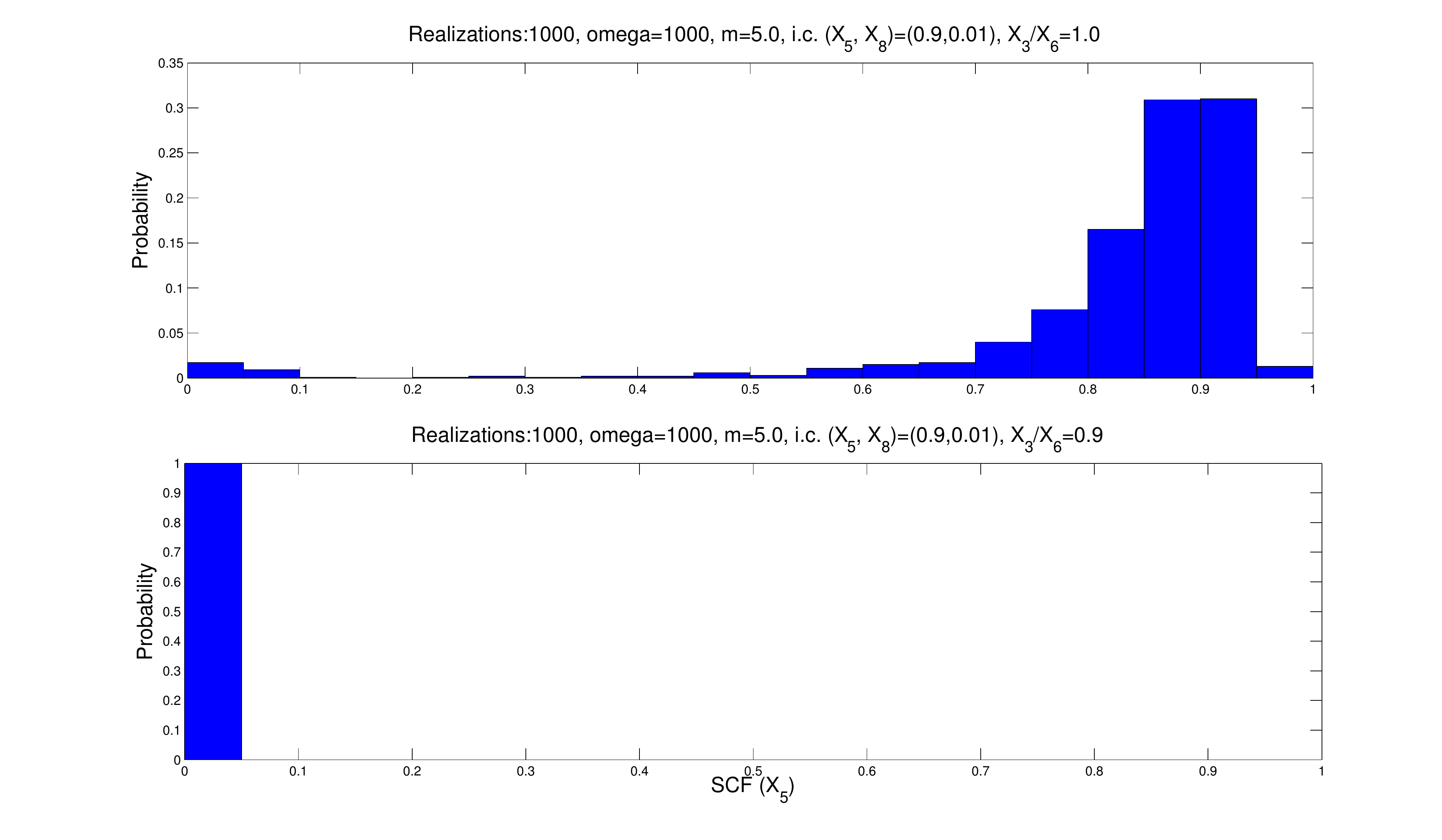}
\caption{Simulation results for the stochastic model of the oxygen-regulated G1/S transition defined by the transition rates given in Table \ref{table:Wr2}. We have plotted the probability $P(X_{5},T)$, where $T=100$, 
with different values of $X_{3}(\tau=0)/X_{6}(\tau=0)$. 1000 realisations and $m$=5.0.}\label{fig:histogram}
\end{center}
\end{figure}

These results are confirmed by direct simulation of the stochastic cell-cycle model (Table \ref{table:Wr2}) using Gillespie's stochastic simulation algorithm \cite{gillespie1976}, see Fig. \ref{fig:histogram}.

\subsection{Scaling theory of the G1/S transition age}\label{sec:scalingg1s}

We finish our analysis of the intracellular dynamics by formulating a scaling theory of one of the fundamental quantities in our multi-scale model, 
namely, the G1/S transition age, $a_{G1/S}(c,p_6,p_3)$, which determines the age-dependent birth rate (see Eq. (\ref{eq:birthratesummary})). In this 
section, we will show that, in spite of the complexity of the SCQSSA formulation of the oxygen-dependent cell-cycle progression model (see Eqs. 
(\ref{eq:finalsystem1xmaintext})-(\ref{eq:finalsystemxmaintext})), $a_{G1/S}$ exhibits remarkable regularities with respect to its dependence on the 
oxygen concentration and the cell-cycle parameters $p_6$ and $p_3$. Such regularities hugely simplify our multi-scale methodology.

\begin{figure}
\begin{center}
$\begin{array}{cc}
\mbox{(a)} & \mbox{(b)} \\
\includegraphics[scale=0.3]{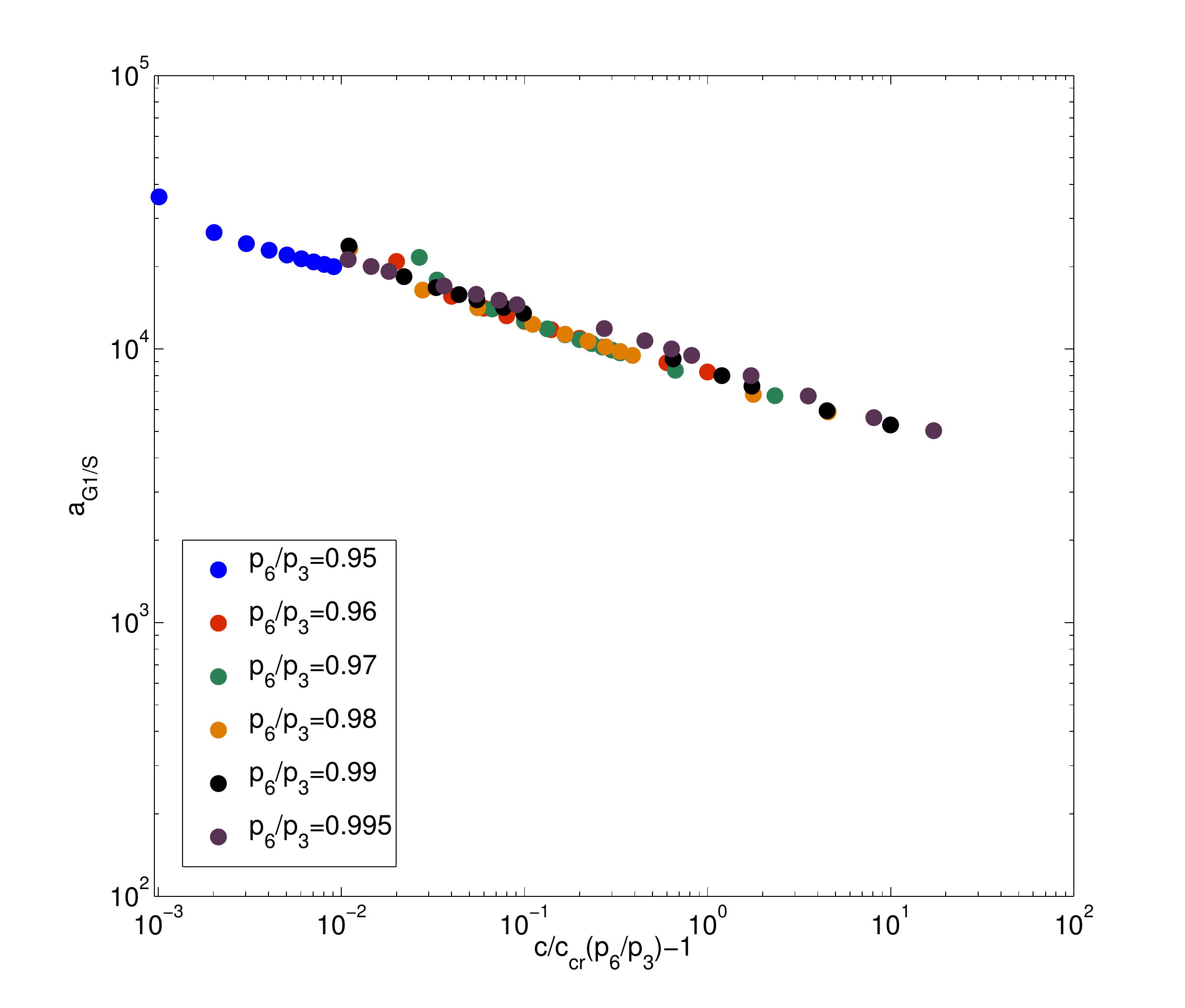} & \includegraphics[scale=0.3]{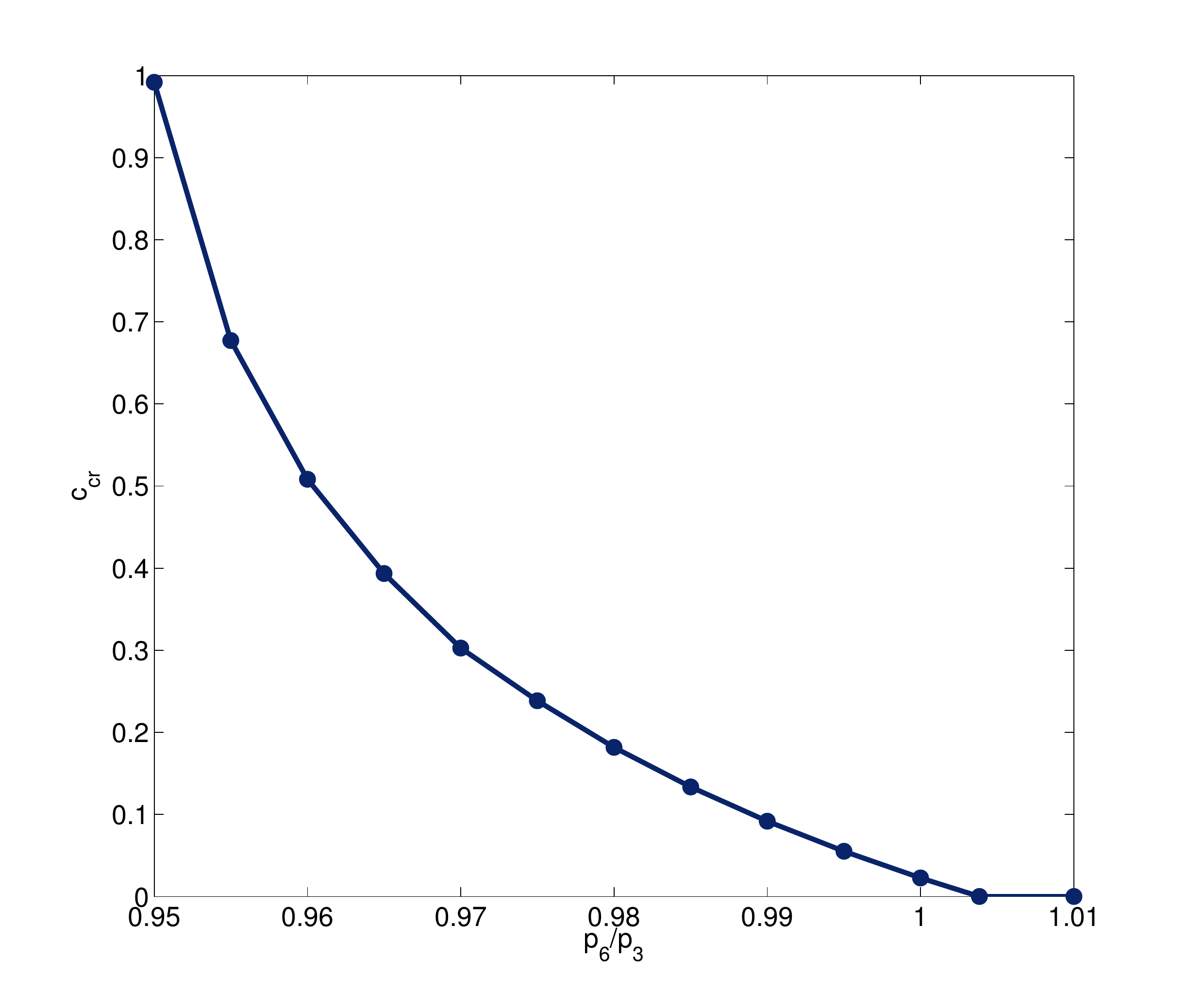} \\
\mbox{(c)} & \mbox{(d)} \\
\includegraphics[scale=0.3]{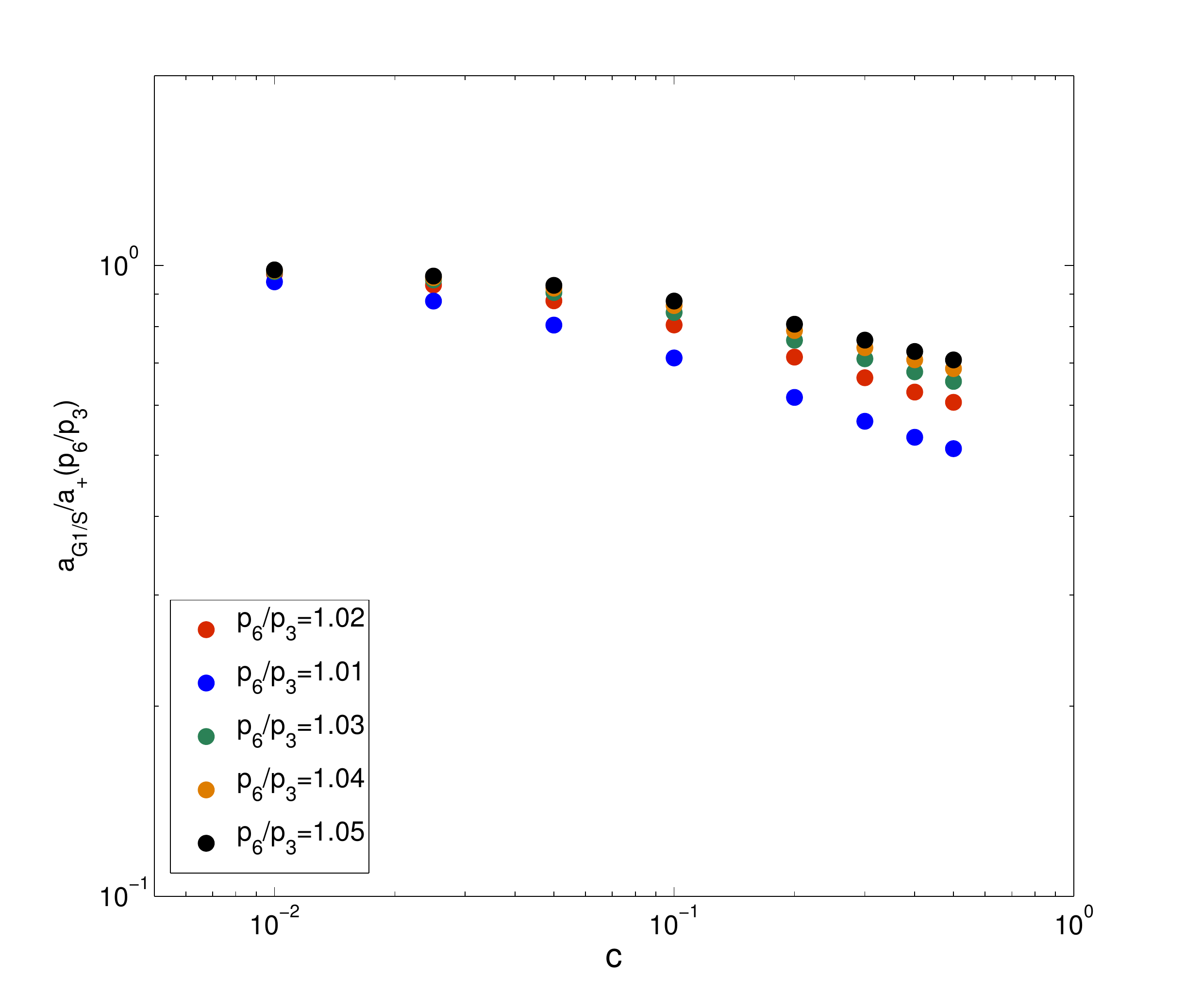} & \includegraphics[scale=0.3]{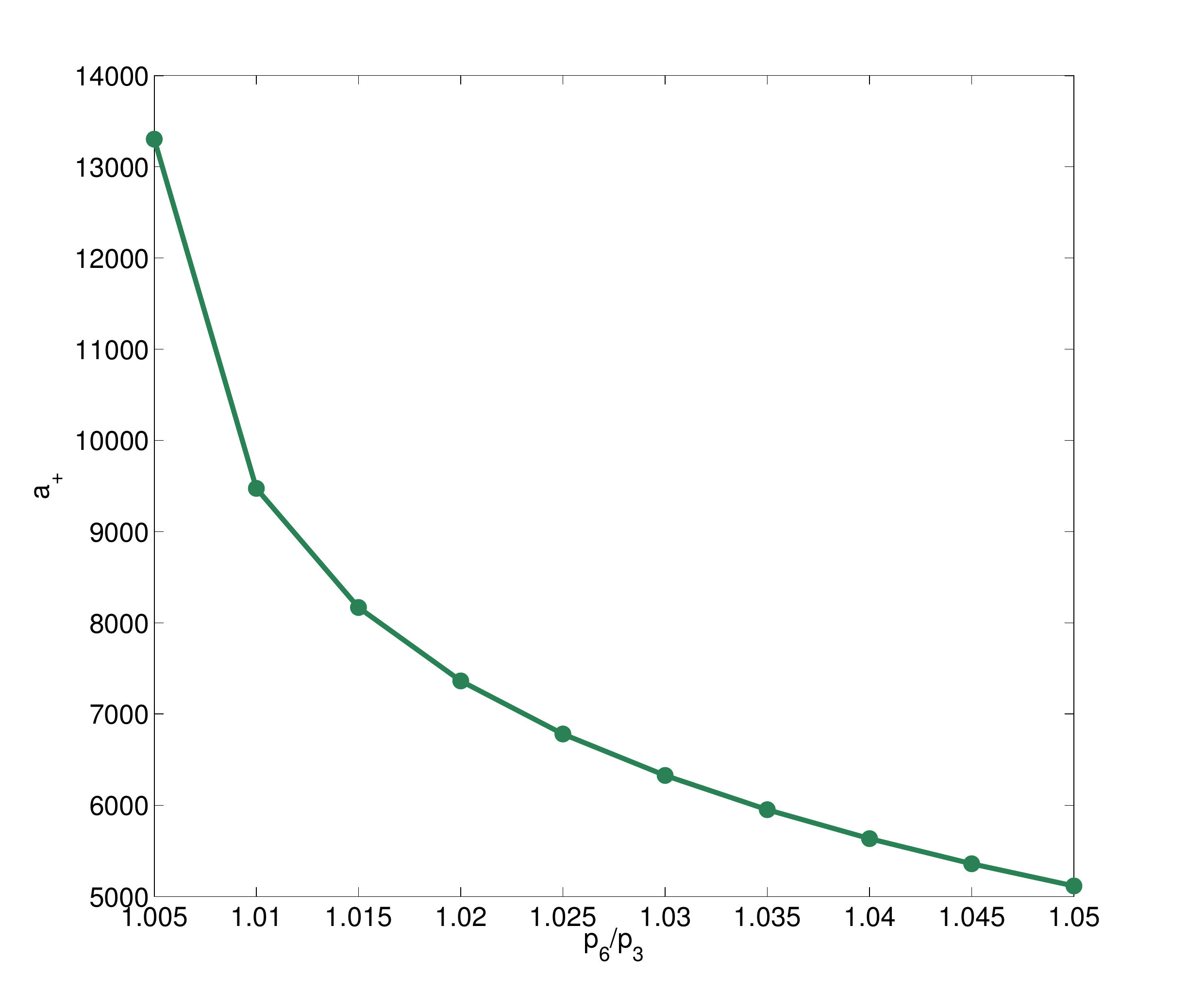} \\
\end{array}$
\caption{Series of plots showing the scaling analysis of the G1/S transition age. Plot (a) shows that below the critical value 
$r_{cr}$, which corresponds to the value of the ratio $p_6/p_3$, above which there is no transition to quiescence (\emph{i.e.} 
if $p_6/p_3>r_{cr}$ the G1/S transition age is finite when $c=0$), $a_{G1/S}(c,p_6,p_3)$ follows an algebraic decay with a universal 
$p_6/p_3$-independent exponent provided that the oxygen, $c$, is rescaled by the critical oxygen concentration, $c_{cr}(p_6/p_3)$. Plot (c) shows 
that if, by contrast, $p_6/p_3>r_{cr}$ $a_{G1/S}(c,p_6,p_3)$ decays exponentially with the oxygen concentration with a characteristic concentration 
$c_0$ which, provided $p_6/p_3$ is larger enough than $r_{cr},$ is $p_6/p_3$-independent. Plot (b) shows how $c_{cr}$ 
varies as $p_6/p_3$ is changed. Similarly, plot (d) shows how $a_+$ varies $p_6/p_3$ changes. Parameter values as given 
in Table \ref{table:parametersBedessem1}.}\label{fig:ag1s-data-collaps}
\end{center}
\end{figure}

We can see this regularity in Figure \ref{fig:ag1s-data-collaps}. Figure \ref{fig:ag1s-data-collaps}(a) shows $a_{G1/S}$ as a function of the ratio $\left(\frac{c}{c_{cr}(p_6/p_3)}-1\right)$, for six different values of $\frac{p_6}{p_3}<r_{cr}$, where $r_{cr}=1$ is the critical value. We see that all graphs fall together on a straight line in log-log space, indicating a power-law dependence of $a_{G1/S}$ on $\left(\frac{c}{c_{cr}(p_6/p_3)}-1\right)$. On the other hand, Figure \ref{fig:ag1s-data-collaps}(c) shows the dependence of the normalized function $\frac{a_{G1/S}\left(c,\frac{p_6}{p_3}\right)}{a_+(p_6/p_3)}$, on $c$, for five values $\frac{p_6}{p_3}>r_{cr}$. Here, $a_+(p_6/p_3)=a_{G1/S}(c=0,p_6/p_3)$. We see that there is good agreement between the five values $\frac{p_6}{p_3}$ for small $c$, with the disagreement increasing with increasing $c$. Thus, a good scaling approximation for $a_{G1/S}$ is given by
\begin{equation}\label{eq:ag1sscaling}
a_{G1/S}\left(c,\frac{p_6}{p_3}\right)\simeq\left\{\begin{array}{l}a_{+}\left(\frac{p_6}{p_3}\right)e^{-c/c_0}\mbox{ if 
}\frac{p_6}{p_3}>r_{cr}\\a_-\left(\frac{c}{c_{cr}(p_6/p_3)}-1\right)^{-\beta}\mbox{ if 
}\frac{p_6}{p_3}<r_{cr}\end{array}\right.
\end{equation}
\noindent Here, $c_0$, $a_-$, and $\beta$ are constants. According to our analysis of the data presented in Figure \ref{fig:ag1s-data-collaps}, these constants are given by $c_0\simeq 1.1$, $a_-\simeq 8.25\cdot 10^3$ and $\beta\simeq 0.2$. Likewise, $a_+(p_6/p_3)$ is obtained by fitting to the data presented in Figure \ref{fig:ag1s-data-collaps}(d). The critical oxygen concentration for quiescence $c_{cr}$ can be estimated analytically (with parameter values taken 
from Table \ref{table:parametersBedessem1}):

\begin{equation}
c_{cr}\left(\frac{p_6}{p_3}\right)= 
1-\frac{1}{\beta_{1}}\log{\left(\frac{1}{a_{3}H_{0}}\left(a_{1}+\frac{a_{2}d_{2}}{d_{1}[e2f]_{t}}\left(1-\frac{1}{1-a_0(\frac{p_{3}}{p_{6}})^2}
\right)\right)\right)}
\end{equation}

\noindent where $a_1$, $a_3$, $d_1$, $d_2$, $[e2f]_{t}$, and $H_0$ are parameters defined in \ref{sec:appscqssa}, Table 
\ref{table:parametersBedessem1}. The parameter $a_0$ can be estimated as follows. Let $A(c)$ be defined as:

\begin{equation}\label{eq2}
 A(c)=\left(\frac{p_{6}}{p_{3}}\right)^{2}\left(1-\frac{d_{2}}{(d_{2}+d_{1}\frac{a_{1}-a_{3}[H]}{a_{2}})[e2f]_{t}}\right)
\end{equation}

\noindent where $[H]=a_{3}H_{0}e^{\beta_{1}(1-c)}$. $a_{0}=A(c_{bif})$, where $c_{bif}$ is the critical value of the oxygen 
concentration for which the saddle node bifurcation occurs, \emph{i.e.} the critical value for which the number of real, positive solutions of the 
equation:

\begin{equation}
 e_{1}(1-x)(J_{4}+x)(b_{2}+b_{3}x)=A(c)e_{2}b_{1}m_{*}[e2f]_{t}x(1+J_{3}-x)
\end{equation}

\noindent goes from 3 to 1. The parameters $b_i$, $e_i$, and $J_i$ are defined in \ref{sec:appscqssa}, Table 
\ref{table:parametersBedessem1}

\section{Cellular scale: Multi-scale Master Equation and path integral formulation}\label{sec:MSME}

We start by summarising the formulation of the multi-scale Master Equation (MSME) for the population dynamics model in terms of an age-structured stochastic process \cite{guerrero2013}. First, we consider a simple age-dependent birth-and-death process where $n(a,t)$ stands for the number of cells of age $a$ at time $t$. Both $a$ and $t$ are dimensionless according to the scaling prescribed in Table \ref{tab:rescale}, \ref{sec:appscqssa}. The time variable is $t\rightarrow k_7ESt$ and $a$ is defined after Eq. (\ref{eq:finalsystemxmaintext}). The offspring of such cells with age $a=0$.

\begin{eqnarray}\label{eq:probbalance}
\nonumber P(n,a+ \delta a,t+ \delta t)= && W(n+1,a,t) \delta t P(n+1,a,t)\\
&& +(1-W(n,a,t) \delta t)P(n,a,t),
\end{eqnarray}

\noindent where $W(n(a),a,t)= (\nu+b(a))n(a)$ (see Table \ref{tab:ageBD}), $\mu$ is the (age-independent) death rate, and the age-dependent birth rate, $b(a)$, is given by:

\begin{equation}\label{eq:birthrate}
b(a)=\left\{\begin{array}{l}0\mbox{ if }q_8(a)<\mbox{CycE$_T$}\\\tau_p^{-1}\mbox{ if }q_8(a)\geq\mbox{CycE$_T$}\end{array}\right. 
\end{equation}

\noindent where $q_8(a)$ is the generalised coordinate associated with the concentration of CycE which must exceed a threshold value, CycE$_T$ for the cell-cycle to progress beyond the G$_1$/S transition. Before this transition occurs, cells are not allowed to divide. 

\begin{table}[htb]
\centerline{
\begin{tabular}{llll}
Event & Reaction & Transiton rate, $W_k(n,a,t)$ & $r_k$        \\\hline
Birth & $\begin{array}{c}n(a)\rightarrow n(a)-1\\\varnothing\rightarrow n(a=0)=2\end{array}$ & $W_1(n,a,t)=b(a)n(a)$ & $r_1=-1$   \\
Death & $n(a)\rightarrow n(a)-1$ & $W_2(n,a,t)=\mu n(a)$ & $r_2=-1$   
\end{tabular}} 
\caption{This table summarises the elementary events involved in the age-dependent birth-and-death process. Cells of age $a$ produce offspring at a rate proportional to the age dependent birth rate, $b(a)$, Eq. (\ref{eq:birthrate}). We are assuming that upon cell division both cells are reset to $a=0$. Therefore, upon proliferation, one cell is removed form the population of age $a$ and two cells are added to the population of age $a=0$. For simplicity, death is assumed to be age-independent and to occur at a rate proportional to the death rate, $\mu$.}\label{tab:ageBD}
\end{table}

By re-arranging Eq. (\ref{eq:probbalance}) and taking the limit $\delta t=\delta a \rightarrow 0$, we obtain:

\begin{eqnarray}\label{eq:ME}
\nonumber \frac{\partial P(n,a,t)}{\partial t} + \frac{\partial P(n,a,t)}{\partial a} = && W(n+1,a,t)P(n+1,a,t)\\
&& -W(n,a,t)P(n(a),a,t).
\end{eqnarray}

\noindent where $W(n,a,t)=W_1(n,a,t)+W_2(n,a,t)$ (see Table \ref{tab:ageBD}). Eqs. (\ref{eq:ME}) needs to be supplemented with the appropriate boundary condition at $a=0$, $P(n_0,a=0,t)$. We proceed by first 
considering the number of births that occur within the age group $a=a_j$ during a time interval of length $\delta t$. Since we are assuming that our 
stochastic model is a Markov process where, within each age group, birth and death occur independently and with exponentially distributed waiting 
times, the number of births, $B(a_j)$, is distributed according to a Poisson distribution:

\[P(B(a_j)=b_j\vert \delta t)=e^{-b(a_j)n(a_j,t)\delta t}\frac{(b(a_j)n(a_j,t)\delta t)^{b{_j}}}{b!}.\]

\noindent The total number of births delivered by the whole population during $\delta t$, $B$, is $B=\sum_{i\in I(t)}B(a_j)$. Its probability density is therefore given by:

\begin{equation}\label{eq:convolution}
P(B=b_0)= \displaystyle \sum _{\{b_{j}\}} \left( \prod_j    P_{j}(B(a_j)=b_j)            \right) \delta_{ \sum_{j\in I(t)} b_j,b_0}. 
\end{equation}

\noindent Since Eq. (\ref{eq:convolution}) is a convolution, using the well-known property of the probability generating function, the generating function associated with $P(B)$, $G_B(p,t)$ is given by \cite{grimmett2001}:

\[G_B(p,t)=\prod_{j\in I(t)}G_j(p,t),\]

\noindent where $G_j(p,t)$ is the generating function associated with $P_j(B(a_j))$:

\[G_j(p,t)=e^{b(a_j)n(a_j)\delta t(p-1)}.\]

\noindent Therefore, by taking $\delta t=\delta a \rightarrow 0$

\begin{equation}\label{eq:birthgf}
G_B(p,t)=e^{(p-1)\left(\sum_{a}b(a)n(a)\right)\delta t}.
\end{equation}

\noindent Eq. (\ref{eq:birthgf}) determines $P(n_0,a=0,t)$ since $P(n_0,a=0,t)=P(B=n_0/2,t)$ and

\begin{equation}\label{eq:pofB}
P(B=b_0,t)=\frac{1}{b_0!}\left.\frac{\partial^{b_0}G_B}{\partial p^{b_0}}\right\vert_{p=0}.
\end{equation}

\subsection{Numerical method}\label{sec:ageSSA}

Before proceeding further, we briefly describe the numerical methodology that we use to simulate the stochastic multi-scale model. In essence, the 
numerical method is an extension of the hybrid stochastic simulation algorithm used in \cite{guerrero2015a} to accommodate the age 
structure of the cell populations we deal with here \cite{guerrero2013}. For simplicity, we restrict our description of the algorithm to a homogeneous cell population. Its generalisation to heterogeneous populations composed of a variety of cellular types is straightforward.

Similarly to the procedure described in \cite{guerrero2013}, we start by defining the population vector ${\cal N}(t)=\{n_{a_j}(t), j\in {\cal 
J}(t)\}$, where ${\cal J}(t)$ is the set of indexes which label all those age groups which, at time $t$, are represented within the population, \emph{i.e.} 
all those age groups such that $n(a=a_j,t)>0$. Having defined ${\cal N}(t)$, we summarise the numerical algorithm:

\begin{enumerate}
\item Set initial conditions: $c(t=0)=c_0$, $n(a,t=0)=f(a)\Rightarrow{\cal N}(t)=\{f_{a_j}, j\in {\cal 
J}(t=0)\}$, $N(t=0)=\int_{0}^{\infty}f(a)da$. We also set the value of the ratio of the SCF-regulating enzymes $p_6/p_3$.
\item At some later time $t$, the system is characterised by the quantities $c(t)$, ${\cal N}(t)$, ${\cal J}(t)$, and $N(t)$
\item Generate two random numbers, $z_1$ and $z_2$, uniformly distributed in the interval $(0,1)$
\item Use $z_1$ to calculate the exponentially distributed waiting time to the next event: $\tau=\frac{1}{a_0(t)}\log\frac{1}{z_1}$ where 
$a_0(t)=\sum_{j\in{\cal J}(t)}\left(W_1(n_{a_j})+W_2(n_{a_j})\right)$. The rates $W_1$ and $W_2$ are given in Table \ref{tab:ageBD}
\item Update the oxygen concentration at $c(t+\tau)$ by solving the associated ODE Eq. (\ref{eq:oxygen}) between $(t,t+\tau)$ taking $c(t)$ as 
initial condition. We use a 4 stage Runge-Kutta solver
\item Update the age-dependent birth rate for each $j\in{\cal J}$, \emph{i.e.} 

$b(a)=\tau_p^{-1}H\left(a-a_{G1/s}(c(t+\tau),p_6/p_3)\right)$. The quantity
$a_{G1/s}(c(t+\tau),p_6/p_3)$ is given by Eq. (\ref{eq:ag1sscaling}) 
\item Use $z_2$ to determine which event occurs (\emph{i.e.} birth or death within the $j$th sub-population) at time $t+\tau$: event $l$ occurs with probability

\[\sum_{k=1}^{l}{\cal W}_{k}\leq z_2a_0(t)<\sum_{k=l+1}^{J}{\cal W}_{k},\]

where $J=\mbox{card}({\cal J}(t))$ and ${\cal W}_{2(j-1)+1}=W_1(n_{a_j})$ and ${\cal W}_{2(j-1)+2}=W_2(n_{a_j})$
\item If the randomly chosen event is $l=2(j-1)+1$ (\emph{i.e.} cell proliferation within age group $j$), then $n_{a_j+\tau}(t+\tau)=n_{a_j}(t)-1$, $n_{a_k+\tau}(t+\tau)=n_{a_k}(t)$ for all $k\neq j$, and $n_{a=0}(t+\tau)=n_{a=0}(t)+2$
\item If the randomly chosen event is $l=2(j-1)+2$ (\emph{i.e.} cell death within age group $j$), then $n_{a_j+\tau}(t+\tau)=n_{a_j}(t)-1$, $n_{a_k+\tau}(t+\tau)=n_{a_k}(t)$ for all $k\neq j$
\item Finally, we update the set ${\cal J}(t+\tau)$, \emph{i.e.} the set of age groups for which $n_{a_j+\tau}(t+\tau)>0$
\item Steps 3 to 10 are repeated until some stopping condition (e.g. $t\geq T$) is satisfied
\end{enumerate}

Note that Step 5 does not involve the use a stochastic method of integration of the ODE which rules the time evolution of the oxygen concentration, Eq. (\ref{eq:oxygen}). This is due to the fact that, within the time interval $(t,t+\tau]$, the population stays constant, so that Eq. (\ref{eq:oxygen}) can be solved by means of a non-stochastic solver. 

\subsection{Steady-state of a homogeneous population: mean-field analysis}\label{sec:MFStSt}

Before proceeding further, in order to check the numerical algorithm proposed in Section \ref{sec:ageSSA}, we analyse how it compares with results regarding the steady-state of the mean-field limit of a homogeneous (\emph{i.e.} composed by one cellular type only) population \cite{hoppensteadt1975}. The mean-field equations associated with the stochastic multi-scale model are:

\begin{eqnarray}\label{eq:residentmf}
\nonumber && \frac{dc}{dt}=S-kNc, \\
&& \frac{\partial Q}{\partial t}+\frac{\partial Q}{\partial a}=-(\nu+b(a))Q
\end{eqnarray}

\noindent with boundary condition:

\[Q(a=0,t)=2\int b(a)Q(a,t)da.\]

\noindent and birth rate given by:

\[b(a)=\tau_p^{-1}H\left(a-a_{G1/S}(c,p_6/p_3)\right)\]

\noindent with $a_{G1/S}(c,p_6/p_3)$ is given by Eq. (\ref{eq:ag1sscaling})

According to \cite{hoppensteadt1975}, in order to ascertain whether a steady-state solution, \emph{i.e.} whether the system settles onto an age distribution where the proportion of cells of each age does not change, we seek for a separable solution: $Q(a,t)=A(a)T(t)$. Defining $\mu(a)=\nu+b(a)$, we obtain:

\begin{equation}
\frac{1}{T}\frac{dT}{dt}=-\frac{1}{A}\left(\frac{dA}{da}+A\mu(a)\right)=\sigma 
\end{equation}

\noindent with $\sigma=$cnt. to be determined. $Q(a,t)$ is therefore given by:

\begin{equation}\label{eq:com2}
Q(a,t)=A(a=0)\exp\left(\sigma(t-a)-\int_0^a\mu(y)dy\right) 
\end{equation}

\noindent The value of the parameter $\sigma$ is obtained by means of the characteristic equation obtained by introducing Eq. (\ref{eq:com2}) with $a=0$ into the boundary condition:

\begin{equation}\label{eq:chareq}
1=2\int_0^{\infty}b(a)\exp\left(-\sigma a-\int_0^a\mu(y)dy\right)da 
\end{equation}



After some algebra, the characteristic equation Eq. (\ref{eq:chareq}) reads:

\begin{equation}\label{eq:peq}
2\frac{\tau_{p}^{-1}e^{-(\sigma+\nu)a_{G1/S}}}{\sigma+\nu+\tau_{p}^{-1}}=1 
\end{equation}

\noindent From Eq. (\ref{eq:peq}) we obtain the condition for the system to be in equilibrium, \emph{i.e.} $\sigma=0$. Substituting $\sigma=0$ in Eq. (\ref{eq:chareq}):

\begin{equation}\label{eq:R0}
R_0\equiv 2 \frac{\tau_{p}^{-1}e^{-\nu a_{G1/S}}}{\nu+\tau_{p}^{-1}}=1
\end{equation}

\noindent where $R_0$ is the average number of offspring per cell at equilibrium: if $R_0>1$ the system grows exponentially, $R_0<1$ the system dwindles, and if $R_0=1$ the population remains constant. The equilibrium condition $R_0=1$ alllows us to find the value of $a_{G1/S}$ for which such equilibrium exists:

\begin{equation}\label{eq:aG1S}
a_{G1/S}(c_{\infty},p_6/p_3)=-\frac{1}{\nu}\log\left(\frac{\tau_{p}(\nu+\tau_{p}^{-1})}{2}\right) 
\end{equation}

\noindent For this quantity to be positive $\tau_{p}\nu<1$ must hold. This condition states that for a steady state be reached, the average waiting time to division after the G1/S transition, $\tau_p$, must be smaller than the average life span of the cell, $\nu^{-1}$. Eq. (\ref{eq:aG1S}) determines the stationary value of the oxygen concentration, $c_{\infty}$.

The long-time dynamics of Eqs. (\ref{eq:residentmf}) can therefore be summarised as follows: given the values of $\tau_p$, $\nu$, and the cell-cycle parameters (see Table \ref{table:parametersBedessem1} in \ref{sec:appscqssa}), the population evolves and consumes oxygen until, the oxygen concentration reaches a steady value $c_{\infty}$. At this point, the population of resident cells has settled onto a steady state where its age structure does not change. The total number of cells is also constant and given by (see Eqs. (\ref{eq:residentmf})):

\begin{equation}\label{eq:steadyNr}
N_{\infty}=\frac{S}{kc_{\infty}} 
\end{equation}

\begin{figure}
\begin{center}
$\begin{array}{cc}
\mbox{(a)} & \mbox{(b)} \\
\includegraphics[scale=0.3]{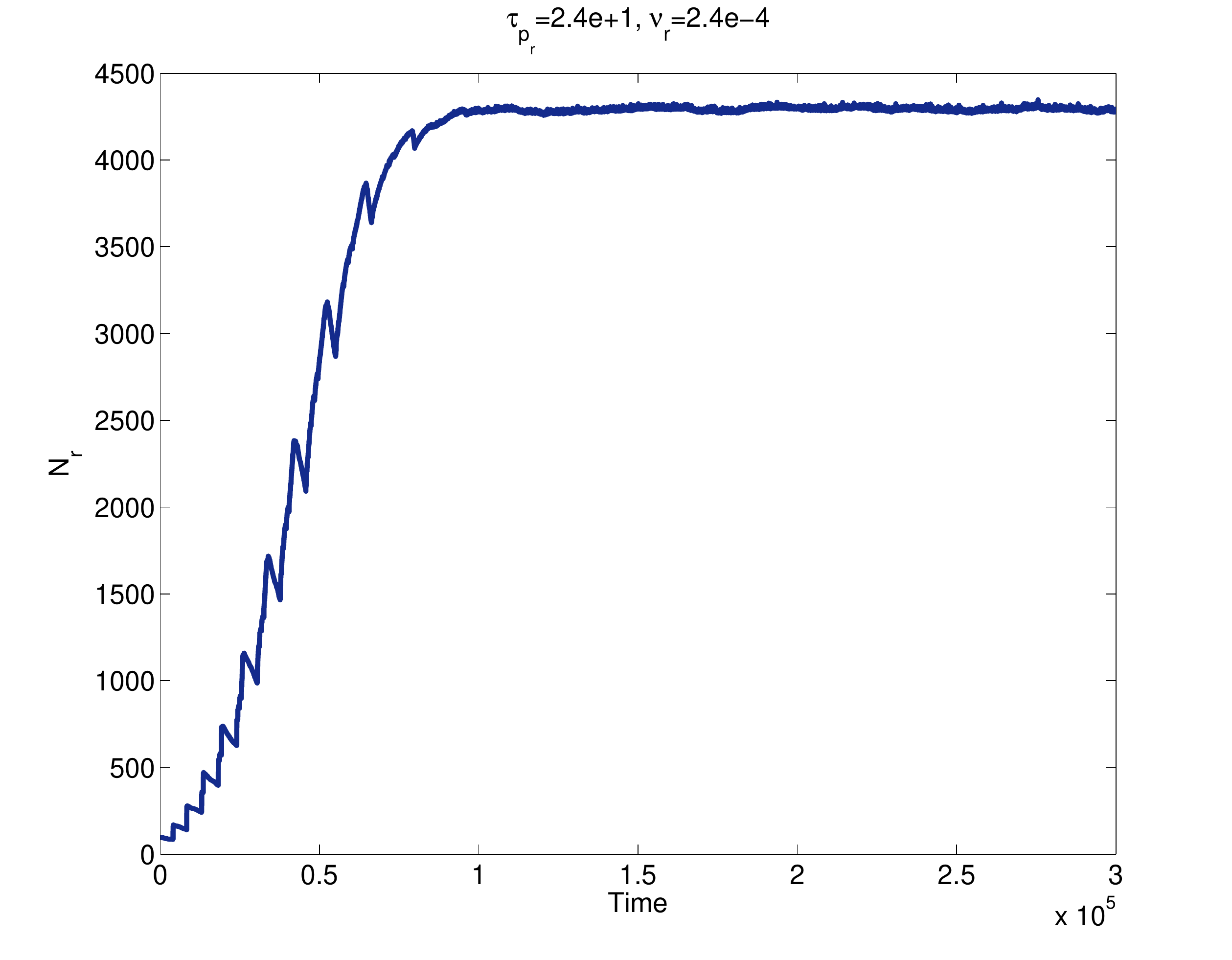} & \includegraphics[scale=0.3]{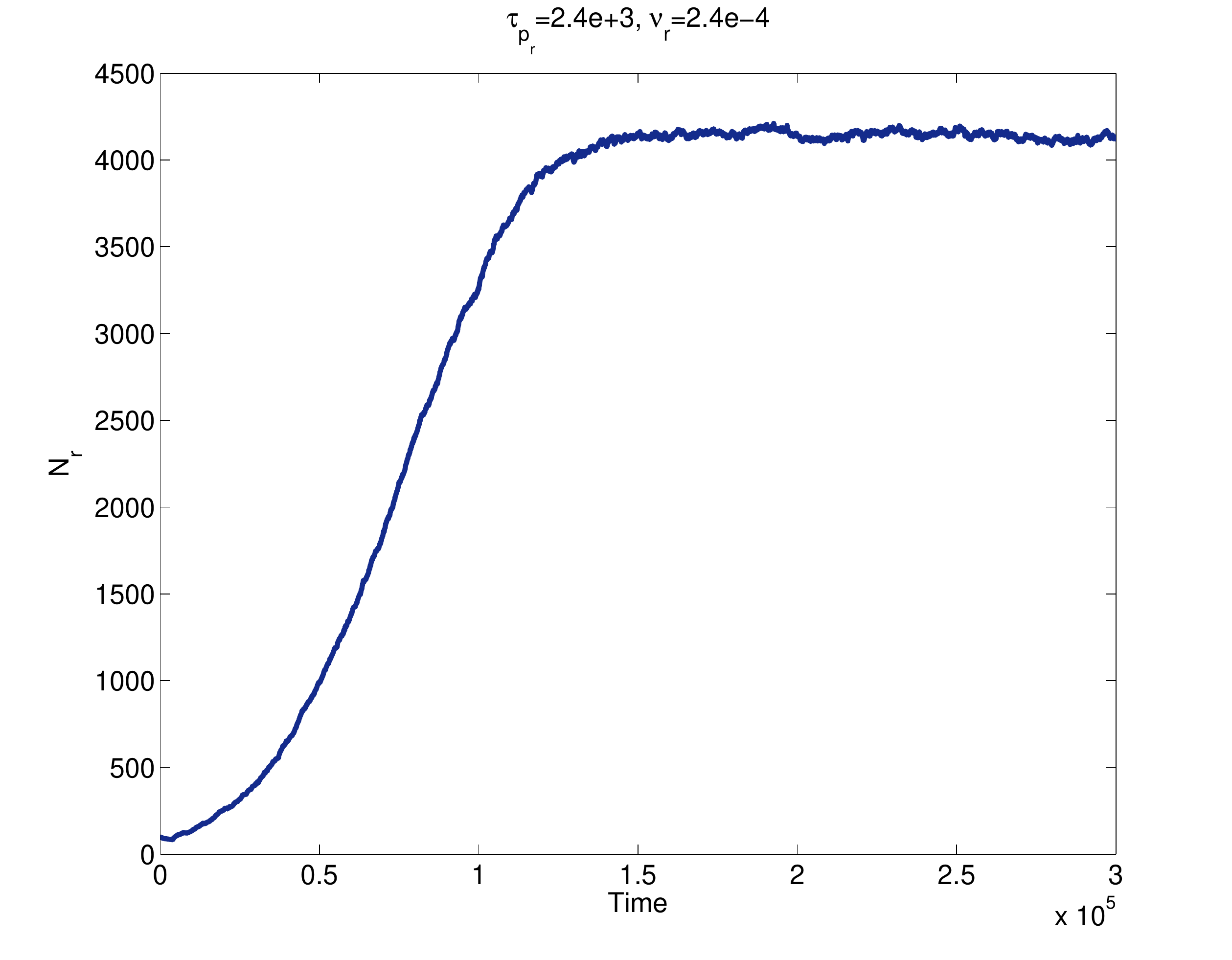}\\
\mbox{(c)} & \mbox{(d)} \\
\includegraphics[scale=0.3]{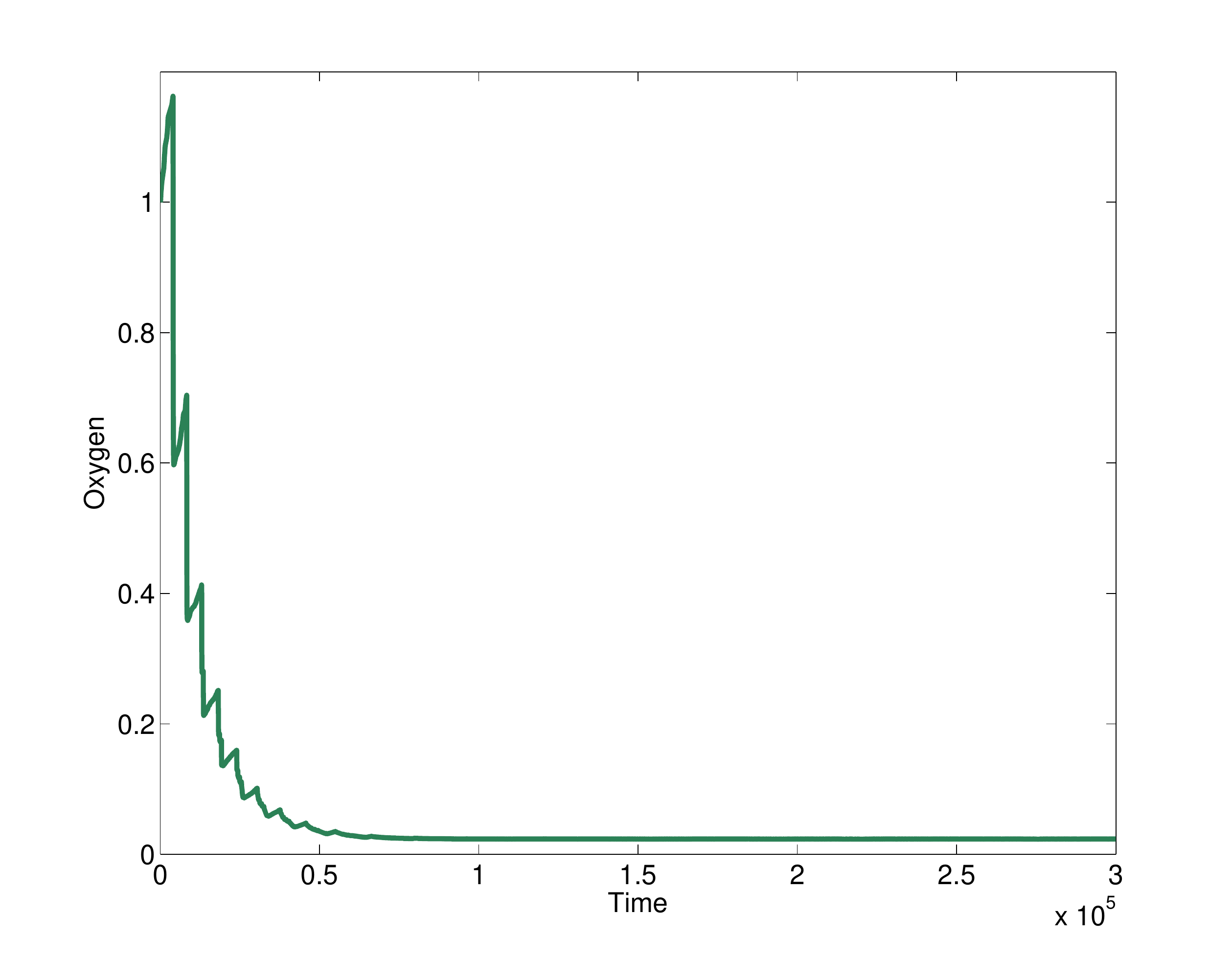} & \includegraphics[scale=0.3]{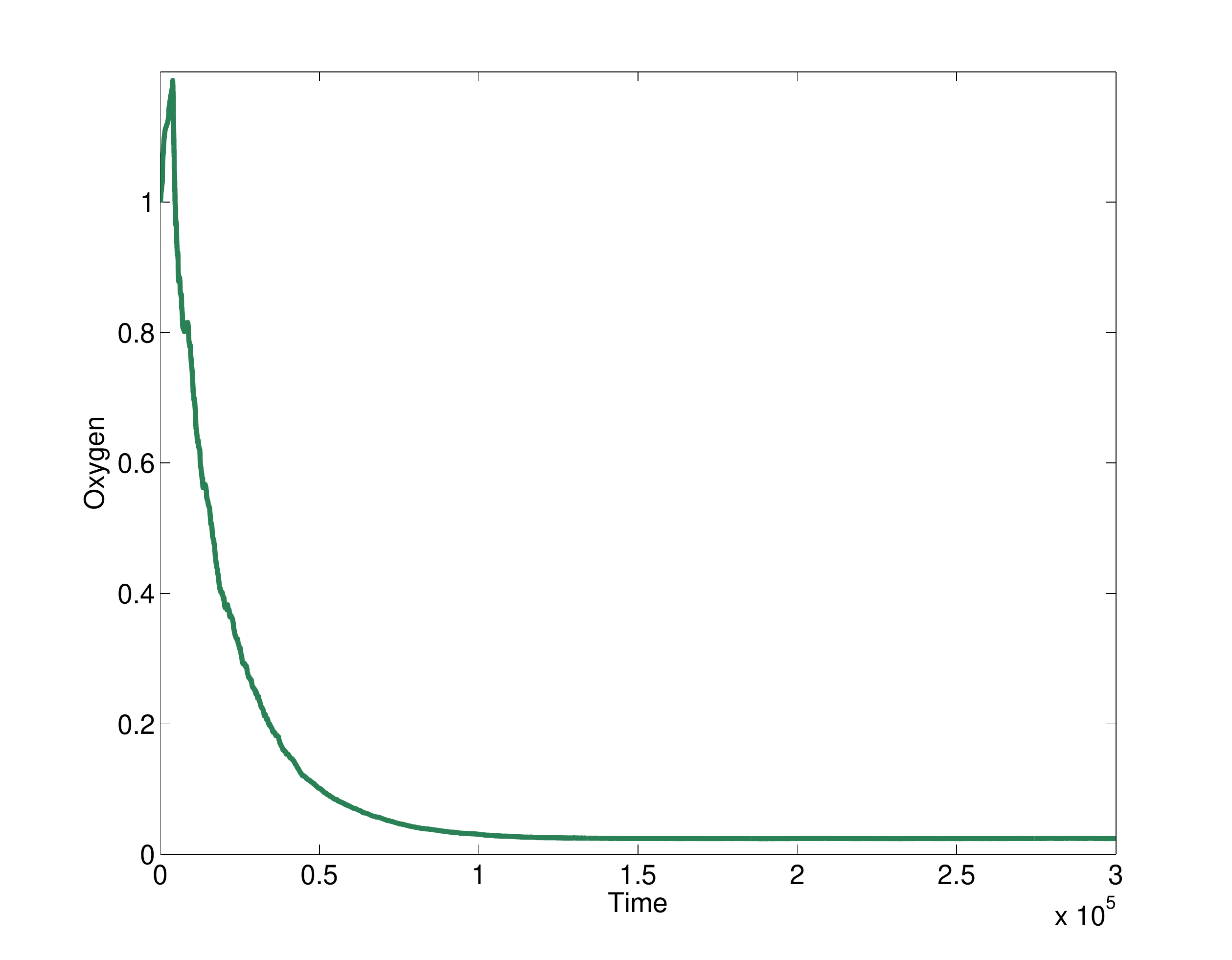}
\end{array}$ 
\caption{Plots showing simulation results of the stochastic multi-scale dynamics of a cell population. These plots show how, in agreement with our steady-state analysis, the population evolves until it reaches a 
steady-state where the population of resident cells fluctuates around its associated carrying capacity Eq. (\ref{eq:steadyNr}). Colour code: blue 
lines show the total resident cell population at time $t$, $N(t)$ (panels (a) \& (b)). Green lines (panels (c) \& (d)) show the associated oxygen concentration, $c(t)$. The results shown 
in this figure correspond to a single realisation of the process. Parameter values: $\nu=2.4\cdot 10^{-4}$, 
$S=1.57\cdot 10^{-2}$, $k=1.57\cdot 10^{-4}$, $\tau_{p}=2.4\cdot 10^1$ in panels (a) \& (c), and $\tau_{p}=2.4\cdot 10^3$ in panels (b) \& (d).}\label{fig:residentpop}
\end{center}
\end{figure}

In order to verify the age-structured SSA proposed in Section \ref{sec:ageSSA}, we compare its results with the mean-field predictions, which should be in agreement with the stochastic behaviour of the system for large values of the carrying capacity, $N_{\infty}$. Results are shown in Fig. 
\ref{fig:residentpop}. We observe that, as predicted by our steady-state analysis, the stochastic simulations show how the resident population goes 
through an initial (oxygen-rich) phase of exponential growth. As the population grows, oxygen is depleted and the resident population eventually 
saturates onto a number of cells which fluctuates around the mean-field prediction of the carrying capacity (Eq. (\ref{eq:steadyNr})). 

\begin{figure}
\begin{center}
$\begin{array}{cc}
\mbox{(a)} & \mbox{(b)} \\
\includegraphics[scale=0.3]{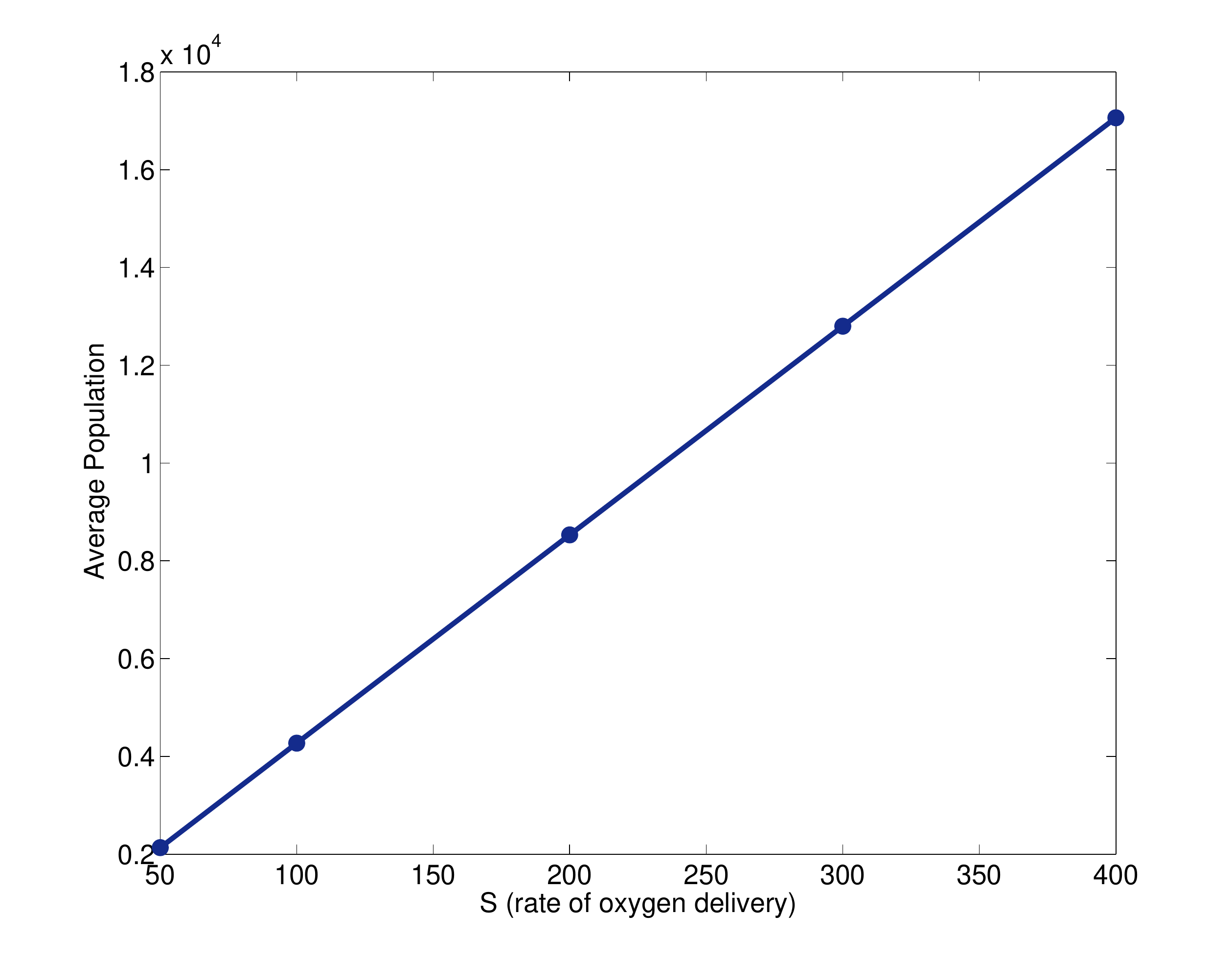} & \includegraphics[scale=0.3]{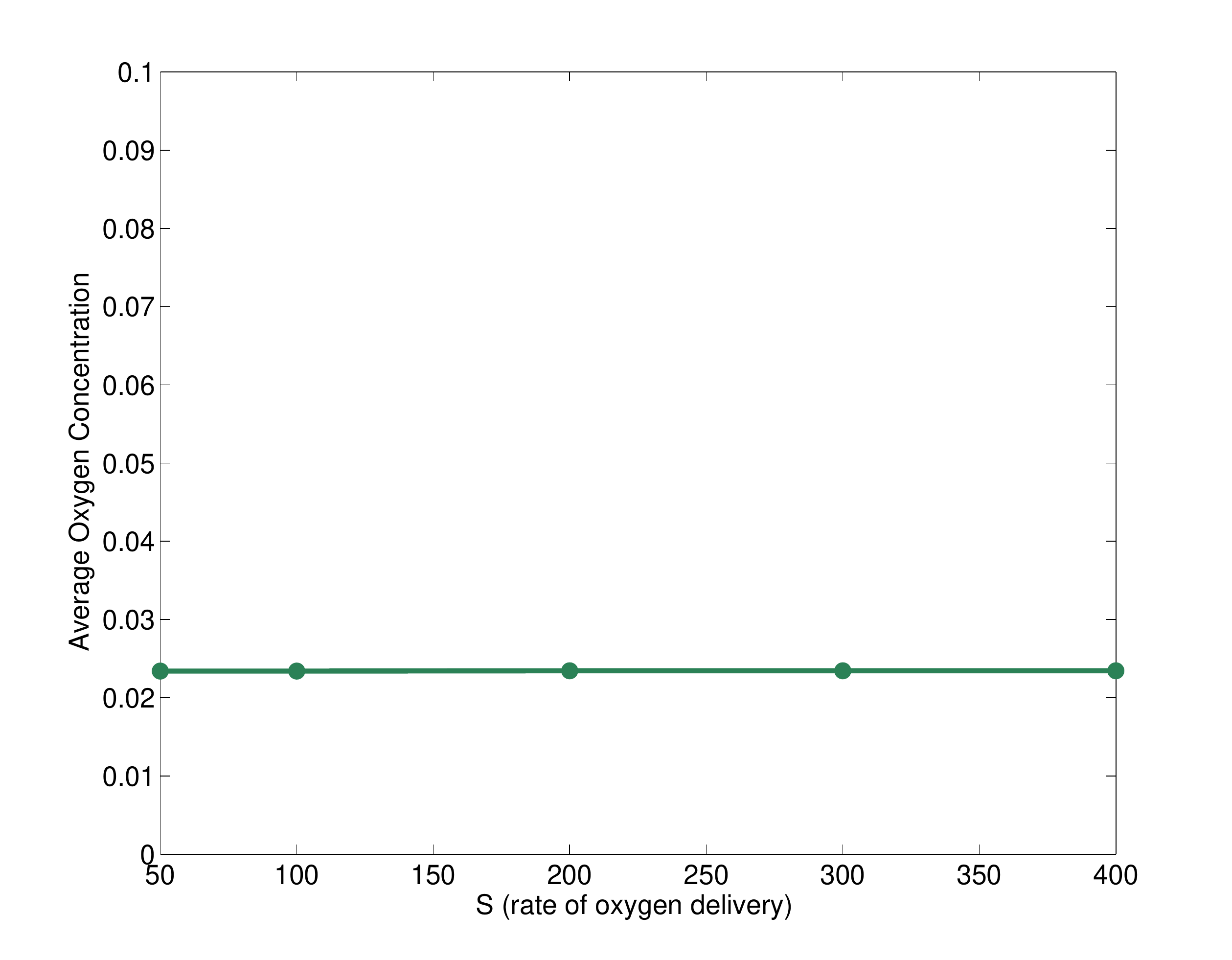}\\
\end{array}$ 
\caption{Plots showing simulation results of the stochastic multi-scale dynamics of a cell population. These plots show how, in agreement with our steady-state analysis, the average population at equilibrium increases linearly as the rate of oxygen delivery, $S$, is changed (plot (a)). Also, in accordance with our mean-field theory, the average, steady-state oxygen concentration remains unchanged as $S$ increases. Parameter values: $\nu=2.4\cdot 10^{-4}$, $k=1.57\cdot 10^{-4}$, $\tau_{p}=2.4\cdot 10^1$ in panels (a) \& (c), and $\tau_{p}=2.4\cdot 10^3$ in panels (b) \& (d). Average has been performed over 100 realisations.}\label{fig:meanfield}
\end{center}
\end{figure}

Further verification of the validity of our numerical methodology is provided in Fig. \ref{fig:meanfield}. Our mean-field theory predicts that, everything else remaining unchanged, the average steady-state population should increase linearly with the rate of oxygen delivery, $S$ (see Eq. \ref{eq:steadyNr}). By contrast, the average equilibrium oxygen concentration does not depend on $S$, as shown in Eq. (\ref{eq:aG1S}), and, consequently, it should stay constant upon increasing the rate of oxygen delivery. Fig. \ref{fig:meanfield} shows that our simulations agree with these mean-field predictions. 

\section{Quasi-neutral competition within heterogeneous populations}\label{sec:iec}

A problem of fundamental importance in several biological and biomedical contexts is that of a population composed by a heterogeneous mixture of coexisting cellular types. A particularly relevant example of such situation is that of cancer, where heterogeneity within the cancer cell population is assumed to be a major factor in the evolutionary dynamics of cancer as well as the emergence of drug resistance \cite{gatenby2003a,maley2006a,gillies2012,greaves2012}. Within this context, we are interested in (i) exploring the stability of a co-existing heterogeneous population, and (ii) the effects such heterogeneity has on the long-term effects a cell-cycle dependent therapy.

The the origins of heterogeneity of cancer cell populations is normally attributed to genetic variability arising from chromosomal instability and increased mutation rate \cite{maley2006a}. Our discussion of the model of the G1/S transition (see Section \ref{sec:intra}) suggests a different source of heterogeneity associated with stochastic effects due to intrinsic fluctuations. We have shown, by means of both the SCQSSA analysis and stochastic simulations, that the relative abundance of SCF-regulating enzymes, which is quantified in our analysis by the ratio $p_6/p_3$ (see Sections \ref{sec:timingg1s}, \ref{sec:scalingg1s} and \ref{sec:scqssa}). In Section \ref{sec:timingg1s} we have shown that the timing of the G1/S transition, and, consequently, the overall birth rate is strongly affected by changes in this quantity. In view, of this we associate heterogeneity to a distribution of the abundance of such enzymes, whereby we associate cell phenotypes with different values of the ratio $p_6/p_3$. We further 
assume that this heterogeneity is hereditary, \emph{i.e.} daughter cells inherit the value of the ratio $p_6/p_3$ from their mother.   

\subsection{Competition between two sub-populations}

To proceed further, we consider the case of the birth-and-death dynamics of a heterogeneous population composed by two sub-populations, $n_1(a,t)$ and $n_2(a,t)$, competing by a common resource, $c(t)$, which regulates the rate of progression of the cell-cycle of each cell type. The stochastic dynamics of the whole population is determined by the associated multi-scale master equation:  

\begin{eqnarray}\label{eq:MECompetition}
\nonumber \frac{\partial P(n_1,n_2,a,t)}{\partial t} + \frac{\partial P(n_1,n_2,a,t)}{\partial a} = && W_1(n_1+1,n_2,a,t)P(n_1+1,n_2,a,t)\\
\nonumber && +W_2(n_1,n_2+1,a,t)P(n_1,n_2+1,a,t) \\
\nonumber && -(W_1(n_1,n_2,a,t)+W_2(n_1,n_2,a,t))P(n_1,n_2,a,t). \\
\end{eqnarray}

\noindent where $W_1(n_1,n_2,a,t)= (\nu_1+b_1(a))n_1(a,t)$ and $W_2(n_1,n_2,a,t)= (\nu_2+b_2(a))n_1(a,t)$, $\nu_1$ and $b_1(a)$ are the (age-independent) death rate and the birth rate of the resident population, and $\nu_2$ and $b_2(a)$ are the (age-independent) death rate and the birth rate of the invader. The quantities $b_1(a)$ and $b_2(a)$ are determined by the (oxygen-dependent) rate of cell-cycle progression of each sub-population:

\begin{eqnarray}
\label{eq:birthrateres}
&&b_1(a)=\tau_{p_1}^{-1}H\left(a-a_{G1/S}(c,p_{6_1}/p_{3_1})\right), \\
\label{eq:birthrateinv}
&& b_2(a)=\tau_{p_2}^{-1}H\left(a-a_{G1/S}(c,p_{6_2}/p_{3_2})\right)
\end{eqnarray}

\noindent $a_{G1/S}(c,p_6/p_3)$ is given by Eq. (\ref{eq:ag1sscaling}). The concentration of resource (oxygen), $c(t)$, is determined by the following ODE:

\begin{equation}\label{eq:oxygencompetition}
\frac{dc}{dt}=S-(k_1N_1+k_2N_2)c 
\end{equation}

\noindent where $N_i(t)=\int_0^{\infty}n_i(a,t)da$ for $i=1,2$.

The associated initial conditions are given by:

\[P(n_1,n_2,a,t=0)=\delta(n_1(a,t=0)-n_{0_1}(a))\delta(n_2(a,t=0)-n_{0_2}(a))\]

In the remaining of this Section we will consider two cellular populations, resident and invader. Each of these phenotypes are determined in terms of 
the values four parameters. The resident cells are characterised by two population-dynamics parameters, namely, the average time to division after 
the G1/S transition, $\tau_{p_1}$, and the death rate, $\nu_1$. We consider two further parameters, $p_{3_1}$ and $p_{6_1}$, 
associated with the cell-cycle progression dynamics of the resident cells (see Section \ref{sec:semiclassg1s}). Similarly, the invader is characterised 
by the corresponding parameters: $\tau_{p_2}$, $\nu_2$, $p_{3_2}$ and $p_{6_2}$  

\subsection{Mean-field coexistence versus quasi-neutral stochastic competition}\label{sec:quasineutral}

We proceed to analyse the conditions under which two populations are capable of long-term coexistence. In particular, we analyse a scenario 
in which the mean-field description predicts long term coexistence between within a heterogeneous population leads to mutual exclusion of all strands 
but one through so-called quasi-neutral stochastic competition \cite{lin2012,kogan2014,guerrero2015a}. 

The starting point for our study is the mean-field analysis carried out in Section \ref{sec:MFStSt}. According to these results, (mean-field) 
populations evolve until a concentration of oxygen, $c_{\infty}$, is reached so that the associated replication number $R_{0_i}(c_{\infty})=1$, $i=1,2$. The replication number of either population is given by:

\[R_{0_1}= 2 \frac{\tau_{p_1}^{-1}e^{-\nu_1a_{G1/S_1}(c_{\infty},p_{6_1}/p_{3_1})}}{\nu_1+\tau_{p_1}^{-1}}\mbox{ and 
}R_{0_2}= 2 \frac{\tau_{p_2}^{-1}e^{-\nu_2a_{G1/S_2}(c_{\infty},p_{6_2}/p_{3_2})}}{\nu_2+\tau_{p_2}^{-1}}\]

\noindent Therefore, our theory predicts that, provided that there exists $c_{\infty}$ such that:

\begin{eqnarray}\label{eq:coexcond}
\nonumber && a_{G1/S_1}(c_{\infty},p_{6_1}/p_{3_1})=-\frac{1}{\nu_1}\log\left(\frac{\tau_{p_1}(\nu_1+\tau_{p_1}^{-1})}{2}
\right), \\
&& a_{G1/S_2}(c_{\infty},p_{6_2}/p_{3_2})=-\frac{1}{\nu_2}\log\left(\frac{\tau_{p_2}(\nu_2+\tau_{p_2}^{-1})}{2}
\right), 
\end{eqnarray}
 
\noindent is satisfied, long-term coexistence ensues, since the whole system (oxygen, resident and invader) is able to evolve to a state where 
both populations are in equilibrium (in the sense that $R_{0_i}(c_{\infty})=1$ for both populations) with the same concentration of oxygen. 
Furthermore, the number of resident cells, $N_1$, and the number of invaders, $N_2$, satisfy:

\begin{equation}\label{eq:carryingcapacitycoex}
k_1N_1+k_2N_2=\frac{S}{c_{\infty}} 
\end{equation}

\noindent Thus, the mean-field theory predicts that there exist a continuous of fixed points. The eventual convergence on to a particular point 
along the line of fixed points Eq. (\ref{eq:carryingcapacitycoex}) depends on the initial conditions. Eq. (\ref{eq:carryingcapacitycoex}) can be 
further simplified by assuming $k_1=k_2=k$, in which case $N_1+N_2=K$, where $K\equiv S/(kc_{\infty})$ is the carrying capacity. 

This mean-field scenario is the basis for the study of the long-term stochastic dynamics of two populations which satisfy Eqs. (\ref{eq:coexcond}), 
which are equivalent to $R_{0_1}=R_{0_2}=1$. The reproduction number is the average number of offspring per cell. We know from 
elementary considerations \cite{grimmett2001} that the value of such quantity allows us to classify birth-death/branching processes. If $R_0>1$ the 
population grows, on average, exponentially and has a finite probability of eventual survival. In this case, the process is referred to as 
super-critical. If $R_0<1$ the population undergoes exponential decline (on average) and the extinction probability is equal to 1. The last case, in 
which $R_0=1$, the so-called critical case, the population, on average, stays constant. However, due to effects of noise, extinction occurs with 
probability 1, with the probability of survival up to time $t$ asymptotically tends to $P_S(t)\sim t^{-1}$ \cite{kimmel2002}. In our case, both the 
resident and the invader undergo a critical stochastic dynamics, where, once the steady state has established itself, the population evolves 
very close to the mean-field line of fixed points $N_1+N_2=K$ until fixation of one of the species (and, consequently, extinction of the other) 
occurs. 

\begin{figure}
\begin{center}
\includegraphics[scale=0.5]{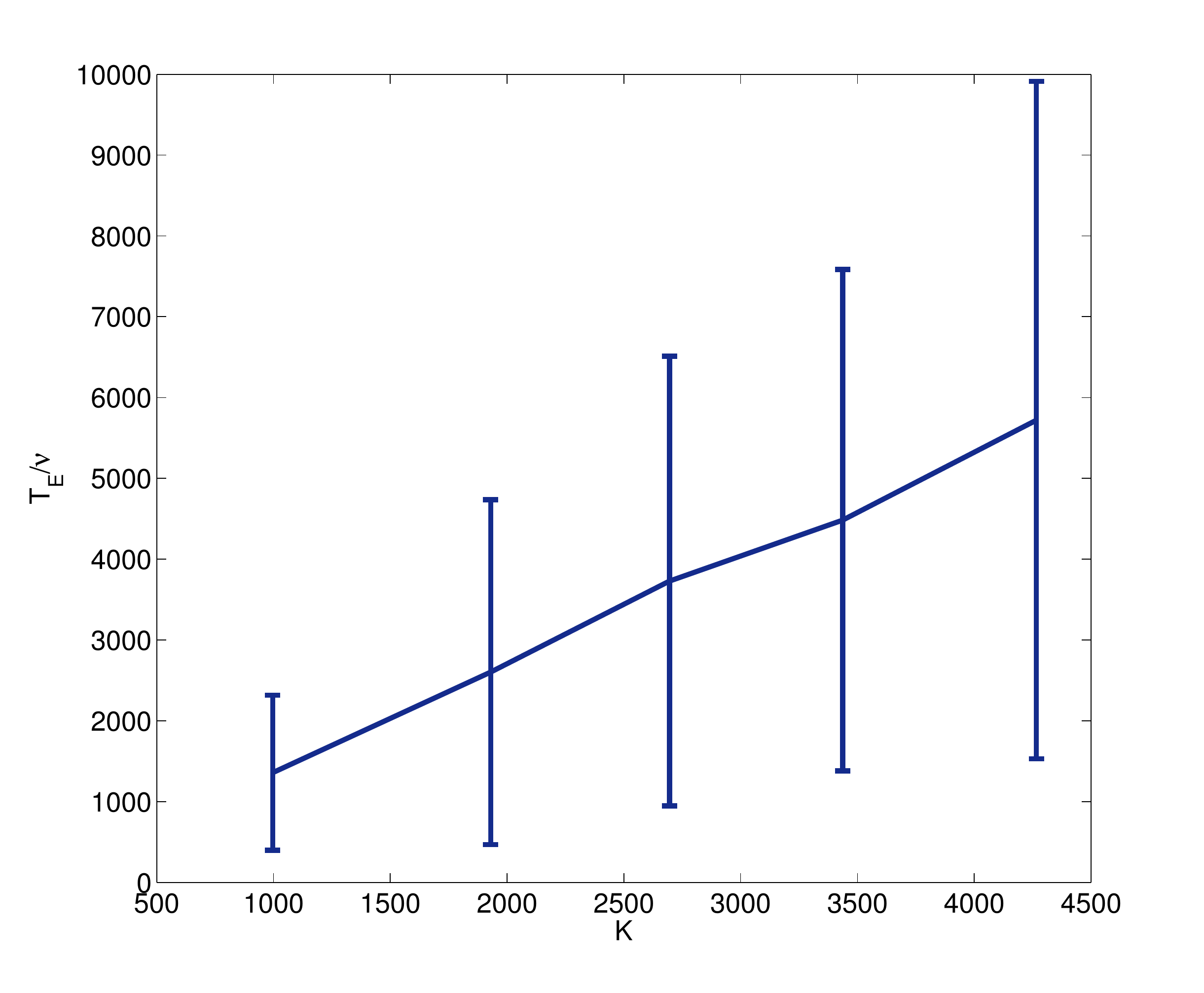} 
\caption{Simulation results corresponding to the competition between the sub-populations of a heterogeneous populations. We show how the average 
extinction time of either population, $T_E$, varies as the carrying capacity, $K$, changes. We observe that the dependence is algebraic (square 
root). For simplicity, the two sub-populations are assumed to be identical (\emph{i.e.} with the same characteristic parameter 
values for both the intracellular dynamics (cell-cycle) and the population-level dynamics (birth and death rates)). The 
carrying capacity $K=\frac{S}{k_1c_{\infty}}$ by varying the death rates of both populations. The values of are $\nu=1.0\cdot 10^{-4},\,0.83\cdot 
10^{-4},\,0.73\cdot 10^{-4},\,0.625\cdot 10^{-4},\,0.417\cdot 10^{-4}$, which correspond to $K=0.9969\cdot 10^{3}, 1.9301\cdot 10^{3}, 2.6956\cdot 10^{3}, 3.4367\cdot 10^{3}, 4.2661\cdot 10^{3}$, 
respectively. Averages are done over 500 realisations of the hybrid 
stochastic model.}\label{fig:quasineutral}
\end{center} 
\end{figure}

In order to numerically check this scenario, we could proceed to estimate the survival probability at time $t$, $P_S(t)$. However, since this 
quantity exhibits a fat-tail behaviour, this would be computationally costly. A more efficient method is to resort to the asymptotic of the 
extinction time with system size, which in this case can be identified with the carrying capacity, $K$ \cite{hidalgo2015}. Typically, a quasi-neutral 
competition is associated with an algebraic dependence of the average extinction time of either population, $T_E$, on the system size, in this case 
determined by the carrying capacity, $K$ \cite{doering2005,lin2012,kogan2014}. In Fig. \ref{fig:quasineutral} we plot simulation results for the 
competition between two identical populations. In particular, we study how the average extinction time of either population, $T_E$, varies as the 
carrying capacity is changed. We observe that this quantity exhibits a linear dependence on the carrying capacity:

\begin{equation}\label{eq:extinctiontimecritical} 
T_E\sim K, 
\end{equation}

Our scenario produces the same qualitative results as Lin et al. \cite{lin2012} and Kogan et al. \cite{kogan2014} who studied the average 
extinction time of birth-and-death processes engaged in quasi-neutral competition. In both papers, the average extinction time was reported to depend linearly on system size.



\section{Study of the effects of cell-cycle-dependent therapy}\label{sec:therapy}

In Section \ref{sec:iec} we have analysed how the dependence of the cell-cycle progression on the concentration of SCF-activating and SCF-inactivating enzymes allows us to engineer heterogeneous populations where invasion and coexistence may occur. In this Section, we further explore the ability of inducing quiescence by varying the ratio between SCF-activating and SCF-inactivating enzymes, this time in connection with the ability of such populations to withstand the effects of cell-cycle-dependent therapy \cite{powathil2012,gabriel2012,billy2013,powathil2013}. 

In particular, we consider an scenario where two cellular populations coexist. Initially, one of these populations consists of a set of cells actively progressing through the cell-cycle which have reached a steady state characterised by the mean-field equilibrium Eqs. (\ref{eq:R0})-(\ref{eq:steadyNr}). The second population consists of quiescent cells whose cell-cycle is locked into the G0 phase and therefore do not proliferate. These cells are further assumed to undergo apoptosis at a very slow rate. More specifically, the ratio SCF-activating and SCF-inactivating enzymes in the quiescent cell population is such that, for the steady-state level of oxygen for the active cells (see Eq. (\ref{eq:aG1S})), cells are locked into the G1-fixed point (see Fig. \ref{fig:bif_pe1_pe2}).

In this Section we show that the presence of a quiescent population within a heterogeneous population may lead to resistance to cell-cycle-dependent therapy. In particular, we show that, whereas such therapies effectively reduce or even eradicate the active cell population, the feedback between the therapy-induced decrease in cell numbers and the associated increase in oxygen availability can yield to the quiescent population to enter the active state and thus regrow the population. In this sense, we claim that the quiescent population has a \emph{stem-cell-like} effect whereby, under the action of therapeutic agent, can repopulate the system \cite{alarcon2010b}.

\subsection{Mean-field analysis}\label{sec:mftherapy}

We start our mean field analysis by considering a heterogeneous population composed by cells of two types: type 1 and type 2 cells. Type 1 consists of cells with values of $p_3\equiv p_{3_1}$ and $p_6\equiv p_{6_1}$ (see Section \ref{sec:semiclassg1s}) such that cells are actively progressing through the cell-cycle. Type 2 cells are characterised by values of $p_3\equiv p_{3_2}$ and $p_6\equiv p_{6_2}$ so that they are locked in G0 (\emph{i.e.} not cycling). The associated mean-field dynamics are given by:

\begin{eqnarray}\label{eq:mfnother}
\label{eq:mfnother1} && \frac{dc}{dt}=S-k_1(N_1+N_2)c, \\
\label{eq:mfnother2} && \frac{\partial Q_1}{\partial t}+\frac{\partial Q_1}{\partial a}=-(\nu_1+b_1(a))Q_1\\
\label{eq:mfnother2} && \frac{\partial Q_2}{\partial t}+\frac{\partial Q_2}{\partial a}=-\nu_2Q_2
\end{eqnarray}

\noindent with boundary conditions:

\[Q_1(a=0,t)=2\int b_1(a)Q_1(a,t)da,\;Q_2(a=0,t)=0.\]

\noindent $N_1(t)$ and $N_2(t)$ are the total cell population of type 1 and type 2 cells, respectively:

\[N_1(t)=\int Q_1(a,t)da,\;N_2(t)=\int Q_2(a,t)da.\]

\noindent For simplicity, we assume that both populations consume oxygen at the same rate, $k_1$. We further assume that $\nu_2\ll\nu_1$, \emph{i.e.} type 2 (quiescent) cells die at a much slower rate than type 1 (active) cells. In Section \ref{sec:MFStSt}, we have already analysed under which conditions Eqs. (\ref{eq:mfnother1})-(\ref{eq:mfnother2}) reach a steady state:

\begin{equation}\label{eq:R01}
R_{0_1}\equiv 2 \frac{\tau_{p_1}^{-1}e^{-\nu_1a_{G1/S_1}}}{\nu_1+\tau_{p_1}^{-1}}=1
\end{equation}

\noindent where $R_{0_1}$ is the average number of offspring per cell of type 1 at equilibrium. The associated equilibrium value of $a_{G1/S_1}$ is then given by:

\begin{equation}\label{eq:aG1S1}
a_{G1/S_1}(c_{\infty},p_{6_1}/p_{3_1})=-\frac{1}{\nu_1}\log\left(\frac{(\tau_{p_1}\nu_1+1)}{2}\right) 
\end{equation}

\noindent which determines the steady-state value of the oxygen concentration $c_{\infty}$. At this point, the population of resident cells has settled onto a steady state where its age structure does not change. The total number of cells is also approximately constant and given by:

\begin{equation}\label{eq:steadyN1}
N_{1_{\infty}}+N_{2}\simeq\frac{S}{k_1c_{\infty}} 
\end{equation}

\noindent where we have used the fact that $\nu_2\ll\nu_1$. The cell-cycle parameters $p_{3_2}$ and $p_{6_2}$ have been chosen so that, for $c=c_{\infty}$, $a_{G1/S_2}(c_{\infty},p_{6_2}/p_{3_2})\rightarrow\infty$, \emph{i.e.} type 2 cells are initially locked into G0 (see Fig. \ref{fig:bif_pe1_pe2}). 

Once the population reaches this therapy-free quasi-equilibrium state, we assume that a therapy which only acts on proliferating cells is administered. Examples of such therapies abound in cancer treatment and can take the form of cell-cycle specific drugs or radiotherapy \cite{powathil2012,gabriel2012,billy2013,powathil2013}. We characterise the efficiency of the therapy by the so-called survival fraction, $F_S$, \emph{i.e.} the percentage of cells which survive the prescribed dose. For example, in radiotherapy $F_S$ is usually taken to be given by the linear quadratic model: $\log F_S=-\left(\alpha D+\beta D^2\right)$ where $D$ stands for the radiation dosage expressed in Grays and $\alpha$ and $\beta$ are cell type-specific parameters. In the present context, we do not specify any particular form of therapy and we simply take $F_S\in[0,1)$. Initially, the therapy only affects type 1 cells (since type 2 are not proliferating). Therapy affects the birth and death rates of the type 1 population, which now read:

\begin{eqnarray}
\label{eq:br1ther} && b_{T_1}(a)=\tau_{p_1}^{-1}F_SH(a-a_{G1/S_1}(c,p_{3_1},p_{6_1}))=F_sb_{1}(a)\\  
\label{eq:dr2ther} && \nu_{T_1}(a)=\nu_{1}+(1-F_s)b_{1}(a)
\end{eqnarray}

The resulting mean-field equation is given by:

\begin{eqnarray}
\label{eq:mfther1} && \frac{\partial Q_1}{\partial t}+\frac{\partial Q_1}{\partial a}=-(\nu_1+b_1(a))Q_1\\
\label{eq:mfther2} && Q_1(a=0,t)=2F_S\int b_1(a)Q_1(a,t)da
\end{eqnarray}

\noindent The action of therapeutic agent on the active population initially induces a decline of the population, which, in turn, involves an increase in the available oxygen concentration. The latter has the effect of accelerating the rate of progression of type 1 cells through the cell cycle. In the absence of the quiescent population, eventually both effects would find a balance and the population of active cells would settle onto a new equilibrium characterised by:

\begin{equation}\label{eq:R0T1}
R_{0_{T_1}}\equiv 2 F_S\frac{\tau_{p_1}^{-1}e^{-\nu_1a_{G1/S_{T_1}}}}{\nu_1+\tau_{p_1}^{-1}}=1
\end{equation}

\noindent or, equivalently:

\begin{equation}\label{eq:aG1ST1}
a_{G1/S_{1}}(c_{T_{\infty_1}},p_{3_1},p_{6_1})=-\frac{1}{\nu_1}\left(-\log F_S+\log\left(\frac{(\tau_{p_1}\nu_1+1)}{2}\right)\right) 
\end{equation}

\noindent where $c_{T_{\infty_1}}$ is the equilibrium oxygen concentration for the type 2 population with therapy. Note that $a_{G1/S_{1}}(c_{T_{\infty_1}},p_{3_1},p_{6_1})<a_{G1/S_1}(c_{\infty},p_{6_1}/p_{3_1})$ and therefore $c_{T_{\infty_1}}>c_{\infty}$ since $a_{G1/S}$ is a decreasing function of the oxygen concentration. Re-oxygenation during cell-cycle dependent therapy, in particular, radiotherapy, has been predicted by other models \cite{kempf2015}.

Consider now the effect of this process on the type 2 cell population which is initially quiescent. We know that hypoxia-induced arrest of the cell cycle is reversible \cite{alarcon2004,bedessem2014}, \emph{i.e.} upon increase of the concentration of oxygen quiescent cells may re-enter the cell cycle and become proliferating. Re-entry of quiescent cells into the cell cycle is predicated upon a sufficient increase in the oxygen contraction: $c>c_{H_2}(p_{3_2},p_{6_2})$, where the critical oxygen concentration for type 2 cells, $c_{H_2}$, depends on the momenta $p_{3_2}$ and $p_{6_2}$ or, equivalently, on the concentration of SCF-activating and SCF-inactivating enzymes. Taking this property into account, one can devise a scenario in which $c_{T_{\infty_1}}>c_{H_2}(p_{3_2},p_{6_2})>c_{\infty}$, i.e the initial oxygen concentration is such that type 2 cells are quiescent, but, as the therapy is administered and proceeds to act upon the type 1 cells, the oxygen concentration increases until it reaches its critical re-
entry concentration. At this point, type 2 cells abandon quiescence and become active and competition between type 1 and type 2 cells ensues.

In order to assess the long-time behaviour of the system, we first study the equilibrium of the type 2 cell population upon re-entry into cell-cycle progression. Its mean-field dynamics is given by:

\begin{eqnarray}
\label{eq:mf2ther1} && \frac{\partial Q_2}{\partial t}+\frac{\partial Q_2}{\partial a}=-(\nu_2+b_2(a))Q_1\\
\label{eq:mf2ther2} && Q_2(a=0,t)=2F_S\int b_2(a)Q_2(a,t)da
\end{eqnarray}

\noindent where $b_2(a)=\tau_{p_2}^{-1}H(a-a_{G1/S_2}(c,p_{3_2},p_{6_2}))$ and $c$ is determined by Eq. (\ref{eq:mfnother1}). Recall that, upon re-entering cell-cycle progression, type 2 cells are no longer immune to the therapy. Although in general the survival fraction is type-dependent, for simplicity we assume that $F_S$ has the same value for both cell types. The equilibrium condition is once again given in terms of the associated reproduction number, \emph{i.e.} $R_{0_{T_2}}=1$ which yields:

\begin{equation}\label{eq:aG1ST2}
a_{G1/S_{2}}(c_{T_{\infty_2}},p_{3_2},p_{6_2})=-\frac{1}{\nu_2}\left(-\log F_S+\log\left(\frac{(\tau_{p_2}\nu_2+1)}{2}\right)\right) 
\end{equation}

Since $\nu_2\ll\nu_1$, we have that $a_{G1/S_{2}}(c_{T_{\infty_2}},p_{3_2},p_{6_2})\gg a_{G1/S_{1}}(c_{T_{\infty_1}},p_{3_1},p_{6_1})$ (see Eqs. (\ref{eq:aG1ST1}) and (\ref{eq:aG1ST2})). The latter inequality implies that the equilibrium oxygen concentration for type 2 cell, $c_{T_{\infty_2}}$, is such that $c_{T_{\infty_2}}<c_{T_{\infty_1}}$. It is easy to argue that, in these conditions, the type 2 population out-competes the type 1 cells: for $c_{T_{\infty_2}}<c<c_{T_{\infty_1}}$, the growth rate of type 1 cells is positive whereas the growth rate of the type 2 cell population is negative (see Eq. (\ref{eq:peq})). This implies that, upon application of therapy and provided that $c_{T_{\infty_1}}>c_{H_2}(p_{3_2},p_{6_2})>c_{\infty}$ is satisfied, the type 1 population declines and it is replaced by the type 2 population.

\subsection{Critical dosage}\label{sec:critdose}

In order to gain some degree of control over the behaviour described in the previous section, it would be useful to provide an estimate of the critical dosage above which therapy-induced re-oxygenation is capable of activating quiescent cells. We characterise the therapy dose by means of the critical survival fraction, $F_{S_C}$. Recall that the characteristic equation for the oxygen-dependent growth rate, $\sigma_1(c)$, of the population of active cells is given by:

\begin{equation}\label{eq:chareqFS}
2F_S\frac{\tau_{p_1}^{-1}e^{-(\sigma_1+\nu_1)a_{G1/S_1}(c)}}{\sigma_1+\nu_1+\tau_{p_1}^{-1}}=1 
\end{equation}

\noindent In order to activate the quiescent population the oxygen concentration must raise above the critical value $c_{H_2}(p_{3_2},p_{6_2})$. For the oxygen concentration to grow above this threshold $\sigma_1(c_{H_2})<0$ so that the active cell population continues to decline thus allowing the oxygen concentration to keep on raising. Therefore the critical value $F_{S_C}$ is such that $\sigma_1(c_{H_2})=0$, i.e.

\begin{equation}\label{eq:FSC}
F_{S_C}=\frac{\nu_1+\tau_{p_1}^{-1}}{2\tau_{p_1}^{-1}e^{-\nu_1a_{G1/S_1}(c_{H_2}(p_{3_2},p_{6_2})}} 
\end{equation}

\noindent If $F_S<F_{S_C}$ the decrease in the active cell population is enough to provide enough oxygen for the quiescent population to become 
active. 

This analysis implies that in heterogeneous populations which include quiescent sub-populations, the effect of cell-cycle-dependent therapy does not eradicate the population. Rather the following two scenarios are possible. If $c_{H_2}(p_{3_2},p_{6_2})>c_{T_{\infty_1}}$, type 2 cells do not become activated and become eventually extinct, and the type 1 cells settle onto the steady state prescribed by Eq. (\ref{eq:aG1ST1}). If, by contrast, $c_{T_{\infty_1}}>c_{H_2}(p_{3_2},p_{6_2})>c_{\infty}$ is satisfied, type 2 cells out-compete type 1 cells and the system, composed entirely of type 2 cells, settles onto the steady state prescribed by Eq. (\ref{eq:aG1ST2}). In this sense, quiescent cells have \emph{stem-cell-like behaviour}, in the sense that they can repopulate the system. In this second scenario, therapy is not completely without virtue since therapy drives the system to be taken over by a slower-cycling (less aggressive) phenotype.  

\subsection{Simulation results}

\begin{figure}
\begin{center}
$\begin{array}{cc}
\mbox{(a)} & \mbox{(b)} \\
\includegraphics[scale=0.3]{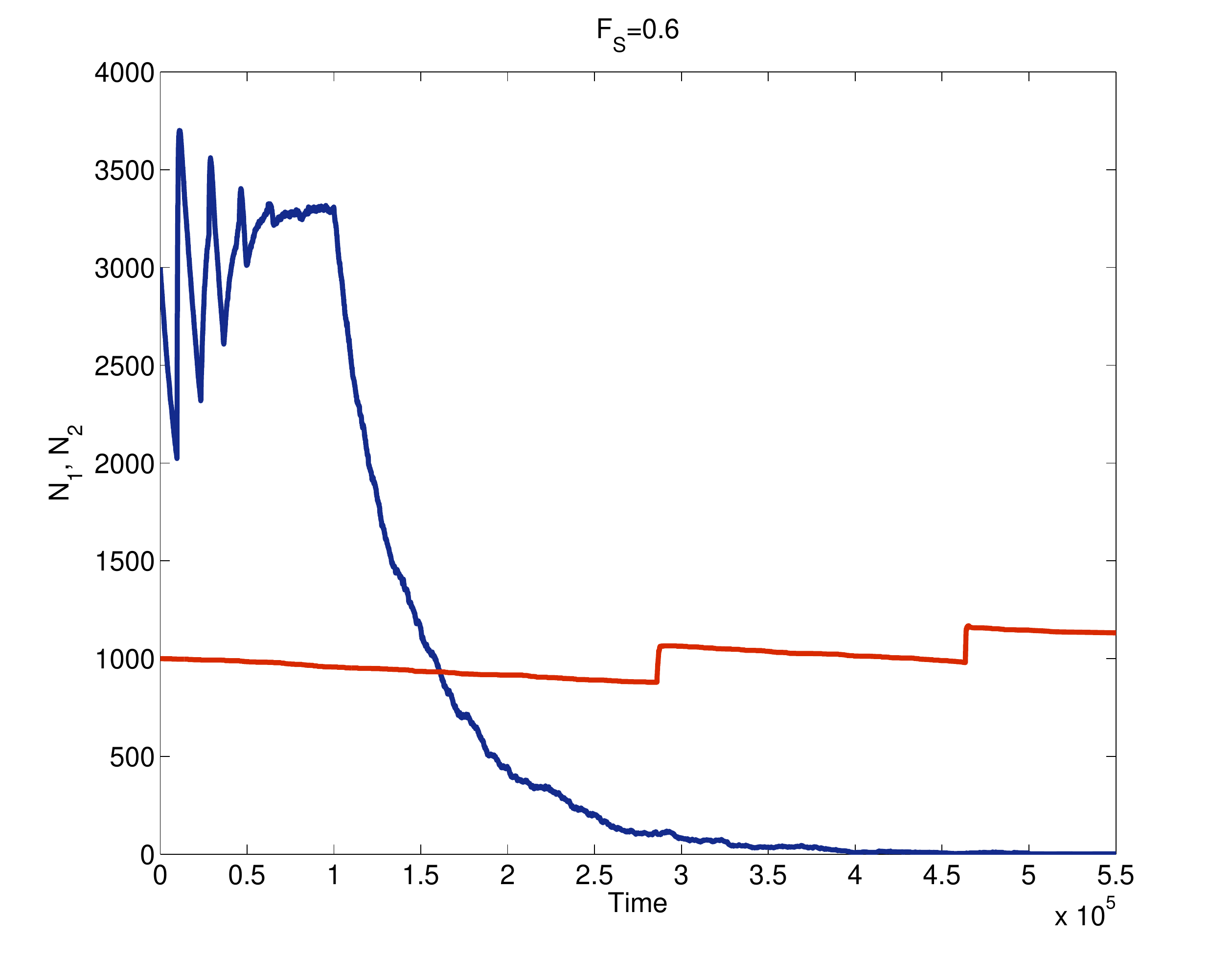} & \includegraphics[scale=0.3]{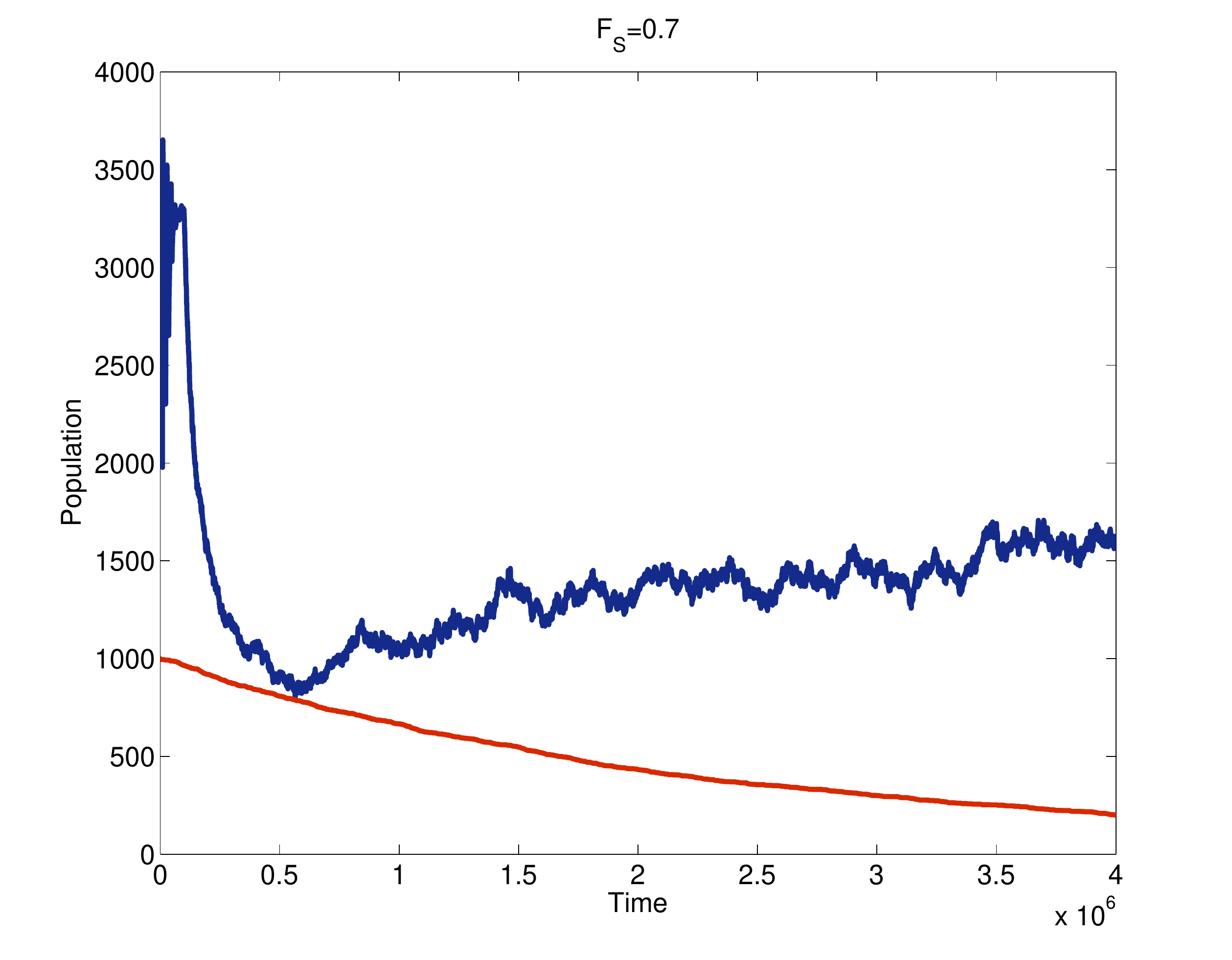} \\
\mbox{(c)} & \mbox{(d)} \\
\includegraphics[scale=0.3]{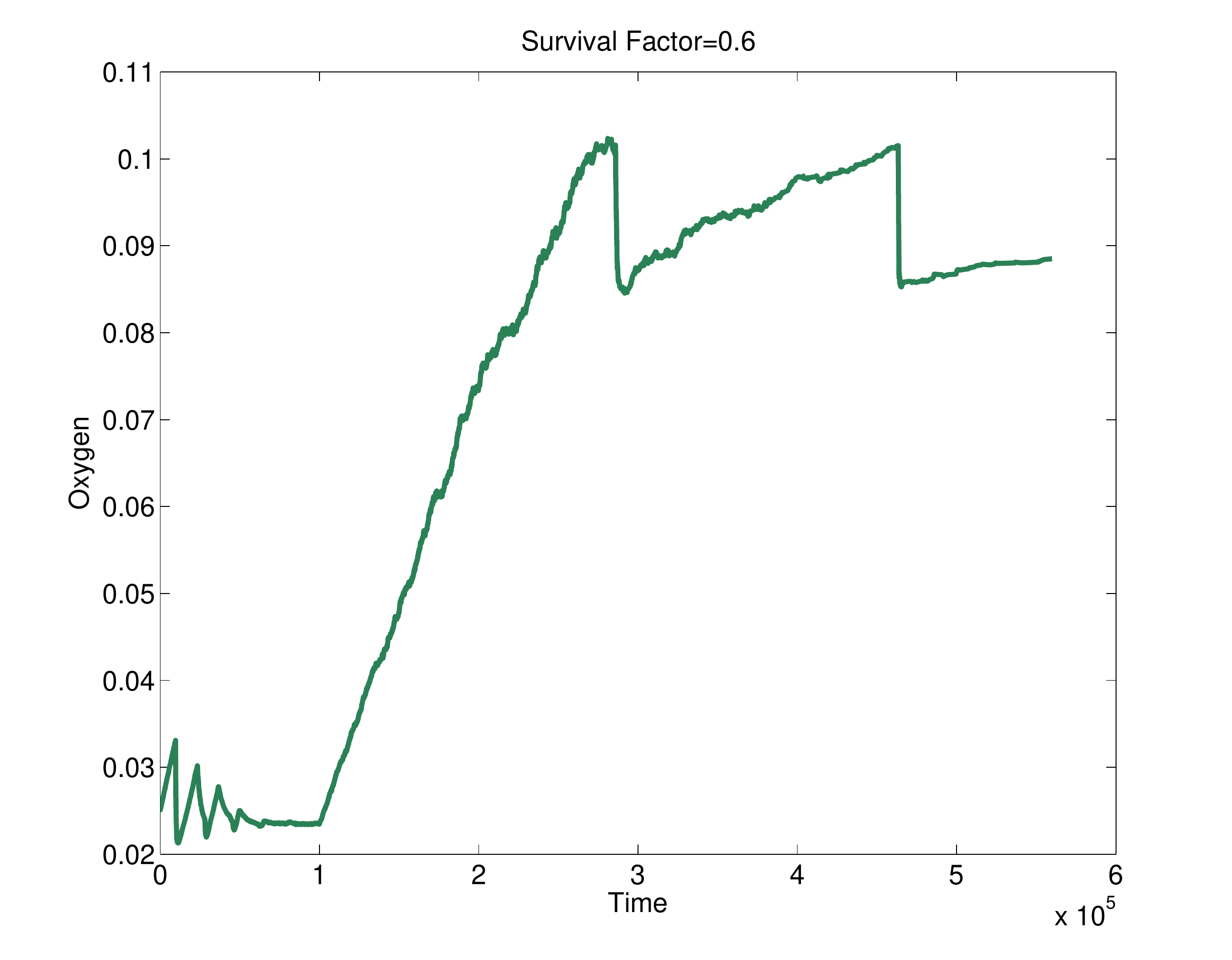} & \includegraphics[scale=0.3]{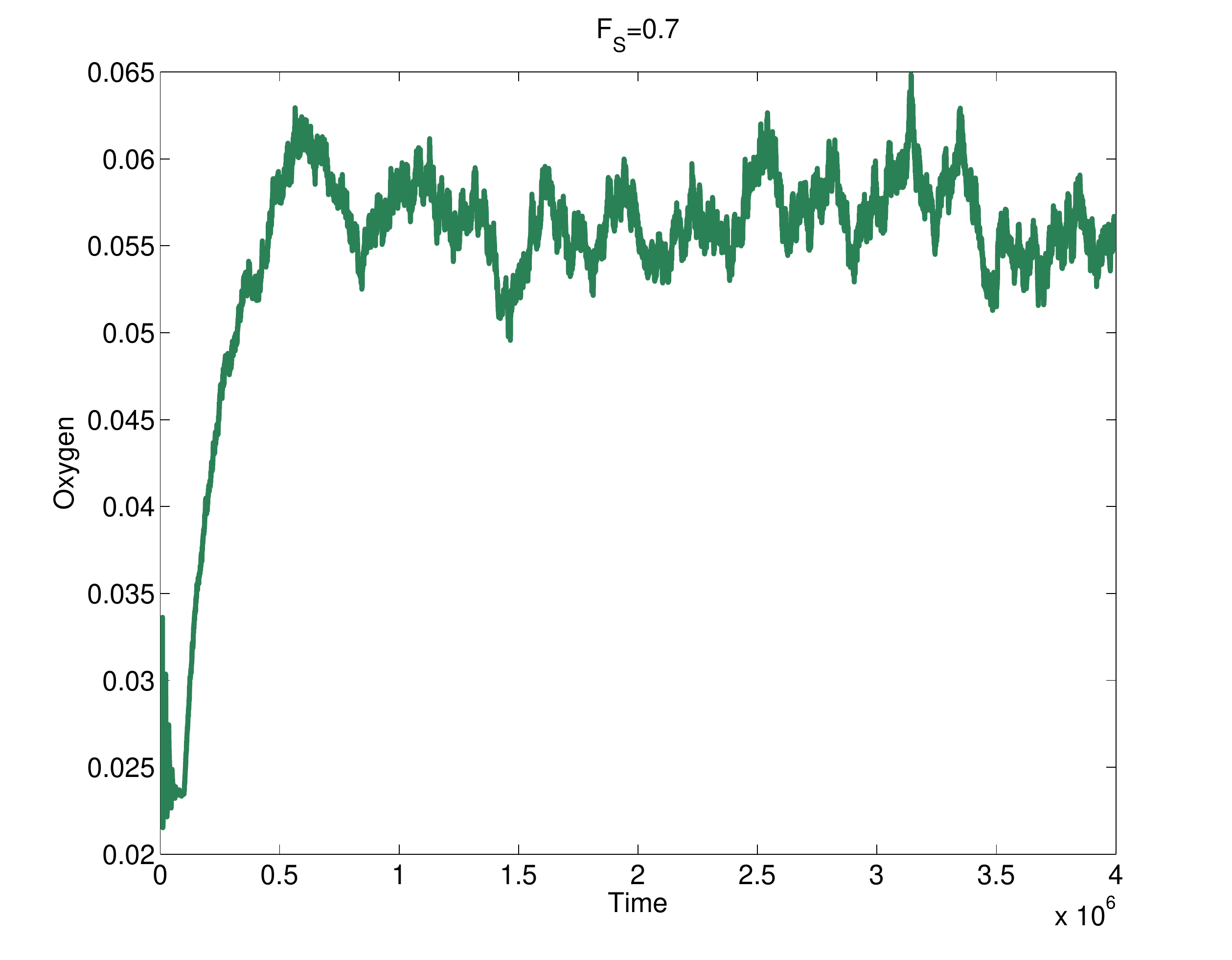} \\
\end{array}$
\caption{Stochastic simulation results showing typical realisations associated with rescue of quiescence cells (plots (a) \& (c)) and recovery of the 
proliferating population (plots (b) \& (d)) upon application of a cell-cycle dependent therapy. The efficiency of the therapy is characterised by 
the survival fraction, $F_S$. Quiescence rescue is achieved when the survival fraction is set to a value which falls below the critical threshold, Eq. 
(\ref{eq:FSC}). If $F_S<F_{S_C}$ (plots (a) \& (c)), the cell killing triggered by the therapy is enough to re-oxygenate the population above the 
activation threshold of the quiescent cells. By contrast, if $F_S>F_{S_C}$ (plots (b) \& (d)), re-oxygenation is not 
enough to rescue latent cells from quiescence. Parameter values: $\nu_1=4.167\cdot 10^{-5}$, $\nu_1=4.167\cdot 
10^{-7}$, $\tau_{p_1}=\tau_{p_2}=2.1\cdot 10^{-3}$, $\frac{p_{6_1}}{p_{3_1}}=1$, $\frac{p_{6_1}}{p_{3_1}}=0.989$. The subindex ``1'' corresponds to 
the active population whilst the subindex ``2'' denotes quantities associated with the quiescent population.  The critical 
oxygen (as defined in Sections \ref{sec:quiescence} \& \ref{sec:scalingg1s}) is $c_{cr_1}=0.023$ for the active cells and 
$c_{cr_2}=0.1$ for the quiescent cells. Colour code: I all of the panels in this figure, blue (red) lines correspond to the time evolution of the total number of proliferating (quiescent) cells and gree lines, to the time evolution of the oxygen concentration.}\label{fig:rescue-realisations}
\end{center}
\end{figure}

In order to check the accuracy of the mean-field analysis carried out in Section \ref{sec:critdose} regarding the critical survival fraction for 
rescue from quiescence. We start by showing (Fig. \ref{fig:rescue-realisations}) two typical realisations of the stochastic population dynamics which 
illustrate the rescue mechanism. In this simulations, we first let the active population settle on to its steady state. We then apply a sustained 
therapy with constant survival fraction. A more aggressive treatment ($F_S=0.6$ in Fig. \ref{fig:rescue-realisations}) greatly affects the active 
population: the amount of active cells killed by the therapy induces re-oxygenation of the population above the critical oxygen level for activation 
of the quiescent population whereupon the quiescent cells become proliferating. In Figs. \ref{fig:rescue-realisations}(a) \& (c), we show that upon 
activation of the quiescent population, a competition between both populations ensues, which eventually leads to extinction of the active population. 
A less aggressive therapy ($F_S=0.7$ in Fig. \ref{fig:rescue-realisations}) also induces death of the active population and re-oxygenation. However, 
in this case, the latter is not intense enough to induce activation of the quiescent cells (see Fig. \ref{fig:rescue-realisations}(d)) and therefore 
the active cells will repopulate the system as the quiescent population stays on its course to eventual extinction, as shown in Fig. 
\ref{fig:rescue-realisations}(b). 

\begin{figure}
\begin{center}
\includegraphics[scale=0.4]{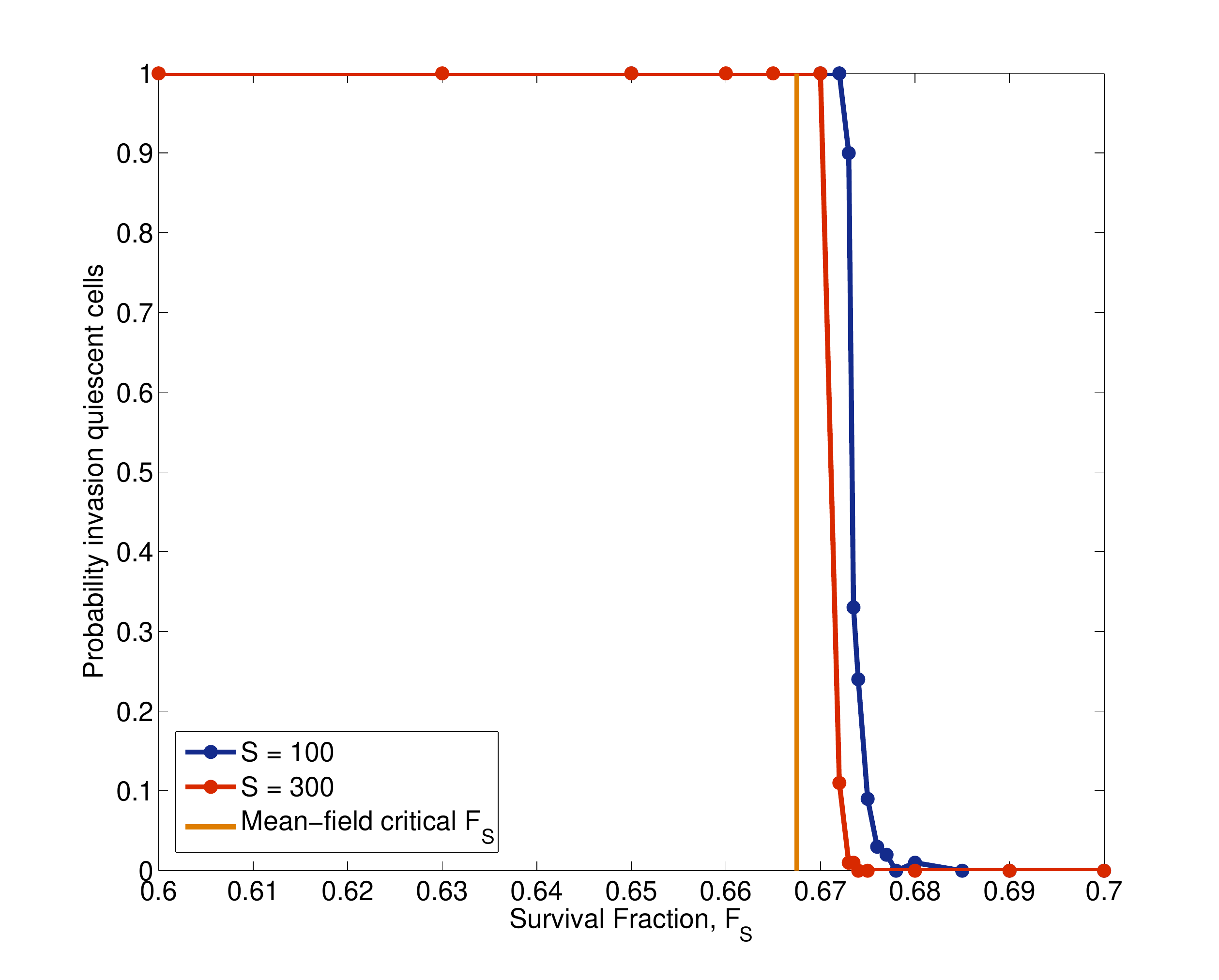}
\caption{This figure shows stochastic simulation results regarding the variation of the probability of fixation of the quiescence population varies 
the efficiency of a cell-cycle dependent therapy changes. The efficiency of the therapy is measured in terms of the survival fraction $F_S$. The 
orange vertical line represents the mean-field, theoretical critical survival fraction (Eq. \ref{eq:FSC}). We observe that as 
the carrying capacity of the system increases, which in these simulations is achieved by increasing the rate of oxygen 
supply, $S$, the results of the stochastic simulations tend towards the mean-field prediction. Parameter values: $\nu_1=4.167\cdot 10^{-5}$, 
$\nu_1=4.167\cdot 
10^{-7}$, $\tau_{p_1}=\tau_{p_2}=2.1\cdot 10^{-3}$, $\frac{p_{6_1}}{p_{3_1}}=1$, $\frac{p_{6_1}}{p_{3_1}}=0.989$. The subindex ``1'' corresponds to 
the active population whilst the subindex ``2'' denotes quantities associated with the quiescent population.  The critical 
oxygen (as defined in Sections \ref{sec:quiescence} \& \ref{sec:scalingg1s}) is $c_{cr_1}=0.023$ for the active cells and 
$c_{cr_2}=0.1$ for the quiescent cells. Each data point corresponds to the average over 500 realisations of the stochastic 
population dynamics. }\label{fig:fixationprob}
\end{center} 
\end{figure}

Fig. \ref{fig:fixationprob} shows simulation results for the variation of probability of fixation of the quiescent population as the survival 
fraction of the therapy, $F_S$ changes. Our simulation results show qualitative agreement with our mean-field theory (see Sections 
\ref{sec:mftherapy} \& \ref{sec:critdose}): as the survival fraction increases (\emph{i.e.} the therapy becomes less efficient), the probability of 
fixation abruptly decreases from almost certainty of fixation to almost certainty of extinction. We observe that our mean-field theoretical predicts 
a critical value for $F_S$ slightly smaller than the observed when fluctuations due to finite size effects are present. However, we observe that, as 
the carrying capacity of the system is increased, the critical value of $F_S$ converges to the mean-field value. 

\section{Discussion}\label{sec:discussion}

In this paper, we have presented and studied a stochastic multi-scale model of a heterogeneous, resource-limited cell population. This model accounts for a stochastic intracellular dynamics (in this particular case, a model of the oxygen-regulated G1/S transition) and an age-structured birth-and-death process for the cell population dynamics. Both compartments are coupled by (i) a model for the time variation of resource (oxygen) abundance which regulates the rate of cell-cycle progression, and (ii) a model of the age-dependent birth rate which carries out the coupling between the intracellular and the cellular compartments (see Fig. \ref{fig:scheme} for a schematic representation of the model and Section \ref{sec:summary} for a summary of the model formulation).  

Our analysis of the stochastic dynamics of the oxygen-regulated G1/S transition, which is a generalisation of the mean-field model presented in \cite{bedessem2014}, has revealed a number of previously unreported properties related to the presence of fluctuations. In particular, the optimal path theory and the quasi-steady state approximation allows us to explore the effect of the SCF-regulating enzymes on the timing of the G1/S transition. The relative abundance of SCF-activating and inhibiting enzymes regulates the rate at which cells reach the G1/S transition: excess of SCF-activating enzyme can delay the transition and even rendering the cell quiescent regardless of oxygen concentration beyond the predictions of the mean-field model (see Sections \ref{sec:timingg1s} \& \ref{sec:quiescence}). Furthermore, we have shown that the effects on timing of the G1/S transition of the relative abundance of SCF-activating enzyme give rise to a scaling form dependence of the age to the transition, $a_{G1/S}$, whereby 
this quantity is a function of $a_{G1/S}(c,p_6/p_3)$ which takes the form of Eq. (\ref{eq:ag1sscaling}) (see Section \ref{sec:scalingg1s}). Taken together, these results imply that stochasticity in the intracellular dynamics naturally generates variability within the population: an otherwise homogeneous population presents a distribution of birth rates induced by variability in the relative abundance of SCF-activating enzyme within the population of cells. 

This variation in the duration of the cell-cycle allows us to analyse the dynamics of a stochastic heterogeneous population under resource limitation conditions. To this end, we consider populations formed by sub-populations of cells characterised by differing relative abundance of SCF-activating and inhibiting enzymes. We further assume that this heterogeneity is heritable (\emph{i.e.} daughter cells inherited the ratio of SCF-regulating enzymes from their mother). In this scenario, we have shown sub-populations within heterogeneous population engage in quasi-neutral competition: sub-populations of cells get extinct in an average time which is of the order of the carrying capacity of the system (see Section \ref{sec:quasineutral}). In the context of modelling cancer cells populations, where heterogeneity is a main contributor to the complex dynamics of cancer \cite{anderson2006,maley2006a,anderson2008,gillies2012,greaves2012}, this result is of relevance since it allows us to estimate the rate at which sub-populations or clones disappear from the tumour. The fact that this rate is proportional to the inverse of the carrying capacity reveals a highly dynamical scenario where clones are quickly decaying and being replaced within the tumour. This can have a profound impact in a variety of evolutionary phenomena such as emergence of drug resistant phenotypes. From the modelling perspective, we should note that quasi-neutral competition is a purely stochastic scenario: the mean-field limit predicts coexistence between the corresponding cell types.  

We have further explored the issue of emergence of drug resistant cell types by analysing a case study in which a quiescent population can be rescued from latency by the application of a cell-cycle dependent therapy. Examples of such therapies are radiotherapy or cytotoxic drugs designed to target cells in specific stages of cell cycle progression \cite{powathil2012}. In the particular example analysed in Section \ref{sec:therapy}, we have shown that in mixed population composed by active (proliferating) and quiescent cells, if the drug is not efficient enough (characterised in terms of its associated survival fraction, $F_S$), quiescent cells are rescued from latency and eventually reach fixation within the population, \emph{i.e.} the activated quiescent cells out-compete the original active cells until the latter population becomes extinct \cite{alarcon2010b}. It is noteworthy the fact that in this process of quiescence rescue the original cancer population is replaced by a much more resistant population as the activated quiescent cells are less sensitive to the therapy than the original active cells.   

Our results show that the methods and models presented in this paper are of great potential importance for the analysis of the complex dynamics of heterogeneous populations under resource limitation, in particular for the study of emergence of drug resistance in heterogeneous cancer cell populations. Several important issues have been left out of the present work. A major contributor to heterogeneity within a cancer cell population is spatial heterogeneity \cite{anderson2006,gillies2012} which is closely related to micro-environmental heterogeneity. A further issue which should be analysed in depth concerns the scaling properties of the age to the G1/S transition (see Section \ref{sec:scalingg1s}), in particular whether this is a general property of the cell-cycle dynamics or rather a specific attribute of the stochastic model presented here. A thorough analysis of these issues falls beyond the scope of the present paper and are left for future research.    

\paragraph*{Acknowledgements}\mbox{ }R.C. and T.A. acknowledge the Ministry of Economy and Competitivity (MINECO) for funding under grant MTM2015-71509-C2-1-R and 
Generalitat de Catalunya for funding under grant 2014SGR1307. R.C. acknowledges AGAUR-Generalitat de Catalunya for funding under its doctoral 
scholarship programme. P.G. thanks the Wellcome Trust for financial support under grant 098325. T.A. acknowledges support from the Ministry of Economy 
\& Competitivity (MINECO) for funding awarded to the Barcelona Graduate School of Mathematics under the ``Mar\'{\i}a de Maeztu'' programme, 
grant number MDM-2014-0445. F.S. acknowledges the support of the NCI under grant number 5U01CA177799.

\appendix

\section{Summary of the semi-classical quasi-steady state approximation}\label{sec:scqssa}

In this Appendix, we give a summary of the semi-classical quasi-steady state approximation (SCQSSA). For a more detailed presentation of the method, the reader is referred to 
\cite{alarcon2014,delacruz2015}. 

Following \cite{alarcon2014}, we formulate the QSS approximation for the asymptotic solution of the CME obtained by means of large deviations/WKB 
approximations \cite{kubo1973,alarcon2007,touchette2009}. An alternative way to analyse the dynamics of continuous-time Markov processes on a discrete 
space of states is to derive an equation for the generating function, $G(p_1,\dots,p_n,t)$ of the corresponding probabilistic density:

\begin{equation}
G(p_1,\dots,p_n,t)=\sum_x p_1^{X_1}p_2^{X_2}\cdots p_n^{X_n}P(X_1,\dots,X_n,t)
\end{equation}

\noindent where $P(X_1,\dots,X_n,t)$ is the solution of the CME Eq. (\ref{eq:cmebedessem}). $G(p_1,\dots,p_n,t)$ satisfies a partial 
differential equation (PDE) which can be derived from the CME. This PDE is the basic element of the so-called momentum representation of 
the Master Equation \cite{doi1976,peliti1985,assaf2006,assaf2010,kang2013}. Analytic solutions for the generating function PDE are seldom available 
and one needs to resort to approximate solutions, which are commonly obtained by means of the WKB asymptotic \cite{assaf2010}. More specifically, 
the (linear) PDE that governs the evolution of the generating function can be written as:

\begin{equation}\label{eq:charfuncPDE}
\frac{\partial G}{\partial t}=H_k\left(p_1,\dots,p_n,\partial_{p_1},\dots,\partial_{p_n}\right)G(p_1,\dots,p_n,t) 
\end{equation}

\noindent where the operator $H_k$ is determined by the reaction rates of the CME Eq. (\ref{eq:cmebedessem}). Furthermore, the solution to this
equation must satisfy the normalisation condition $G(p_1=1,\dots,p_n=1,t)=1$ for all $t$. The operator $H$ is obtained 
by multiplying both sides of the CME by $\prod_{i=1}^{n}p_i^{X_i}$ and summing up over all the possible values of $(X_1,\dots,X_n)$.

Eq. (\ref{eq:charfuncPDE}) is a Schr\"odinger-like equation and, therefore, there is a plethora of methods at our disposal in order to analyse it. In 
particular, when the fluctuations are (assumed to be) small, it is common to use WKB methods \cite{kubo1973,alarcon2007,gonze2002}. This approach is 
based on the WKB-like Ansatz that $G(p_1,\dots,p_n,t)=e^{-S(p_1,\dots,p_n,t)}$. By substituting this Ansatz in Eq. (\ref{eq:charfuncPDE}) we obtain 
the following Hamilton-Jacobi equation for the function $S(p_1,\dots,p_n,t)$:

\begin{equation}\label{eq:hamjac}
\frac{\partial S}{\partial t}=-H_k\left(p_1,\dots,p_n,\frac{\partial S}{\partial p_1},\dots,\frac{\partial S}{\partial p_n}\right)
\end{equation}

Instead of directly tackling Eq. (\ref{eq:hamjac}), we will use the so-called semi-classical approximation. We use the Feynman 
path-integral representation which yields a solution to Eq. (\ref{eq:charfuncPDE}) of the type
\cite{peliti1985,feynman2010,kubo1973,dickman2003,elgart2004,tauber2005}:

\begin{equation}\label{eq:pathintegral}
G(p_1,\dots,p_n,t)=\int_0^t e^{-S(p_1,\dots,p_n,Q_1,\dots,Q_n)}{\cal D}Q(s){\cal D}p(s),
\end{equation}

\noindent where ${\cal D}Q(s){\cal D}p(s)$ indicates integration over the space of all possible trajectories and
$S(p_1,\dots,p_n,Q_1,\dots,Q_n)$ is given by \cite{kubo1973}:

\begin{eqnarray}\label{eq:actionintegral}
\nonumber S(p_1,\dots,p_n,Q_1,\dots,Q_n)=&&-\int_0^t\left(H_k(p_1,\dots,p_n,Q_1,\dots,Q_n)+\sum_{i=1}^nQ_i(s)\dot{p}_i(s)\right)ds \\
&& + \sum_{i=1}^nS_{0,i}(p_i,Q_i),
\end{eqnarray}

\noindent The position operators in the momentum representation have been defined as $Q_i\equiv \partial_{p_i}$ with the commutation relation $[Q_i,p_j]=S_{0,i}\delta_{i,j}$. $S_{0,i}(p_i,Q_i)$ corresponds to the action associated with the generating function of the probability distribution function of the initial value of each variable, $X_i(t=0)$, which are assumed to be independent random variables. The so-called semi-classical approximation consists of approximating the path integral in Eq. (\ref{eq:pathintegral}) by

\begin{equation}\label{eq:geomopt}
G(p_1,\dots,p_n,t)=e^{-S(p_1,\dots,p_n,t)}
\end{equation}

\noindent where $p_1(t),\dots,p_n(t)$ are now the solutions of the Hamilton equations, \emph{i.e.} the orbits which maximise the action $S$:

\begin{eqnarray}
\label{eq:hameqs-p} && \frac{dp_i}{dt}=-\frac{\partial H_k}{\partial Q_i}\\
\label{eq:hameqs-q} && \frac{dQ_i}{dt}=\frac{\partial H_k}{\partial p_i} 
\end{eqnarray}

\noindent where the pair ($Q_i$,$p_i$) are the generalised coordinates corresponding to chemical species $i=1,\dots,n$. These
equations are (formally) solved with boundary conditions\cite{elgart2004} $Q_i(0)=x_{i}(0)$, $p_i(t)=p_i$, where $x_i(0)$ is the initial number of
molecules of species $i$.

Eqs. (\ref{eq:hameqs-p})-(\ref{eq:hameqs-q}) are the starting point for the formulation of the semi-classical quasi-steady state approximation
(SCQSSA) \cite{alarcon2014,delacruz2015}. In order to proceed further, we assume, as per the Briggs-Haldane treatment of the Michealis-Menten model 
for enzyme kinetics \cite{briggs1925,keener1998}, that the species involved in the system under scrutiny are divided into two groups according to 
their characteristic scales. More specifically, we have a subset of chemical species whose numbers, $X_i$, scale as:

\begin{equation}
X_i=Sx_i, 
\end{equation}

\noindent where $x_i=O(1)$, whilst the remaining species are such that their numbers, $X_j$, scale as:

\begin{equation}
X_j=Ex_j, 
\end{equation} 

\noindent where $x_j=O(1)$. Key to our approach is the fact that $S$ and $E$ must be such that:

\begin{equation}\label{eq:epsilon}
\epsilon=\frac{E}{S}\ll 1. 
\end{equation}

\noindent We further assume \cite{delacruz2015} that the generalised coordinates, $Q_i$, scale in the same fashion as the corresponding variable 
$X_i$, i.e.

\begin{equation}
Q_i=Sq_i, 
\end{equation}

\noindent where $q_i=O(1)$. We refer to the variables belonging to this subset as \emph{slow variables}. Similarly,

\begin{equation}
Q_j=Eq_j, 
\end{equation}  
 
\noindent where $q_j=O(1)$, which are referred to as \emph{fast variables}. Moreover, we assume that the moment coordinates, $p_i$, are all
independent of $S$ and $E$, and therefore remain invariant under rescaling.

Under this scaling for the generalised coordinates, we define the following scale transformation for the Hamiltonian in Eq. (\ref{eq:actionintegral}):

\begin{equation}\label{eq:scaledham}
H_k(p_1,\dots,p_n,Q_1,\dots,Q_n)=k_JS^kE^lH_{\kappa}(p_1,\dots,p_n,q_1,\dots,q_n)
\end{equation}

\noindent where $J$ identifies the reaction with the largest order among all the reactions that compose the dynamics and $k_J$ is the
corresponding rate constant. In the case of the stochastic model of the G1/S transition $J=7$ (see Table \ref{table:Wr2}), as this reaction is 
order 3 whereas all the others are of lower order. The exponents $k$ and $l$ correspond to the number of slow and fast variables involved in the 
transition rate $W_{7}$, respectively. In this particular case we have $k=2$ and $l=1$. The last step is to rescale the time variable so that a 
dimensionless variable, $a$, is defined such that:

\begin{equation}\label{eq:scaledtime}
a=k_JS^{k-1}E^lt
\end{equation}

It is now a trivial exercise to check that, upon rescaling, Eqs. (\ref{eq:hameqs-p})-(\ref{eq:hameqs-q}) read

\begin{eqnarray}
\label{eq:hameqs-p-slow} && \frac{dp_i}{da}=-\frac{\partial H_{\kappa}}{\partial q_i},\\
\label{eq:hameqs-q-slow} && \frac{dq_i}{da}=\frac{\partial H_{\kappa}}{\partial p_i}, 
\end{eqnarray}

\noindent for the slow variables. By contrast, rescaling of the Hamilton equations corresponding to the subset of fast variables leads to:

\begin{eqnarray}
\label{eq:hameqs-p-fast} && \epsilon\frac{dp_j}{da}=-\frac{\partial H_{\kappa}}{\partial q_j},\\
\label{eq:hameqs-q-fast} && \epsilon\frac{dq_j}{da}=\frac{\partial H_{\kappa}}{\partial p_j}, 
\end{eqnarray} 

\noindent where $\epsilon$ is defined in Eq. (\ref{eq:epsilon}). The QSS approximation consists on assuming that $\epsilon\frac{dp_j}{da}\simeq 0$
and $\epsilon\frac{dq_j}{da}\simeq 0$ in Eqs. (\ref{eq:hameqs-p-fast})-(\ref{eq:hameqs-q-fast}), 

\begin{eqnarray}
\label{eq:hameqs-p-QSSA} && -\frac{\partial H_{\kappa}}{\partial q_j}=0,\\
\label{eq:hameqs-q-QSSA} && \frac{\partial H_{\kappa}}{\partial p_j}=0, 
\end{eqnarray} 

\noindent resulting in a differential-algebraic system of
equations which provides us with the semi-classical quasi-steady state approximation (SCQSSA).  

\section{Semi-classical quasi-steady state analysis of the stochastic model of the G1/S transition}\label{sec:appscqssa}

In this appendix we derive the equations of the SCQSSA of the stochastic model of the G1/S transition as defined by the Master Equation Eq. 
(\ref{eq:cmebedessem}) and the associated transition rates shown in Table \ref{table:Wr2}. See \cite{alarcon2014,delacruz2015} for details of the 
procedure. The associated generating function satisfies the following PDE:

\begin{eqnarray}\label{eq:appc1}
\frac{\partial G}{\partial t}=H(p_1,\dots,p_{10},\partial_{p_1},\dots,\partial_{p_{10}})G(p_1,\dots,p_{10},t) 
\end{eqnarray}

\noindent where the operator $H(p_1,\dots,p_{10},Q_1,\dots,Q_{10})\equiv H(p,Q)$, where $Q_i\equiv \partial_{p_i}$ which satisfies the commutation 
relation $[Q_i,p_j]=S_{0,i}\delta_{i,j}$, is given by:

\begin{equation}\label{eq:hamspli}
H(p,Q)=H_{1}(p,Q)+H_{2}(p,Q)+H_{3}(p,Q)+H_{4}(p,Q)+H_{5}(p,Q)+H_{6}(p,Q)
\end{equation}

\noindent where the full Hamiltonian $H(p,Q)$ has been separated into 6 parts, each corresponding to one of the reactions shown in Fig. 
\ref{fig:bedessemreactions}. Thus, 

\begin{equation}\label{eq:actscf}
H_{1}(p,Q)=k_{4}(p_{4}-p_{2}p_{3})Q_{2}Q_{3}+k_{5}(p_{2}p_{3}-p_{4})Q_{4}+k_{6}(p_{3}p_{5}-p_{4})Q_{4},
\end{equation}

\noindent and 

\begin{equation}\label{eq:inactscf}
H_{2}(p,Q)=k_{9}p_{8}(p_{2}p_{6}-p_{7})Q_{7}Q_{8}+k_{7}p_{8}(p_{7}-p_{5}p_{6})Q_{5}Q_{6}Q_{8}+k_{8}p_{8}(p_{5}p_{6}-p_{7})Q_{7}Q_{8},
\end{equation}

\noindent correspond to enzymatic activation and (CycE-mediated) inactivation of SCF ($X_5$), respectively. $H_3(p,Q)$ through to $H_6(p,Q)$ are 
given by: 

\begin{equation}\label{eq:degx1}
H_{3}(p,Q)=(k_{1}-k_{2})(p_{1}-1)+k_{3}(1-p_{1})Q_{1},
\end{equation}

\begin{equation}\label{eq:degx8}
H_{4}(p,Q)=k_{10}m(p_{8}-1)p_{10}Q_{10}(1-\frac{p_{9}}{[e2f]_{tot}}Q_{9})+(1-p_{8})Q_{8}(k_{11}+k_{12}p_{5}Q_{5}),
\end{equation}

\begin{equation}\label{eq:degx9}
H_{5}(p,Q)=k_{13}(p_{9}-1)+(1-p_{9})(k_{14}Q_{9}+k_{15}p_{1}Q_{1}Q_{9}),
\end{equation}

\begin{equation}\label{eq:degx10}
H_{6}(p,Q)=k_{16}(p_{10}-1)+k_{17}(1-p_{10})Q_{10},
\end{equation}

\noindent which are the Hamiltonians associated with synthesis and degradation of CycD ($X_1$), CycE ($X_8$), Rb ($X_9$), and E2F ($X_{10}$), 
respectively.

\begin{table}[htb]
\centerline{
\begin{tabular}{lll}
Dimensionless variables &\vline& Dimensionless parameters         
\\
\hline
$a=k_{7}E S t$  &\vline& $\epsilon=E/S $
\\
$q_{1}=Q_{1}/S$ &\vline& $\widehat{[e2f]}_{tot}=[e2f]_{tot}/S$
\\
$q_{2}=Q_{2}/S $ &\vline& $\kappa_{1}=k_{1}/(k_{7}ES^2)$
\\
$q_{3}=Q_{3}/E$ &\vline& $\kappa_{2}={k_{2}}/({k_{7}E S^2})$
\\
$q_{4}=Q_{4}/E$&\vline&  $\kappa_{3}={k_{3}}/({k_{7}E S})$
\\
 $q_{5}=Q_{5}/S $ &\vline& $\kappa_{4}=k_{4}/(k_{7}S)$ 
\\ 
$q_{6}=Q_{6}/E$ &\vline& $\kappa_{5}={k_{5}}/({k_{7}S^2})$  
\\
$q_{7}=Q_{7}/E$ &\vline&   $\kappa_{6}={k_{6}}/({k_{7}S^2})$
\\
$q_{8}=Q_{8}/S$ &\vline& $\kappa_{8}={k_{8}}/({k_{7}S })$
\\
$q_{9}=Q_{9}/S$ &\vline& $\kappa_{9}={k_{9}}/({k_{7}S})$
\\
$q_{10}=Q_{10}/S$ &\vline& $\kappa_{10}={k_{10}}/({k_{7}ES})$
\\
&\vline& $\kappa_{11}={k_{11}}/({k_{7}ES})$
\\
&\vline& $\kappa_{12}={k_{12}}/({k_{7}E})$
\\
&\vline& $\kappa_{13}={k_{13}}/({k_{7}ES^2})$
\\
&\vline& $\kappa_{14}={k_{14}}/({k_{7}E S})$
\\
&\vline& $\kappa_{15}={k_{15}}/({k_{7}E})$
\\
&\vline& $\kappa_{16}={k_{16}}/({k_7 E S^2})$
\\
&\vline& $\kappa_{17}={k_{17}}/({k_{7}E S})$
\end{tabular}} 
\caption{Dimensionless variables and parameters corresponding. 
}\label{tab:rescale}
\end{table} 

By re-scaling the coordinate-like variables $Q_i$ according to the scaling shown in Table \ref{tab:rescale} (see Section \ref{sec:semiclassg1s} and 
reference \cite{delacruz2015}) and defining the dimensionless time as $a=k_{7}E S t$, the associated Hamilton equations, Eqs. 
(\ref{eq:hameqs-p})-(\ref{eq:hameqs-q}) 

\begingroup
\allowdisplaybreaks
\begin{eqnarray}
 \frac{dq_{1}}{da}&=&\kappa_{1}-\kappa_{2}-\kappa_{3}q_{1}+\kappa_{15}(1-p_{9})q_{1}q_{9}
\\
 \frac{dq_{2}}{da}&=&-\kappa_{4}p_{3}q_{2}q_{3}+\kappa_{5}p_{3}q_{4}+\kappa_{9}p_{8}p_{6}q_{7}q_{8}
\\
 \epsilon\frac{dq_{3}}{da}&=&-\kappa_{4}p_{2}q_{2}q_{3}+(\kappa_{5}p_{2}+\kappa_{6}p_{5})q_{4}\label{eq:appcq3}
\\
 \epsilon \frac{dq_{4}}{da}&=&\kappa_{4}q_{2}q_{3}-(\kappa_{5}+\kappa_{6})q_{4}
\\
 \frac{dq_{5}}{da}&=&\kappa_{6}p_{3}q_{4}-p_{8}p_{6}q_{5}q_{6}q_{8}+\kappa_{8}p_{8}p_{6}q_{7}q_{8}+\kappa_{12}q_{5}q_{8}(1-p_{8})
 \\
 \epsilon\frac{dq_{6}}{da}&=&\kappa_{9}p_{8}p_{2}q_{7}q_{8}-p_{8}p_{5}q_{5}q_{6}q_{8}+\kappa_{8}p_{8}p_{5}q_{7}q_{8}
\\
\epsilon \frac{dq_{7}}{da}&=&-\kappa_{9}p_{8}q_{7}q_{8}+p_{8}q_{5}q_{6}q_{8}-\kappa_{8}p_{8}q_{7}q_{8}\label{eq:appcq7}
\\
 \frac{dq_{8}}{da}&=&(\kappa_{9}  (p_{2}p_{6}-p_{7})+\kappa_{8} (p_{5}p_{6}-p_{7}))q_{7}q_{8}+(p_{7}-p_{5}p_{6})q_{5}q_{6}q_{8}  +\\
&&+\kappa_{10}mp_{10}q_{10}\left({1}-\frac{p_{9}q_{9} }{\widehat{[e2f]}_{tot}}\right)-\kappa_{11}q_{8}-\kappa_{12}p_{5}q_{5}q_{8}
 \\
 \frac{dq_{9}}{da}&=&\kappa_{10}m(1-p_{8})p_{10}\frac{q_{9}q_{10}}{\widehat{[e2f]}_{tot}}+\kappa_{13}-\kappa_{14}q_{9}-\kappa_{15}p_{1}q_{1}q_{9}
\\
 \frac{dq_{10}}{da}&=&\kappa_{10}m(p_{8}-1)q_{10}(1-\frac{p_{9}q_{9} }{\widehat{[e2f]}_{tot}})+ \kappa_{16}-\kappa_{17}q_{10}
\\
\frac{dp_{1}}{da}&=&-\kappa_{3}(1-p_{1})-\kappa_{15}(1-p_{9})p_{1}q_{9}
\\
\frac{dp_{2}}{da}&=&-\kappa_{4}(p_{4}-p_{2}p_{3})q_{3}\label{eq:appcp2}
\\
\epsilon \frac{dp_{3}}{da}&=&-\kappa_{4}(p_{4}-p_{2}p_{3})q_{2}
\\
\epsilon \frac{dp_{4}}{da}&=&-\kappa_{5}(p_{2}p_{3}-p_{4})-\kappa_{6}(p_{3}p_{5}-p_{4})
\\
\label{eq:appcp5}\frac{dp_{5}}{da}&=&-p_{8}(p_{7}-p_{5}p_{6})q_{6}q_{8}-\kappa_{12}(1-p_{8})p_{5}q_{8}
\\
 \epsilon \frac{dp_{6}}{da}&=&-p_{8}(p_{7}-p_{5}p_{6})q_{5}q_{8}
\\
 \epsilon \frac{dp_{7}}{da}&=&-(\kappa_{9}(p_{2}p_{6}-p_{7})+\kappa_{8}(p_{5}p_{6}-p_{7}))p_{8}q_{8}
\\  
\frac{dp_{8}}{da}&=&-(\kappa_{9}(p_{2}p_{6}-p_{7})+\kappa_{8}(p_{5}p_{6}-p_{7}))p_{8}q_{7}-p_{8}(p_{7}-p_{5}p_{6})q_{5}q_{6} \hspace{2cm}
\\
&&-{(1-p_{8})(\kappa_{11}+\kappa_{12}p_{5}q_{5} )}
\\
  \frac{dp_{9}}{da}&=&-\kappa_{10}m(1-p_{8})p_{10} \frac{p_{9}q_{10}}{\widehat{[e2f]}_{tot}}-(1-p_{9})(\kappa_{14}+\kappa_{15}p_{1}q_{1})
\\
  \frac{dp_{10}}{da}&=&-\kappa_{10}m(p_{8}-1)p_{10}({1}-\frac{p_{9}q_{9}}{\widehat{[e2f]}_{tot}})-\kappa_{17}(1-p_{10})
\end{eqnarray}
\endgroup

\noindent where $\epsilon=E/S\ll 1$ and the re-scaled parameters $\kappa_i$ are given in Table \ref{tab:rescale}. Applying the QSSA to the equations for the momenta $p_3$, $p_4$, $p_6$ and $p_7$ associated with the fast variables, we obtain:

\begin{equation}\label{eq:appqss1}
\epsilon \frac{dp_{3}}{da}=0 \Rightarrow p_{4}=p_{2}p_{3}, \quad \epsilon \frac{dp_{4}}{da}=0 \Rightarrow p_{4}=p_{5}p_{3}
\end{equation}
\begin{equation}\label{eq:appqss2}
\epsilon \frac{dp_{6}}{da}=0 \Rightarrow p_{7}=p_{5}p_{6}, \quad \epsilon \frac{dp_{7}}{da}=0 \Rightarrow p_{7}=p_{2}p_{6}
\end{equation}

\noindent which implies that $p_{2}=p_{5}$, which, in turn, yields that $p_8=1$ (see Eqs. (\ref{eq:appcp2}) and (\ref{eq:appcp5})). Furthermore, the 
QSSA applied to the equations for the generalised coordinates associated with the SCF-activating, $q_3$, Eq. (\ref{eq:appcq3}), and the 
enzyme-active SCF complex, $q_7$, Eq. (\ref{eq:appcq7}), yields:

\begin{equation}\label{eq:appqss3}
\epsilon \frac{dq_{3}}{da}=0 \Rightarrow \kappa_5q_4-\kappa_4q_2q_3=-\kappa_6q_{4},\quad \epsilon \frac{dq_{6}}{da}=0 
\Rightarrow \kappa_{8}q_{7}-q_{5}q_{6}=-\kappa_{9}q_{7},
\end{equation}

\noindent which implies that $q_{2}+q_{5}=p_c$. Furthermore, since $q_{3}+q_{4}=p_{e_1}$ and $q_{6}+q_{7}=p_{e_2}$ are satisfied, we obtain that:

\begin{equation}\label{eq:appqss4}
q_{4}=\frac{p_{e_1}q_{2}}{q_{2}+\frac{\kappa_{5}+\kappa_{6}}{\kappa_{4}}}, \quad q_{7}=\frac{p_{e_2}q_{5}}{q_{5}+(\kappa_{8}+\kappa_{9})}
\end{equation}

Finally, using Eqs. (\ref{eq:appqss1})-(\ref{eq:appqss4}), the SCQSSA equations for the stochastic model of the G1/S transition are:

\begin{eqnarray} \label{eq:finalsystem1x}
\frac{dq_{1}}{da}&=&\kappa_{1}-\kappa_{2}-\kappa_{3}q_{1}+\kappa_{15}(1-p_{9})q_{1}q_{9}
 \\
\frac{dq_{5}}{da}&=&\frac{\kappa_{6}p_{3}p_{e_1}(p_c-q_{5})}{(p_c-q_{5})+\frac{\kappa_{5}+\kappa_{6}}{\kappa_{4}}}-\frac{\kappa_{9}p_{8}p_{6}p_{e_2}q_{5}q_{8}}{q_{5}+(\kappa_{8}+\kappa_{9})}  \label{eq:scf2}
\\
\frac{dq_{8}}{da}&=&\kappa_{10}mp_{10}q_{10}\left(1-\frac{p_{9}q_{9}}{\widehat{[e2f]}_{tot}}\right)-\kappa_{11}q_{8}-\kappa_{12}p_{5}q_{5}q_{8}   \label{eq:cycE2}
\\
\frac{dq_{9}}{da}&=&\kappa_{10}m(1-p_{8})p_{10}\frac{q_{9}q_{10} }{\widehat{[e2f]}_{tot}}+\kappa_{13}-\kappa_{14}q_{9}-\kappa_{15}p_{1}q_{1}q_{9}
\\
 \frac{dq_{10}}{da}&=&\kappa_{10}m(p_{8}-1)q_{10}(1-\frac{p_{9}q_{9}}{\widehat{[e2f]}_{tot}})+ \kappa_{16}-\kappa_{17}q_{10}
\\
 q_{2}&=&p_c-q_{5}
\\
 q_{4}&=&\frac{p_{e_1}q_{2}}{q_{2}+\frac{\kappa_{5}+\kappa_{6}}{\kappa_{4}}}, \quad q_{3}+q_{4}=p_{e_1}
\\
q_{7}&=&\frac{p_{e_2}q_{5}}{q_{5}+(\kappa_{8}+\kappa_{9})}, \quad q_{6}+q_{7}=p_{e_2}
\\
\frac{dp_{1}}{da}&=&-\kappa_{3}(1-p_{1})-\kappa_{15}(1-p_{9})p_{1}q_{9}
\\
\frac{dp_{9}}{da}&=&-\kappa_{10}m(1-p_{8})p_{10} \frac{p_{9}q_{10}}{\widehat{[e2f]}_{tot}}-(1-p_{9})(\kappa_{14}+\kappa_{15}p_{1}q_{1})
\\
\frac{dp_{10}}{da}&=&-\kappa_{10}m(p_{8}-1)p_{10}({1}-\frac{p_{9}q_{9}}{\widehat{[e2f]}_{tot}})-\kappa_{17}(1-p_{10})
\\
p_{2}&=&p_{5}
\\
p_{4}&=&p_{2}p_{3}
\\
\label{eq:finalsystemx}
p_{7}&=&p_{2}p_{6}
\end{eqnarray}

If we set $p_{i}=1$ in Eqs. (\ref{eq:finalsystem1x})-(\ref{eq:finalsystemx}), we recover the mean-field model proposed in \cite{bedessem2014}. This limit allows us to determine the parameter values of the stochastic model (see \cite{alarcon2014,delacruz2015} for details). This equivalence is shown in Table \ref{table:parametersBedessem1}. The parameter values of the mean-field model \cite{bedessem2014} are given in Table \ref{table:parametersBedessem}

\begin{table}[htb]
\centerline{
\begin{tabular}{ll}
Parameters	& Parameters
\\ \hline
$k_{1}=a_1*S$ &  $k_{11}=b_2$
\\
$k_{2}=a_3*S*[H]$ & $k_{12}=b_3/S$
\\
$k_{3}=a_2$ &  $k_{13}=d_2*S$
\\
$k_{5}=J_2*k_{4} S-k_{6}$ & $k_{14}=d_2$
\\
$k_{6}=e_1 S/E$& $k_{15}=d_{1}/S$
\\
$k_{8}=J_1*S-k_{9}$ &$k_{16}=g_1 [E2F]_{tot}S$
\\
$k_{9}=e_2/E$ & $k_{17}=g_1$
\\
$k_{10}=b_1$ &
\\
\hline
\end{tabular}} 
\caption{Table showing the equivalence between the parameters of the stochastic model of the G1/S transition (see Table \ref{table:Wr2}) and the parameters of the associated mean-field model as formulated by Bedessem \& Stephanou \cite{bedessem2014}.}\label{table:parametersBedessem1}
\end{table}

\begin{table}[htb]
\centerline{
\begin{tabular}{lll}
Parameter 	& Value 	& Reference         \\\hline
$a_1$ 		& 0.51		&  \cite{bedessem2014} \\
$a_2$		& 1		&\cite{bedessem2014} \\
$a_3 H_0$	& 0.0085		& \cite{bedessem2014} \\
$b_1$		& 0.018		& \cite{bedessem2014}\\
$b_2$		& 0.5 		& \cite{bedessem2014} \\
$b_3$		& 1 		& \cite{bedessem2014} \\
$d_1$		& 0.2		& \cite{bedessem2014} \\
$d_2$		& 0.1		& \cite{bedessem2014}\\
$e_1$		& 1		& \cite{bedessem2014}\\
$e_2$		& 14		& \cite{bedessem2014}\\
$m_\star$	& 10		& \cite{tyson2001} \\
$J_3$, $J_4$	& 0.04		& \cite{tyson2001} \\
$g_{1}$		& 0.016		& \cite{bedessem2014}  \\
$[E2F]_{tot}$	& 1		& \cite{bedessem2014}  \\
$S$  &10 & \\
$E$      & 1 & \\
$r_{cr}$ & 0.04& \\
\hline
\end{tabular}} 
\caption{Parameters values of the mean-field model of the hypoxia regulated G1/S transition proposed by Bedessem \& Stephanou \cite{bedessem2014}.}\label{table:parametersBedessem}
\end{table}
\begin{table}[htb]
\centerline{
\begin{tabular}{lll}
Parameter 	& Initial condition 	& Reference         \\\hline
$CycD$ 		& 0.1			& \cite{bedessem2014} \\
$SCF$		& 0.9			& \cite{bedessem2014} \\
$Rb$		& 1.0		& \cite{bedessem2014} \\
$E2F$		& 0.1			& \cite{bedessem2014}\\
$m$		& 5.0 			& \cite{bedessem2014} \\
\hline
\end{tabular}} 
\caption{Initial conditions used in simulation system (\ref{eq:finalsystem1xmaintext})-(\ref{eq:finalsystemxmaintext}) -Figure \ref{fig:g1stransition} }  \label{table:initialBedessem}
\end{table} 


\begin{thebibliography}{100}
\expandafter\ifx\csname url\endcsname\relax
  \def\url#1{\texttt{#1}}\fi
\expandafter\ifx\csname urlprefix\endcsname\relax\def\urlprefix{URL }\fi
\expandafter\ifx\csname href\endcsname\relax
  \def\href#1#2{#2} \def\path#1{#1}\fi

\bibitem{kitano2004}
H.~Kitano, {Cancer as a robust system: implications for anticancer therapy},
  Nature Rev. Cancer 4 (2004) 227--235.

\bibitem{alarcon2005}
T.~Alarc{\'o}n, H.~M. Byrne, P.~K. Maini, {A multiple scale model of tumour
  growth}, Multiscale Model. Sim. 3 (2005) 440--475.

\bibitem{ribba2006}
B.~Ribba, T.~Collin, S.~Schnell, {A multi-scale mathematical model of cancer,
  and its use in analysing irradiation therapy}, Theor. Biol. Med. Model. 3
  (2006) 7.

\bibitem{macklin2009}
P.~Macklin, S.~McDougall, A.~R.~A. Anderson, M.~A.~J. Chaplain, V.~Cristini,
  J.~Lowengrub, {Multi-scale modelling and non-linear simulation of vascular
  tumour growth}, J. Math. Biol. 58 (2009) 765--798.

\bibitem{osborne2010}
J.~M. Osborne, A.~Walter, S.~K. Kershaw, G.~R. Mirams, A.~G. Fletcher,
  P.~Pathmanathan, D.~Gavaghan, O.~E. Jensen, P.~K. Maini, H.~M. Byrne, {A
  hybrid approach to multi-scale modelling of cancer}, Phil. Trans. R. Soc. A
  368 (2010) 5013--5028.

\bibitem{deisboeck2011}
T.~S. Deisboeck, Z.~Wang, P.~Macklin, V.~Cristini, {Multi-scale cancer
  modelling}, Annu. Rev. Biomed. Eng. 13 (2011) 125--155.

\bibitem{powathil2013}
G.~G. Powathil, D.~J.~A. Adamson, M.~A.~J. Chaplain, {Towards predicting the
  response of a solid tumour to chemotherapy and radiotherapy treatments:
  clinical insights from a computational model}, PLoS Comp. Biol. 9 (2013)
  e1003120.

\bibitem{jagiella2016}
N.~Jagiella, B.~Muller, M.~Muller, I.~E. Vignon-Clementel, D.~Drasdo,
  {Inferring Growth Control Mechanisms in Growing Multi-cellular Spheroids of
  NSCLC Cells from Spatial-Temporal Image Data}, PLoS Comput. Biol. 12 (2016)
  e1004412.

\bibitem{smith2004}
N.~P. Smith, D.~P. Nickerson, E.~J. Crampin, P.~J. Hunater, {Multi-scale
  computational modelling of the heart}, Acta Numerica 13 (2004) 371--431.

\bibitem{mcculloch2009}
A.~D. McCulloch, {Systems biology and multi-scale modelling of the heart}, in:
  {Proceedings of the biomedical science \& engineering conference, 2009, BSEC
  2009, First Annual ORNL.}, University of California Press, 2009.

\bibitem{hand2010}
P.~E. Hand, B.~E. Griffith, {Adaptive multi-scale model for simulating cardiac
  conduction}, Proc. Natl. Acad. Sci. USA 107 (2010) 14603--14608.

\bibitem{land2013}
S.~Land, S.~A. Niederer, W.~E. Locuh, O.~M. Sejersted, N.~P. Smith,
  {Integrating multi-scale data to create a virtual physiological mouse heart},
  J. R. Soc. Interface Focus 3 (2013) 20120076.

\bibitem{schnell2008}
S.~Schnell, P.~K. Maini, S.~A. Newman, T.~J. Newman, {Multiscale modelling of
  developmental systems}, Elsevier, London, UK, 2008.

\bibitem{oates2009}
A.~C. Oates, N.~Gorfinkel, M.~Gonzalez-Gaitan, C.~P. Heisenberg, {Quantitative
  approaches in developmental biology}, Nature Rev. Gen. 10 (2009) 517--530.

\bibitem{hester2011}
S.~D. Hester, J.~M. Belmonet, J.~S. Gens, S.~G. Clendenon, J.~A. Glazier, {A
  multi-cell, multi-scale model of vertebrate segmentation and somite
  formation}, PLoS Comput. Biol. 7 (2011) e1002155.

\bibitem{setty2012}
Y.~Setty, {Multi-scale computational modelling of developmental biology},
  Bioinformatics 28 (2012) 2022--2028.

\bibitem{walpole2013}
J.~Walpole, J.~A. Papin, S.~M. Peirce, {Multi-scale computational models of
  complex biological systems}, Annu. Rev. Biomed. Eng. 15 (2013) 137--154.

\bibitem{jiang2005}
Y.~Jiang, J.~Pjesivac-Grbovic, C.~Cantrell, J.~P. Freyer, {A multi-scale model
  for avascular tumour growth}, Biophys. J. 89 (2005) 3884--3894.

\bibitem{owen2009}
M.~R. Owen, T.~Alarc{\'o}n, H.~M. Byrne, P.~K. Maini, {Angiogenesis and
  vascular remodelling in normal and cancerous tissues}, J. Math. Biol. 58
  (2009) 689--721.

\bibitem{preziosi2009}
L.~Preziosi, A.~Tosin, {Multi-phase and multi-scale trends in cancer
  modelling}, Math. Model. Nat. Phenom. 4 (2009) 1--11.

\bibitem{tracqui2009}
P.~Tracqui, {Biophysical models of tumour growth}, Rep. Progr. Phys. 6 (2009)
  056701.

\bibitem{byrne2010}
H.~M. Byrne, {Dissecting cancer through mathematics: From the cell to the
  animal model}, Nature Rev. Cancer 10 (2010) 221--230.

\bibitem{lowengrub2010}
J.~S. Lowengrub, H.~B. Frieboes, F.~Jin, Y.~L. Chuang, X.~Li, P.~Macklin, S.~M.
  Wise, V.~Cristini, {Non-linear modelling of cancer: bridging the gap between
  cells and tumours}, Nonlinearity 23 (2010) R1--R91.

\bibitem{rejniak2010}
K.~A. Rejniak, A.~R.~A. Anderson, {Multi-scale hybrid models of tumour growth},
  Wiley Interdisciplinary Reviews: Systems Biology and Medicine.

\bibitem{perfahl2011}
H.~Perfahl, H.~M. Byrne, T.~Chen, V.~Estrella, T.~Alarc{\'o}n, A.~Lapin, R.~A.
  Gatenby, R.~J. Gillies, M.~C. Lloyd, P.~K. Maini, M.~Reuss, M.~R. Owen,
  {Multiscale Modelling of Vascular Tumour Growth in 3D: The Roles of Domain
  Size and Boundary Conditions}, PLoS One 6 (2011) e14790.

\bibitem{travasso2011}
R.~D.~M. Travasso, E.~C. Poir\'e, M.~Castro, J.~C. Rodriguez-Manzaneque,
  A.~Hernandez-Machado., {Tumour angiogenesis and vascular patterning: a
  mathematical model}, PLoS One 6 (2011) e19989.

\bibitem{durrett2013}
R.~Durrett, {Cancer modelling: a personal perspective}, Notices of the AMS 60
  (2013) 304--309.

\bibitem{szabo2013}
A.~Szabo, R.~M.~H. Merks, {Cellular Potts modeling of tumor growth, tumor
  invasion, and tumor evolution}, Front. Oncol. 3 (2013) 87.

\bibitem{chisholm2015}
R.~H. Chisholm, T.~Lorenzi, A.~Lorz, A.~K. Larsen, L.~N. de~Almeida,
  A.~Escargueil, J.~Clairambault, {Tumor morphology and phenotypic evolution
  driven by selective pressure from the microenvironment}, Cancer Res. 75
  (2015) 930--939.

\bibitem{curtius2015}
K.~Curtius, W.~D. Hazelton, J.~Jeon, E.~G. Luebeck, {A Multiscale Model
  Evaluates Screening for Neoplasia in Barrett's Esophagus}, PLoS Comput. Biol.
  11 (2015) e1004272.

\bibitem{scott2016}
J.~G. Scott, A.~G. Fletcher, A.~R.~A. Anderson, P.~K. Maini, {Spatial Metrics
  of Tumour Vascular Organisation Predict Radiation Efficacy in a Computational
  Model.}, PLoS Comput. Biol. 12 (2016) e1004712.

\bibitem{alarcon2014}
T.~Alarc{\'o}n, {Stochastic quasi-steady state approximations for asymptotic
  solutions of the Chemical Master Equation}, J. Chem. Phys. 140 (2014) 184109.

\bibitem{spill2015}
F.~Spill, P.~Guerrero, T.~Alarc\'on, P.~K. Maini, H.~M. Byrne, Hybrid
  approaches for multiple-species stochastic reaction-diffusion models, J.
  Comp. Phys. 299 (2015) 429--445.

\bibitem{delacruz2015}
R.~de~la Cruz, P.~Guerrero, F.~Spill, T.~Alarc{\'o}n, {The effects of intrinsic
  noise on the behaviour of bistable systems in quasi-steady state conditions},
  J. Chem. Phys. 143 (2015) 074105.

\bibitem{spill2016optimisation}
F.~Spill, P.~K. Maini, H.~M. Byrne, Optimisation of simulations of stochastic
  processes by removal of opposing reactions, The Journal of Chemical Physics
  144~(8) (2016) 084105.

\bibitem{maley2006a}
L.~M.~F. Merlo, J.~W. Pepper, B.~J. Reid, C.~C. Maley, {Cancer as an
  evolutionary and ecological process}, Nature Rev. Cancer 6 (2006) 924--935.

\bibitem{gillies2012}
R.~J. Gillies, D.~Verduzco, R.~Gatenby, {Evolutionary dynamics of
  carcinogenesis and why targeted therapy does not work}, Nature Rev. Cancer 12
  (2012) 487--493.

\bibitem{greaves2012}
M.~Greaves, C.~C. Maley, {Evolutionary dynamics of carcinogenesis and why
  targeted therapy does not work}, Nature 481 (2012) 306--313.

\bibitem{asatryan2016}
A.~D. Asatryan, N.~L. Komarova, {Evolution of genetic instability in
  heterogeneous tumours}, J. Theor. Biol. (2016) In press\href
  {http://dx.doi.org/10.1016/j.jtbi.2015.11.028}
  {\path{doi:10.1016/j.jtbi.2015.11.028}}.

\bibitem{alarcon2003}
T.~Alarc{\'o}n, H.~M. Byrne, P.~K. Maini, {A cellular automaton model of tumour
  growth in an inhomogeneous environment}, J. theor. Biol. 225 (2003) 395--411.

\bibitem{kalluri2006fibroblasts}
R.~Kalluri, M.~Zeisberg, Fibroblasts in cancer, Nature Reviews Cancer 6~(5)
  (2006) 392--401.

\bibitem{grivennikov2010immunity}
S.~I. Grivennikov, F.~R. Greten, M.~Karin, Immunity, inflammation, and cancer,
  Cell 140~(6) (2010) 883--899.

\bibitem{spill2016impact}
F.~Spill, D.~S. Reynolds, R.~D. Kamm, M.~H. Zaman, Impact of the physical
  microenvironment on tumor progression and metastasis, Current Opinion in
  Biotechnology 40 (2016) 41--48.

\bibitem{guerrero2013}
P.~Guerrero, T.~Alarc{\'o}n, {Stochastic multi-scale modelling of cell
  populations: Asymptotic and numerical methods}, Math. Model. Nat. Phen. 10
  (2015) 64--93.

\bibitem{redner2001}
S.~Redner, {A guide to first-time passage processes}, Cambridge University
  Press, Cambridge, U.K., 2001.

\bibitem{bressloff2014b}
P.~C. Bressloff, J.~M. Newby, {Path integrals and large deviations in
  stochastic hybrid systems.}, Phys. Rev. E 89 (2014) 042701.

\bibitem{gardiner2009}
C.~W. Gardiner, {Stochatic methods}, Springer-Verlag, Berlin, Germany, 2009.

\bibitem{Oksendal2003}
B.~Oksendal, {Stochastic Differential Equations}, Springer-Verlag, Berlin,
  Germany, 2003.

\bibitem{gillespie1976}
D.~T. Gillespie, {A general method for numerically simulating the stochastic
  time evolution of coupled chemical reactions}, J. Comp. Phys. 22 (1976)
  403--434.

\bibitem{bedessem2014}
B.~Bedessem, A.~Stephanou, A mathematical model of hif-1-$\alpha$-mediated
  response to hypoxia on the g1/s transition, Math. Biosci. 248 (2014) 31--39.

\bibitem{yao2014}
G.~Yao, {Modelling mammalian cellular quiescence}, Interface Focus 4 (2014)
  20130074.

\bibitem{weinberg2007}
R.~A. Weinberg, {The biology of cancer}, Garland Science, New York, NY, USA,
  2007.

\bibitem{gerard2009}
C.~Gerard, A.~Goldbeter, Temporal self-organisation of the cyclin/cdk network
  driving the mammalian cell cycle, Proc. Natl. Acad. Sci. 106 (2009)
  21643--21648.

\bibitem{gerard2011}
C.~Gerard, A.~Goldbeter, A skeleton model for the network of cyclin-dependent
  kinases driving the mammalian cell cycle, Interface Focus 1 (2011) 24--35.

\bibitem{gerard2015}
C.~Gerard, J.~J. Tyson, D.~Coudreuse, B.~Novak, Cell cycle control by a minimal
  cdk network, PLoS Comput. Biol. 11 (2015) e1004056.

\bibitem{alarcon2004}
T.~Alarc{\'o}n, H.~M. Byrne, P.~K. Maini, {A mathematical model of the effects
  of hypoxia on the cell-cycle of normal and cancer cells}, J. theor. Biol. 229
  (2004) 395--411.

\bibitem{semenza2013}
G.~L. Semenza, {HIF-1 mediates metabolic responses to intratumoural hypoxia and
  oncogenic mutations}, J. Clin. Investigation. 123 (2013) 3664--3671.

\bibitem{bristow2008}
R.~G. Bristow, R.~P. Hill, {Hypoxia, DNA repair and genetic instability},
  Nature Rev. Cancer 8 (2008) 180--192.

\bibitem{koshiji2004}
M.~Koshiji, Y.~Kageyama, E.~A. Pete, I.~Horikawa, J.~C. Barrett, L.~E. Huang,
  {HIF-1a induces cell cycle arrest by functionally counteracting Myc}, EMBO J.
  23 (2004) 1949--1956.

\bibitem{ortmann2014}
B.~Ortmann, J.~Druker, S.~Rocha, {Cell cycle progression in response to oxygen
  levels}, Cell. Mol. Life Sci. 71 (2014) 3569--3582.

\bibitem{goda2003}
N.~Goda, S.~J. Dozier, R.~S. Johnson, {HIF-1 in Cell Cycle Regulation,
  Apoptosis, and Tumor Progression}, Antioxidants \& Redox Signaling 5 (2003)
  467--473.

\bibitem{hubbi2014}
M.~E. Hubbia, D.~M. Gilkes, H.~Hua, Kshitiz, I.~Ahmed, , G.~L. Semenza,
  {Cyclin-dependent kinases regulate lysosomal degradation of hypoxia-inducible
  factor 1a to promote cell-cycle progression}, Proc. Natl. Acad. Sci. 111
  (2014) E3325--E3334.

\bibitem{wen2010}
W.~Wen, J.~Ding, W.~Sun, K.~Wu, B.~Ning, W.~Gong, G.~He, S.~Huang, X.~Ding,
  P.~Yin, L.~Chen, Q.~Liu, W.~Xie, H.~Wang, {Cyclin-dependent kinases regulate
  lysosomal degradation of hypoxia-inducible factor 1a to promote cell-cycle
  progression}, Cancer Res. 70 (2010) 2010--2019.

\bibitem{tyson2001}
J.~J. Tyson, B.~Novak, {Regulation of the Eukaryotic Cell Cycle: Molecular
  Antagonism, Hysteresis, and Irreversible Trasitions}, J. theor. Biol. 210
  (2001) 249--263.

\bibitem{novak2004}
B.~Novak, J.~J. Tyson, {A model for restriction point control of the mammalian
  cell cycle}, J. theor. Biol. 230 (2004) 563--579.

\bibitem{lee2010}
T.~J. Lee, G.~Yao, D.~C. Bennett, J.~R. Nevins, L.~You, {Stochastic E2F
  activation and reconciliation of phenomenological cell-cycle models}, PLoS
  Biology 8 (2010) e1000488.

\bibitem{alberts2002}
B.~Alberts, A.~Johnson, J.~Lewis, M.~Raff, K.~Roberts, P.~Walter, {Molecular
  Biology of the Cell}, 4th Edition, Garland Publishing Inc., New York, 2002.

\bibitem{grimmett2001}
G.~R. Grimmett, D.~R. Stirzaker, {Probability and random processes}, Oxford
  University Press, 2001.

\bibitem{vankampen2007}
N.~G.~V. Kampen, {Stochastic processes in Physics and Chemistry}, Elsevier, The
  Netherlands, 2007.

\bibitem{sanchez2015}
D.~Sanchez-Taltavull, A.~Vieiro, T.~Alarcon, {Stochastic modelling of the
  eradication of the HIV-1 infection by stimulation of latently infected cells
  in patients under highly active anti-retroviral therapy}, J. Math. Biol.
  (Submitted).

\bibitem{guerrero2015a}
P.~Guerrero, H.~M. Byrne, P.~K. Maini, T.~Alarc\'on, {From invasion to latency:
  Intracellular noise and cell motility as key controls of the competition
  between resource-limited cellular populations}, J. Math. Biol. 72 (2015)
  123--156.

\bibitem{hoppensteadt1975}
F.~C. Hoppensteadt, {Mathematical theories of populations: demographics,
  genetics and epidemics}, Society for Industrial and Applied Mathematics,
  1975.

\bibitem{gatenby2003a}
R.~A. Gatenby, T.~L. Vincent, {Application of quantitative models from
  population biology and evolutionary game theory to tumor therapeutic
  strategies}, Mol. Cancer Ther. 2 (2003) 919--927.

\bibitem{lin2012}
Y.~T. Lin, H.~Kim, C.~R. Doering, {Features of fast living: on the weak
  selection for longevity in degenerate birth-death processes}, J. Stat. Phys.
  148 (2012) 647--663.

\bibitem{kogan2014}
O.~Kogan, M.~Khasin, B.~Meerson, D.~Schneider, C.~R. Myers, {Two-strain
  competition in quasineutral stochastic disease dynamics}, Phys. Rev. E 90
  (2014) 042149.

\bibitem{kimmel2002}
M.~Kimmel, D.~E. Axelrod, {Branching processes in Biology}, Springer-Verlag,
  New York, U.S.A., 2002.

\bibitem{hidalgo2015}
J.~Hidalgo, S.~Pigolotti, M.~A. Munoz, {Stochasticity enhances the gaining of
  bet-hedging strategies in contact-process-like dynamics}, Phys. Rev. E 91
  (2015) 032114.

\bibitem{doering2005}
C.~R. Doering, K.~V. Sargsyan, L.~M. Sander, {Extinction Times for Birth-Death
  Processes: Exact Results, Continuum Asymptotics, and the Failure of the
  Fokker--Planck Approximation}, Multiscale Model. Sim. 3 (2005) 283--299.

\bibitem{powathil2012}
G.~G. Powathil, K.~E. Gordon, L.~A. Hill, M.~A.~J. Chaplain, {Modelling the
  effects of cell-cycle heterogeneity on the response of a solid tumour to
  chemotherapy: biological insights from a hybrid multiscale cellular automaton
  model}, J. Theor. Biol. 308 (2012) 1--19.

\bibitem{gabriel2012}
P.~Gabriel, S.~P. Garbett, V.~Quaranta, D.~R. Tyson, G.~F. Webb, {The
  contribution of age structure to cell population responses to targeted
  therapeutics}, J. Theor. Biol. 311 (2012) 19--27.

\bibitem{billy2013}
F.~Billy, J.~Clairambault, {Designing proliferating cell population models with
  functional targets for control by anti-cancer drugs}, Discrete and Continuous
  Dynamical Systems - Series B 18 (2013) 865--889.

\bibitem{alarcon2010b}
T.~Alarc{\'o}n, H.~J. Jensen, {Quiescence: A mechanism for escaping the effects
  of drug on cell populations}, J. R. Soc. Interface 8 (2010) 99--106.

\bibitem{kempf2015}
H.~Kempf, M.~Bleicher, M.~Meyer-Hermann, {Spatio-temporal dynamics of hypoxia
  during radiotherapy}, PLoS One 10 (2015) e0133357.

\bibitem{anderson2006}
A.~R.~A. Anderson, A.~M. Weaver, P.~T. Cummings, V.~Quaranta, {Tumor morphology
  and phenotypic evolution driven by selective pressure from the
  microenvironment}, Cell 127 (2006) 905--915.

\bibitem{anderson2008}
A.~R.~A. Anderson, V.~Quaranta, {Integrative mathematical oncology}, Nature
  Rev. Cancer 8 (2008) 763--773.

\bibitem{kubo1973}
R.~Kubo, K.~Matsuo, K.~Kitahara, {Fluctuation and relaxation of macrovariables},
  J. Stat. Phys. 9 (1973) 51--96.

\bibitem{alarcon2007}
T.~Alarc{\'o}n, K.~M. Page, {Mathematical models of the VEGF receptor and its
  role in cancer therapy}, J. R. Soc. Interface 4 (2007) 283--304.

\bibitem{touchette2009}
H.~Touchette, {The large deviation approach to statistical mechanics}, Phys.
  Rep. 479 (2009) 1--69.

\bibitem{doi1976}
M.~Doi, {Stochastic theory of diffusion-controlled reaction}, J. Phys. A:Math.
  Gen. 9~(9) (1976) 1479.

\bibitem{peliti1985}
L.~Peliti, {Path integral approach to birth-death processes on a lattice}, J.
  Phys. France 46~(9) (1985) 1469--1483.

\bibitem{assaf2006}
M.~Assaf, B.~Meerson, {Spectral formulation and WKB approximation for
  rare-event statistics in reaction systems}, Phys. Rev. E 74 (2006) 041115.

\bibitem{assaf2010}
M.~Assaf, B.~Meerson, P.~V. Sasorov, {Large fluctuations in stochastic
  population dynamics: Momentum space calculations}, J. Stat. Mech. (2010)
  P07018.

\bibitem{kang2013}
H.-W. Kang, T.~G. Kurtz, {Separation of time-scales and model reduction for
  stochastic reaction networks}, The Annals of Applied Probability 23~(2)
  (2013) 529--583.

\bibitem{gonze2002}
D.~Gonze, J.~Halloy, P.~Gaspard, {Biochemical clocks and molecular noise:
  Theoretical study of robustness factors}, J. Chem. Phys. 116 (2002)
  10997--11010.

\bibitem{feynman2010}
R.~P. Feynman, A.~R. Hibbs, Quantum Mechanics and Path Integrals, Dover
  Publications, Mineola, NY, USA, 2010.

\bibitem{dickman2003}
R.~Dickman, R.~Vidigal, {Path integrals and perturbation theory for stochastic
  processes}, Brazilian J. Phys. 33 (2003) 73--93.

\bibitem{elgart2004}
V.~Elgart, A.~Kamenev, {Rare events in reaction-diffusion systems}, Phys. Rev.
  E 70 (2004) 041106.

\bibitem{tauber2005}
U.~C. T\mbox{\"a}uber, M.~Howard, B.~P. Vollmayr-Lee, {Applications of
  field-theoretic renormalisation group methods to reaction-diffusion
  problems}, J. Phys. A: Math. Gen. 38 (2005) R79--R131.

\bibitem{briggs1925}
G.~E. Briggs, J.~B.~S. Haldane, {A note on the kinetics of enzyme action},
  Biochem. J. 19 (1925) 338--39.

\bibitem{keener1998}
J.~Keener, J.~Sneyd, {Mathematical physiology}, Springer-Verlag, New York, NY,
  USA, 1998.

\end{thebibliography}

\end{document}